# B.R.N.O. Contributions #39
# Times of minima

Hoňková, K.[1),*], Juryšek, J.[1),5),*], Lehký M.[1),2)], Šmelcer L.[1),3)], Mašek M.[1),5)], Mazanec J.[1),9)], Hanžl D.[7)], Urbaník M.[1)], Magris M.[8)], Vrašťák M.[1),6)], Walter F.[1),17)], Hladík B.[1)], Medulka T.[10)], Bílek F.[11)], Trnka J.[4)], Jacobsen J.[12)], Benáček J.[9),13)], Kuchťák B.[1)], Audejean M.[14)], Öğmen Y.[15)], Zíbar M.[1)], Fatka P.[9)], Marchi F.[16)], Poddaný S.[1),17)], Quiñones C.[18)], Tapia L.[19)], Scaggiante F.[20)], Zardin D.[20)], Corfini G.[21)], Hájek P.[22)], Lomoz F.[1),30)], Mravik J.[23)], Grnja J.[24)], Campos F.[25)], Čaloud J.[26)], Esseiva N.[27)], Jakš S.[17)], Horník M.[17)], Filip, J.[28)], Uhlář R.[28)], Mina, F.[18)], Artola, R.[18)], Zalazar, J.[18)], Müller D.[17)], Pintr P.[29)], Divišová L.[17)]

1. Variable Star and Exoplanet Section of Czech Astronomical Society
2. Severní 765, 500 03 Hradec Králové, Czech Republic
3. Valašské Meziříčí Observatory, Vsetínská 78, 757 01 Valašské Meziříčí, Czech Republic
4. City Observatory Slaný, Nosačická 1713, 274 01 Slaný, Czech Republic
5. Institute of Physics Czech Academy of Sciences, Na Slovance 1999/2, 182 21 Praha, Czech Republic
6. Kľúčiny 457, 034 01 Liptovská Štiavnica, Slovak Republic
7. Úvoz 118, 602 00 Brno, Czech Republic
8. Ostello scout Alpe Adria, Circolo Culturale Astrofili Trieste, Trieste, Italy
9. Hvězdárna Vyškov, Kroměřížská 721, 682 01 Vyškov, Czech Republic
10. Vihorlatská hvězdáren, Mierová 4, 060 01 Humenné, Slovak Republic
11. TS Observatory, Trocnovská 1188, 374 01 Trhové Sviny, Czech Republic
12. Egeskov Observatory, Syrenvej 6, 7000 Fredericia, Denmark
13. Department of Theoretical Physics and Astrophysics, Masaryk University, Brno, Czech Republic
14. Astronomie en Chinonais Mairie 37500 Chinon, France
15. Karaoğlanoğlu Cad. 63A, Geçitkale, Magosa, North Cyprus
16. V.lo Madonnetta 35, Carbonera (TV), Italy
17. Štefánik Observatory, Petřín 205, 118 46 Praha 1, Czech Republic
18. Observatorio Astronómico Córdoba, Laprida 854, Córdoba, Argentina
19. Estación Astrofísica Bosque Alegre (EABA), Laprida 854, Córdoba, Argentina
20. Gruppo Astrofili Salese Galileo Galilei S.M di Sala, via Tovaglie 13 Noale, Venezia, Italy
21. Giorgio Corfini Observatory, v Santa Maria del Giudice 1542 55100 Lucca, Italy
22. MontePa, 683 41 Pavlovice u Vyškova, Czech Republic
23. Slobodana Jajića 16, Novi Sad 21000, Serbia
24. Maršala Tita 105, Kulpin, Serbia
25. Observatori Puig d, Passatge Bosc, 1, E-08759 Vallirana, Barcelona, Spain
26. V Dolině 211, 760 01 Zlín-Jaroslavice, Czech Republic
27. MJC des Capucins, 18 Rue des Salins, 25300 Pontarlier, France
28. Pohoří, Jílové u Prahy, Czech Republic
29. Svatováclavská 2517, 43801 Žatec, Czech Republic
30. Švermova 441, 264 01 Sedlčany, Czech Republic

*Corresponding authors, nadseni.promenari@astronomie.cz

**Abstract**
This paper presents 1463 times of minima for 455 objects acquired by 46 members and cooperating observers of the Variable Star and Exoplanet Section of the Czech Astronomical Society (B.R.N.O. Observing project). Observations were carried out between October 2013 – September 2014. Some neglected southern eclipsing binaries and newly discovered stars by the observers of project B.R.N.O. are included in the list.

## Introduction

We introduce 1463 times of minima (1429 CCD-based, 42 DSLR[1]-based) of 455 eclipsing binary objects. These timings were acquired by 46 observers from around the world during 2013 – 2014. Observers are mostly members and collaborators of the *Variable Star and Exoplanet Section of Czech Astronomical Society (VSES)* - B.R.N.O. Observing project ("Brno Regional Network of Observers" group; hereafter B.R.N.O.).

This paper contains only new, previously unpublished observations.

CCD and DSLR frames were mostly reduced by *C-MuniPack* code (Motl, 2007), the well-known adaptation of *MuniPack* code (Hroch, 1998), based on *DaoPhot* routines (Stetson, 1987, 1991).

The times of minima were determined by the method which uses phenomenological description of the eclipse shape (Brát et al., 2012; Cagaš & Pejcha, 2012; Chrastina et al., 2014). There is an on-line fitting tool implemented to minima protocols in the var2.astro.cz. The uncertainty of the minimum time (one sigma significance level, column 2 in Table 1), has been derived using bootstrapp method (Brát et al., 2012).

All observations were first submitted to the on-line protocols of B.R.N.O. project[2]. This submission is done by observers themselves. After that, each observation is visually validated by the database administrators (K. Hoňková, J. Juryšek) and only observations with well defined minimum are accepted to the on-line database. Before publication, each minimum is marked as primary or secondary one, according to O-C gateway[3] (Paschke & Brát, 2006). This paper contains only minima of variable stars with previously published light elements (VSX, OEJV, IBVS etc.).

## Description of the paper

There are two sections in this paper. Section 1 contains table with times of minimum, light elements and other related informations. In Section 2, several tables with statistics and details about new variables in CzeV (Brát, 2006) and SvkV catalogues (marked in Table 1) can be found.

The Table 1 contains the times of minima in HJD based on UTC. Each published observation can be viewed in detail following the link under HJD number. There are both light curve and reference CCD frame available. Notes made by observers can be found there as well. If orbital light elements are known (Paschke & Brát, 2006), the O-C value is also given in the Table 1. Based on recent observations, light elements in the O-C gateway are updated continuously by A. Paschke. That is why the Column 4 in Table 4 may contain out of date values (listed O-C values are related to September 2014), but the active link under these values leads to correct O-C diagrams and light elements.

[1] DSLR - Digital Single-Lens Reflex camera
[2] http://var2.astro.cz/brno/protokoly.php
[3] http://var.astro.cz/ocgate/

We would like to point out that the character of a minimum (primary/secondary) depends only on used light elements – especially for EW binaries. All light elements used for character of minima determination were taken from the O-C gateway.

In the Table 1, there are observations of neglected southern eclipsing binaries presented as well. They were obtained mainly by FRAM[4] team (M. Mašek, J. Juryšek and K. Hoňková, with technical support from Institute of Physics ASCR, v. v. i.) in Argentina.

Some of the observations were obtained by the set of equipment[5] borrowed from the VSES. In the Table 1, these observations are descripted as „Set from VSES“. The telescope, situated in Jílové u Prahy (49.895° N, 14.493° E), is remotely controled via internet. The observations with this telescope are currently avaiable for all of VSES members[6].

The 6th column of the Table 1 contains an independent indicator of the timing determination accuracy – the light curve minimum coverage factor – given in fractional form as a ratio between all acquired data points and data points covering decreasing branch of a minimum.

Section 2 contains Table 2 with list of observers with their statistics. Recently some amateur astronomers use DSLR instead of expensive CCD cameras (see last column of Table 1). There is a list of DSLR observers with their statistics in Table 4.

In addition, many new, recently discovered eclipsing binaries were observed. These objects have not final designation in GCVS (Samus et al., 2007-2013). In this paper, they are named using some preliminary names (e.g. CzeV, SvkV, when discoverers are from the Czech Republic or Slovakia), or using catalogue or project numbers (GSC, TYC, USNO, 2MASS, ASAS etc). Each of these new objects are presented in Table 3 with coordinates at equinox J2000.

In addition, Section 2 contains list of individual stars. There is a list of 5 new variable stars discovered by co-authors of this paper and submitted in CzeV or SvkV catalogues. These stars have known, previously published light elements. Their observed minima are listed in Table 1.

---

[4] FRAM - F/(Ph)otometric Robotic Atmospheric Monitor, 30-cm Czech robotic telescope at the Pierre Auger Observatory in Argentina.

[5] Newton RL 150/750 mm, HEQ5 GoTo + SBIG ST-7 + G1-0300 (since September 2014 with photometric filters BVRI)

[6] Please contact L. Šmelcer (chairman of VSES) for further details, lsmelcer@seznam.cz .

## Legend

Section 1

Section 2

Description of the Table 1

- Column 0 – objects designation. If there are any notes about presented object, it is marked with "§"

- Column 1 – HJD of observed minimum (JDhel – 2 400 000) based on coordinated universal time UTC. Five decimal places in HJD are needed for functionality of the hyperlinks to O-C gate.

- Column 2 – uncertainty of minimum time determination in days

- Column 3 – identification of primary (I) or secondary (II) minimum. Although ve give a correct minimum type, it may not correspond to that in the O-C gate in some stars. Such misinterpretation would cause ±5-day shift.

- Column 4 – O-C value (in days), which is given as URL link pointing to the O-C gateway, where figure with highlighted observation (in red color) appears.

- Column 5 – photometric band of CCD observation or DSLR method.

- Column 6 – total number of measurements/number of data on decreasing branch of light curve

- Column 7 – observer's identification

- Column 8 – equipment used for observation.

## Section 1 – Table 1 - Times of minima of eclipsing binaries

| | | | | | | | |
|---|---|---|---|---|---|---|---|
| AA And | | | | | | | |
| 56584.29300 | 0.0005 | I | -0.0035 | CCD+V | 161/97 | Magris M. | Schmidt-Newton 0.25m F/4, CCD MX 716 |
| AD And | | | | | | | |
| 56574.44119 | 0.0004 | II | +0.0015 | CCD+V | 60/42 | Šmelcer L. | Celestron 280/1765 + CCD ST7 |
| 56574.44489 | 0.0003 | II | +0.0052 | CCD+R | 68/45 | Šmelcer L. | Celestron 280/1765 + CCD ST7 |
| 56583.32000 | 0.0002 | II | +0.0046 | CCD+V | 308/227 | Magris M. | Schmidt-Newton 0.25m F/4, Atik Titan |
| | | | show all | | | | |
| BD And | | | | | | | |
| 56569.30581 | 0.0001 | II | -0.2303 | CCD+R | 91/43 | Šmelcer L. | Celestron 355/2460 + CCD G2 1600 |
| 56569.30590 | 0.0001 | II | -0.2302 | CCD+I | 103/55 | Šmelcer L. | Celestron 355/2460 + CCD G2 1600 |
| 56569.30618 | 0.0001 | II | -0.2299 | CCD+V | 71/27 | Šmelcer L. | Celestron 355/2460 + CCD G2 1600 |
| 56861.39823 | 0.0002 | I | +0.0020 | CCD+Clear | 37/11 | Šmelcer L. | Celestron 280/1765 + CCD G2 4000 |
| 56861.39845 | 0.0002 | I | +0.0022 | CCD+V | 39/12 | Šmelcer L. | Celestron 280/1765 + CCD G2 4000 |
| 56861.39880 | 0.0001 | I | +0.0026 | CCD+R | 35/10 | Šmelcer L. | Celestron 280/1765 + CCD G2 4000 |
| | | | show all | | | | |
| CN And | | | | | | | |
| 56590.25975 | 0.0003 | I | -0.0078 | CCD+V | 56/19 | Šmelcer L. | Celestron 355/2460 + CCD G2 1600 |
| 56590.26079 | 0.0003 | I | -0.0068 | CCD+R | 69/20 | Šmelcer L. | Celestron 355/2460 + CCD G2 1600 |
| 56596.27628 | 0.0002 | I | -0.0076 | CCD+R | 204/56 | Šmelcer L. | Newton 254/1200 + CCD G2 402 |
| 56596.27685 | 0.0002 | I | -0.0070 | CCD+V | 206/58 | Šmelcer L. | Newton 254/1200 + CCD G2 402 |
| 56596.27716 | 0.0002 | I | -0.0067 | CCD+B | 182/58 | Šmelcer L. | Newton 254/1200 + CCD G2 402 |
| 56596.50961 | 0.0003 | II | -0.0056 | CCD+R | 204/197 | Šmelcer L. | Newton 254/1200 + CCD G2 402 |
| 56596.51296 | 0.0010 | II | -0.0023 | CCD+V | 206/192 | Šmelcer L. | Newton 254/1200 + CCD G2 402 |
| 56605.30409 | 0.0002 | II | -0.0042 | CCD+V | 216/121 | Šmelcer L. | Newton 254/1200 + CCD G2 402 |
| 56605.30421 | 0.0003 | II | -0.0041 | CCD+R | 233/122 | Šmelcer L. | Newton 254/1200 + CCD G2 402 |
| 56616.17676 | 0.0004 | I | -0.0071 | CCD+R | 89/7 | Šmelcer L. | Newton 254/1200 + CCD G2 402 |
| 56616.17758 | 0.0002 | I | -0.0063 | CCD+V | 91/7 | Šmelcer L. | Newton 254/1200 + CCD G2 402 |
| | | | show all | | | | |
| CzeV267 And | § | | | | | | |
| 56656.37541 | 0.0010 | I | ------- | CCD+Clear | 248/19 | Zíbar M. | Newton 254/903, G2-0402 (VSES), AG(MMys) on SW 1 |
| 56656.51751 | 0.0005 | II | ------- | CCD+Clear | 248/218 | Zíbar M. | Newton 254/903, G2-0402 (VSES), AG(MMys) on SW 1 |
| 56712.37086 | 0.0006 | II | ------- | CCD+Clear | 269/159 | Zíbar M. | Newton 254/903, G2-0402 (VSES), AG(MMys) on SW 1 |
| DS And | | | | | | | |
| 56598.33909 | 0.0009 | I | +0.0020 | CCD+V | 269/129 | Magris M. | Schmidt-Newton 0.25m F/4, Atik Titan |
| EP And | | | | | | | |
| 56582.32263 | 0.0001 | I | -0.0083 | CCD+V | 307/65 | Magris M. | Schmidt-Newton 0.25m F/4, CCD Atik Titan |
| 56585.35193 | 0.0001 | II | -0.0097 | CCD+V | 106/36 | Šmelcer L. | Celestron 355/2460 + CCD G2 1600 |
| 56585.35216 | 0.0001 | II | -0.0095 | CCD+R | 107/34 | Šmelcer L. | Celestron 355/2460 + CCD G2 1600 |
| 56588.38331 | 0.0001 | I | -0.0092 | CCD+V | 55/20 | Šmelcer L. | Celestron 355/2460 + CCD G2 1600 |
| 56588.38357 | 0.0003 | I | -0.0090 | CCD+I | 45/16 | Šmelcer L. | Celestron 355/2460 + CCD G2 1600 |
| 56588.38404 | 0.0001 | I | -0.0085 | CCD+R | 52/19 | Šmelcer L. | Celestron 355/2460 + CCD G2 1600 |
| 56597.47479 | 0.0002 | II | -0.0102 | CCD+V | 111/46 | Šmelcer L. | Celestron 355/2460 + CCD G2 1600 |
| 56597.47490 | 0.0002 | II | -0.0101 | CCD+R | 116/48 | Šmelcer L. | Celestron 355/2460 + CCD G2 1600 |
| 56605.35558 | 0.0001 | I | -0.0096 | CCD+V | 109/57 | Šmelcer L. | Celestron 355/2460 + CCD G2 1600 |
| 56605.35578 | 0.0001 | I | -0.0094 | CCD+R | 111/58 | Šmelcer L. | Celestron 355/2460 + CCD G2 1600 |
| 56614.24772 | 0.0006 | I | -0.0079 | CCD+V | 36/22 | Šmelcer L. | Celestron 355/2460 + CCD G2 1600 |
| 56614.24800 | 0.0004 | I | -0.0076 | CCD+R | 34/19 | Šmelcer L. | Celestron 355/2460 + CCD G2 1600 |
| 56614.24911 | 0.0007 | I | -0.0065 | CCD+I | 31/21 | Šmelcer L. | Celestron 355/2460 + CCD G2 1600 |
| 56615.25459 | 0.0004 | II | -0.0112 | CCD+R | 82/35 | Šmelcer L. | Celestron 355/2460 + CCD G2 1600 |
| 56615.25511 | 0.0003 | II | -0.0107 | CCD+V | 76/36 | Šmelcer L. | Celestron 355/2460 + CCD G2 1600 |
| 56615.25548 | 0.0004 | II | -0.0103 | CCD+I | 81/35 | Šmelcer L. | Celestron 355/2460 + CCD G2 1600 |
| 56616.26728 | 0.0001 | I | -0.0088 | CCD+R | 60/26 | Šmelcer L. | Celestron 355/2460 + CCD G2 1600 |
| 56616.26757 | 0.0002 | I | -0.0086 | CCD+I | 57/26 | Šmelcer L. | Celestron 355/2460 + CCD G2 1600 |
| 56616.26824 | 0.0002 | I | -0.0079 | CCD+V | 57/25 | Šmelcer L. | Celestron 355/2460 + CCD G2 1600 |
| 56624.34838 | 0.0004 | I | -0.0099 | CCD+R | 67/51 | Šmelcer L. | Celestron 355/2460 + CCD G2 1600 |
| 56624.34897 | 0.0004 | I | -0.0094 | CCD+V | 90/66 | Šmelcer L. | Celestron 355/2460 + CCD G2 1600 |
| 56624.34910 | 0.0003 | I | -0.0092 | CCD+I | 60/45 | Šmelcer L. | Celestron 355/2460 + CCD G2 1600 |
| 56629.19615 | 0.0002 | I | -0.0115 | CCD+V | 107/16 | Šmelcer L. | Celestron 355/2460 + CCD G2 1600 |
| 56629.19680 | 0.0002 | I | -0.0109 | CCD+R | 109/18 | Šmelcer L. | Celestron 355/2460 + CCD G2 1600 |
| 56643.34178 | 0.0001 | I | -0.0097 | CCD+V | 224/148 | Šmelcer L. | Newton 254/1200 + CCD G2 402 |
| 56643.34194 | 0.0001 | I | -0.0096 | CCD+R | 231/150 | Šmelcer L. | Newton 254/1200 + CCD G2 402 |
| 56648.39202 | 0.0001 | II | -0.0108 | CCD+R | 222/119 | Šmelcer L. | Newton 254/1200 + CCD G2 402 |
| 56648.39225 | 0.0001 | II | -0.0106 | CCD+V | 205/125 | Šmelcer L. | Newton 254/1200 + CCD G2 402 |
| 56650.41161 | 0.0001 | II | -0.0118 | CCD+V | 223/156 | Šmelcer L. | Newton 254/1200 + CCD G2 402 |
| 56650.41211 | 0.0001 | II | -0.0113 | CCD+R | 225/153 | Šmelcer L. | Newton 254/1200 + CCD G2 402 |
| 56654.25253 | 0.0001 | I | -0.0099 | CCD+R | 124/54 | Šmelcer L. | Newton 254/1200 + CCD G2 402 |
| 56654.25276 | 0.0001 | I | -0.0097 | CCD+V | 123/54 | Šmelcer L. | Newton 254/1200 + CCD G2 402 |
| 56670.21389 | 0.0002 | II | -0.0109 | CCD+R | 234/22 | Šmelcer L. | Newton 254/1200 + CCD G2 402 |
| 56670.21426 | 0.0002 | II | -0.0105 | CCD+V | 231/23 | Šmelcer L. | Newton 254/1200 + CCD G2 402 |
| 56670.41716 | 0.0003 | I | -0.0097 | CCD+R | 234/219 | Šmelcer L. | Newton 254/1200 + CCD G2 402 |
| 56670.41964 | 0.0005 | I | -0.0072 | CCD+V | 231/221 | Šmelcer L. | Newton 254/1200 + CCD G2 402 |
| 56671.22415 | 0.0001 | I | -0.0109 | CCD+V | 143/40 | Šmelcer L. | Newton 254/1200 + CCD G2 402 |
| 56671.22490 | 0.0001 | I | -0.0102 | CCD+R | 145/43 | Šmelcer L. | Newton 254/1200 + CCD G2 402 |
| 56692.23784 | 0.0002 | I | -0.0110 | CCD+B | 70/18 | Šmelcer L. | Newton 254/1200 + CCD G2 402 |
| 56692.23820 | 0.0001 | I | -0.0106 | CCD+R | 77/23 | Šmelcer L. | Newton 254/1200 + CCD G2 402 |
| 56712.24081 | 0.0002 | II | -0.0114 | CCD+V | 82/28 | Šmelcer L. | Newton 254/1200 + CCD G2 402 |
| 56712.24092 | 0.0001 | II | -0.0113 | CCD+R | 86/25 | Šmelcer L. | Newton 254/1200 + CCD G2 402 |
| 56714.26088 | 0.0002 | II | -0.0119 | CCD+V | 57/36 | Šmelcer L. | Newton 254/1200 + CCD G2 402 |
| 56714.26157 | 0.0002 | II | -0.0112 | CCD+R | 55/39 | Šmelcer L. | Newton 254/1200 + CCD G2 402 |
| | | | show all | | | | |
| GZ And | | | | | | | |
| 56569.34796 | 0.0003 | I | -0.0038 | CCD+V | 73/36 | Magris M. | Schmidt-Newton 0.25m F/4, CCD MX 716 |
| 56569.50018 | 0.0002 | II | -0.0041 | CCD+V | 90/37 | Šmelcer L. | Newton 254/1200 + CCD G2 402 |
| 56569.50056 | 0.0001 | II | -0.0037 | CCD+R | 98/43 | Šmelcer L. | Newton 254/1200 + CCD G2 402 |
| 56632.33420 | 0.0002 | II | -0.0037 | CCD+Clear | 38/15 | J. Mravik, J. Grnja | SW 150/750 + EQ6 + SBIG ST7 |
| 56872.53170 | 0.0009 | I | -0.0077 | CCD+V | 139/54 | Magris M. | Schmidt-Newton 0.25m F/4, CCD MX 716 |
| | | | show all | | | | |
| LO And | | | | | | | |
| 56888.45076 | 0.0001 | II | +0.0020 | CCD+V | 207/109 | Magris M. | Schmidt-Newton 0.25m F/4, CCD MX 716 |
| | | | | | | | |
| SvkV020 And | § | | | | | | |
| 56703.28578 | 0.0002 | I | ------- | CCD+Clear | 73/42 | Bílek F. | Newton 0.2m f4,4 CCD Atik 314L+ |
| 56851.46088 | 0.0006 | I | ------- | CCD+R | 38/16 | Vrašťák M. | 0,24m f/5 RL+CCD G2-1600, pointer 80/400+G1-300 |
| 56851.46135 | 0.0012 | I | ------- | CCD+B | 37/16 | Vrašťák M. | 0,24m f/5 RL+CCD G2-1600, pointer 80/400+G1-300 |
| 56851.46154 | 0.0004 | I | ------- | CCD+V | 38/16 | Vrašťák M. | 0,24m f/5 RL+CCD G2-1600, pointer 80/400+G1-300 |
| 56851.46187 | 0.0003 | I | ------- | CCD+I | 36/15 | Vrašťák M. | 0,24m f/5 RL+CCD G2-1600, pointer 80/400+G1-300 |
| | | | | | | | |
| V0449 And | | | | | | | |
| 56898.33414 | 0.0003 | II | -0.0051 | CCD+R | 159/10 | Vrašťák M. | 0,24m f/5 RL+CCD G2-1600, pointer 80/400+G1-300 |
| 56898.50207 | 0.0001 | I | -0.0067 | CCD+R | 159/124 | Vrašťák M. | 0,24m f/5 RL+CCD G2-1600, pointer 80/400+G1-300 |
| | | | show all | | | | |
| V0458 And | | | | | | | |
| 56683.34004 | 0.0004 | I | +0.3006 | CCD+I | 92/52 | Lehký M. | 0.40-m f/5 + CCD G2-1600 + BVRI |
| 56683.34061 | 0.0003 | I | +0.3012 | CCD+V | 93/54 | Lehký M. | 0.40-m f/5 + CCD G2-1600 + BVRI |
| 56683.34071 | 0.0003 | I | +0.3013 | CCD+R | 86/48 | Lehký M. | 0.40-m f/5 + CCD G2-1600 + BVRI |
| | | | show all | | | | |
| V0483 And | | | | | | | |
| 56590.26533 | 0.0003 | I | -0.0211 | CCD+R | 43/22 | Šmelcer L. | Celestron 355/2460 + CCD G2 1600 |
| 56590.26791 | 0.0005 | I | -0.0185 | CCD+V | 43/23 | Šmelcer L. | Celestron 355/2460 + CCD G2 1600 |
| 56590.26817 | 0.0007 | I | -0.0183 | CCD+I | 29/21 | Šmelcer L. | Celestron 355/2460 + CCD G2 1600 |
| 56596.20616 | 0.0003 | I | -0.0216 | CCD+R | 179/13 | Šmelcer L. | Newton 254/1200 + CCD G2 402 |
| 56596.20669 | 0.0005 | I | -0.0211 | CCD+V | 187/13 | Šmelcer L. | Newton 254/1200 + CCD G2 402 |
| 56596.21169 | 0.0006 | I | -0.0161 | CCD+B | 160/14 | Šmelcer L. | Newton 254/1200 + CCD G2 402 |
| 56596.35481 | 0.0001 | II | -0.0219 | CCD+R | 179/101 | Šmelcer L. | Newton 254/1200 + CCD G2 402 |
| 56596.35542 | 0.0002 | II | -0.0213 | CCD+V | 187/103 | Šmelcer L. | Newton 254/1200 + CCD G2 402 |
| 56596.35579 | 0.0005 | II | -0.0210 | CCD+B | 160/98 | Šmelcer L. | Newton 254/1200 + CCD G2 402 |
| 56605.26592 | 0.0002 | II | -0.0228 | CCD+R | 258/94 | Šmelcer L. | Newton 254/1200 + CCD G2 402 |
| 56605.26629 | 0.0001 | II | -0.0224 | CCD+V | 259/93 | Šmelcer L. | Newton 254/1200 + CCD G2 402 |
| 56605.41374 | 0.0002 | I | -0.0231 | CCD+R | 258/235 | Šmelcer L. | Newton 254/1200 + CCD G2 402 |
| 56605.41387 | 0.0003 | I | -0.0229 | CCD+V | 259/239 | Šmelcer L. | Newton 254/1200 + CCD G2 402 |
| 56616.25473 | 0.0003 | II | -0.0255 | CCD+B | 87/55 | Šmelcer L. | Newton 254/1200 + CCD G2 402 |
| 56616.25749 | 0.0002 | II | -0.0227 | CCD+V | 95/61 | Šmelcer L. | Newton 254/1200 + CCD G2 402 |
| 56616.25787 | 0.0002 | II | -0.0223 | CCD+R | 92/61 | Šmelcer L. | Newton 254/1200 + CCD G2 402 |
| | | | show all | | | | |
| V0484 And | | | | | | | |
| 56541.35883 | 0.0010 | I | +0.0068 | CCD+Clear | 140/55 | Jacobsen J. | EQ6, 70 mm refractor, Atik 314 |
| V0488 And | | | | | | | |
| 56567.54743 | 0.0006 | I | +0.2498 | CCD+I | 179/150 | M. Lehký, P. Hajek | 0.20-m f/4 + CCD ST2000XM + I |
| V0512 And | | | | | | | |
| 56568.51470 | 0.0008 | II | +0.0727 | CCD+I | 163/37 | M. Lehký, P. Hajek | 0.20-m f/4 + CCD ST2000XM + I |
| 56569.63372 | 0.0001 | I | +0.0741 | CCD+R | 135/87 | Lehký M. | EQ6 + 0.25-m f/4 + CCD ST7 + R |
| | | | show all | | | | |
| V0518 And | | | | | | | |
| 56668.32602 | 0.0002 | I | -0.0848 | CCD+Clear | 75/45 | Mazanec J. | N400, G2 402 |
| V0527 And | | | | | | | |
| 56596.45239 | 0.0001 | I | +0.0488 | CCD+Clear | 126/74 | Mazanec J. | N400, G2 402 |
| 56610.45640 | 0.0006 | I | +0.0398 | CCD+Clear | 93/36 | Walter F. | RL \"MARK\" 40/406, SBIG ST10XME |
| 56610.46140 | 0.0004 | I | +0.0448 | CCD+I | 84/32 | Walter F. | RL \"MARK\" 40/406, SBIG ST10XME |
| 56610.46214 | 0.0002 | I | +0.0455 | CCD+V | 85/34 | Walter F. | RL \"MARK\" 40/406, SBIG ST10XME |
| | | | show all | | | | |
| V0546 And | | | | | | | |
| 56583.39739 | 0.0003 | II | -0.0026 | CCD+V | 170/95 | Magris M. | Schmidt-Newton 0.25m F/4, CCD MX 716 |
| WZ And | | | | | | | |
| 56597.34809 | 0.0004 | II | +0.0091 | CCD+V | 99/64 | Magris M. | Schmidt-Newton 0.25m F/4, Atik Titan |
| 56872.48394 | 0.0001 | I | +0.0112 | CCD+V | 182/139 | Magris M. | Schmidt-Newton 0.25m F/4, CCD MX 716 |
| | | | show all | | | | |
| V0769 Aql | | | | | | | |
| 56519.46440 | 0.0002 | I | +0.0124 | CCD+Clear | 205/97 | Trnka J. | 200/1 000, ST-9E |
| V1182 Aql | | | | | | | |
| 56853.48897 | 0.0037 | I | +0.0248 | CCD+Clear | 138/103 | Hladík B. | RF F200, ATIK 320E, CG-4 |
| 56857.53534 | 0.0036 | II | +0.0165 | CCD+Clear | 215/148 | Hladík B. | RF F200, ATIK 320E, CG-4 |
| | | | show all | | | | |
| V1315 Aql | | | | | | | |
| 56860.54108 | 0.0002 | I | +0.0010 | CCD+R | 128/70 | CCD group in Upice | 0.6-m RC f/8 + CCD Andor iXon3 888 |
| V1454 Aql | | | | | | | |
| 56824.51273 | 0.0012 | I | -0.0129 | CCD+Clear | 150/54 | Hladík B. | RF D30-F200, ATIK 320E, CG-4 |
| V1470 Aql | | | | | | | |
| 56842.46351 | 0.0011 | I | -0.0236 | CCD+Clear | 202/47 | Hladík B. | RF F200, ATIK 320E, CG-4 |
| V1490 Aql | | | | | | | |
| 56815.37638 | 0.0015 | I | -0.0391 | CCD+Clear | 253/50 | Urbaník M. | ED 80/600,0.5x Reductor, G1 0300 |

V1714 Aql

| | | | | | | | |
|---|---|---|---|---|---|---|---|
| 56460.50026 | 0.0003 | I | -0.0104 | CCD+B | 33/19 | Lehký M. | 0.40-m f/5 + CCD G2-1600 + BVRI |
| 56460.50063 | 0.0001 | I | -0.0100 | CCD+I | 35/21 | Lehký M. | 0.40-m f/5 + CCD G2-1600 + BVRI |
| 56460.50070 | 0.0002 | I | -0.0100 | CCD+V | 33/20 | Lehký M. | 0.40-m f/5 + CCD G2-1600 + BVRI |
| 56460.50072 | 0.0001 | I | -0.0100 | CCD+R | 33/18 | Lehký M. | 0.40-m f/5 + CCD G2-1600 + BVRI |
| 56476.38994 | 0.0003 | II | ------- | CCD+V | 22/10 | Lehký M. | 0.40-m f/5 + CCD G2-1600 + BVRI |
| 56476.39014 | 0.0002 | II | ------- | CCD+I | 26/11 | Lehký M. | 0.40-m f/5 + CCD G2-1600 + BVRI |
| 56476.39050 | 0.0002 | II | ------- | CCD+R | 26/11 | Lehký M. | 0.40-m f/5 + CCD G2-1600 + BVRI |
| 56476.39088 | 0.0004 | II | ------- | CCD+B | 23/11 | Lehký M. | 0.40-m f/5 + CCD G2-1600 + BVRI |
| 56596.28303 | 0.0001 | I | -0.0095 | CCD+Clear | 93/38 | Mazanec J. | N400, G2 400 |
| 56864.47471 | 0.0002 | II | ------- | CCD+I | 53/25 | Lehký M. | 0.40-m f/5 + CCD G2-1600 + BVRI |
| 56864.47482 | 0.0002 | II | ------- | CCD+V | 52/23 | Lehký M. | 0.40-m f/5 + CCD G2-1600 + BVRI |
| 56864.47525 | 0.0002 | II | ------- | CCD+R | 53/25 | Lehký M. | 0.40-m f/5 + CCD G2-1600 + BVRI |
| 56864.47658 | 0.0004 | II | ------- | CCD+B | 47/20 | Lehký M. | 0.40-m f/5 + CCD G2-1600 + BVRI |
| | | | show all | | | | |

V1747 Aql

| | | | | | | | |
|---|---|---|---|---|---|---|---|
| 56813.41046 | 0.0003 | I | +0.0076 | CCD+Clear | 214/119 | Urbaník M. | ED 80/600,0.5x Reductor, G1 0300 |

CD Aqr

| | | | | | | | |
|---|---|---|---|---|---|---|---|
| 56889.38873 | 0.0010 | I | +0.0651 | CCD+Clear | 215/132 | Urbaník M. | ED 80/600,0.5x Reductor, G1 0300 |

CW Aqr

| | | | | | | | |
|---|---|---|---|---|---|---|---|
| 56898.38503 | 0.0003 | I | -0.0102 | CCD+Clear | 151/77 | Urbaník M. | ED 80/600,0.5x Reductor, G1 0300 |

SS Ari

| | | | | | | | |
|---|---|---|---|---|---|---|---|
| 56602.32297 | 0.0003 | I | -0.0026 | CCD+Clear | 487/243 | Magris M. | Schmidt-Newton 0.25m F/4, CCD Atik Titan |

AH Aur

| | | | | | | | |
|---|---|---|---|---|---|---|---|
| 56597.49112 | 0.0011 | II | -0.0159 | CCD+V | 175/74 | Magris M. | Schmidt-Newton 0.25m F/4, CCD MX 716 |
| 56602.43559 | 0.0002 | II | -0.0125 | CCD+Clear | 390/139 | Magris M. | Schmidt-Newton 0.25m F/4, CCD MX 716 |
| 56692.36172 | 0.0002 | II | -0.0139 | CCD+R | 90/35 | Šmelcer L. | Newton 254/1200 + CCD G2 402 |
| 56692.36208 | 0.0002 | II | -0.0136 | CCD+B | 86/35 | Šmelcer L. | Newton 254/1200 + CCD G2 402 |
| 56693.34771 | 0.0004 | II | -0.0161 | CCD+B | 325/97 | Šmelcer L. | Newton 254/1200 + CCD G2 402 |
| 56693.34864 | 0.0002 | II | -0.0152 | CCD+R | 303/89 | Šmelcer L. | Newton 254/1200 + CCD G2 402 |
| 56712.37077 | 0.0002 | I | -0.0163 | DSLR | 250/68 | Walter F. | RF Comet finder 20/137, Canon 350 D |
| | | | show all | | | | |
| AP Aur | | | | | | | |
| 56583.65729 | 0.0003 | II | +0.0151 | CCD+V | 170/115 | Magris M. | Schmidt-Newton 0.25m F/4, CCD Atik Titan |
| BF Aur | | | | | | | |
| 56727.36703 | 0.0006 | II | +0.0097 | CCD+Clear | 282/203 | Urbaník M. | ED 80/600,0.5x Reductor, G1 0300 |
| CG Aur | | | | | | | |
| 56629.40341 | 0.0002 | I | +0.0018 | CCD+R | 147/53 | Šmelcer L. | Newton 254/1200 + CCD G2 402 |
| 56629.40350 | 0.0002 | I | +0.0019 | CCD+V | 158/55 | Šmelcer L. | Newton 254/1200 + CCD G2 402 |
| 56712.42950 | 0.0004 | I | +0.0046 | CCD+R | 98/65 | Šmelcer L. | Newton 254/1200 + CCD G2 402 |
| 56712.43129 | 0.0003 | I | +0.0064 | CCD+V | 90/57 | Šmelcer L. | Newton 254/1200 + CCD G2 402 |
| 56712.43159 | 0.0006 | I | +0.0067 | CCD+B | 95/63 | Šmelcer L. | Newton 254/1200 + CCD G2 402 |
| | | | show all | | | | |
| EM Aur | | | | | | | |
| 56597.42683 | 0.0011 | II | -0.0001 | CCD+V | 199/143 | Magris M. | Schmidt-Newton 0.25m F/4, Atik Titan |
| EP Aur | | | | | | | |
| 56583.48349 | 0.0006 | II | +0.0026 | CCD+V | 236/138 | Magris M. | Schmidt-Newton 0.25m F/4, CCD Atik Titan |
| GI Aur | | | | | | | |
| 56693.25794 | 0.0003 | I | +0.0044 | CCD+R | 140/65 | Lehký M. | EQ6 + 0.25-m f/4 + CCD ST7 + R |
| HL Aur | | | | | | | |
| 56241.53543 | 0.0003 | II | -0.0011 | CCD+V | 35/21 | Lehký M. | 0.40-m f/5 + CCD G2-1600 + BVRI |
| 56569.59139 | 0.0007 | II | -0.0056 | CCD+B | 29/11 | Lehký M. | 0.40-m f/5 + CCD G2-1600 + BVRI |
| 56569.59154 | 0.0004 | II | -0.0055 | CCD+V | 26/12 | Lehký M. | 0.40-m f/5 + CCD G2-1600 + BVRI |
| 56569.59159 | 0.0003 | II | -0.0054 | CCD+R | 27/12 | Lehký M. | 0.40-m f/5 + CCD G2-1600 + BVRI |
| 56569.59194 | 0.0002 | II | -0.0051 | CCD+I | 27/12 | Lehký M. | 0.40-m f/5 + CCD G2-1600 + BVRI |
| 56630.28464 | 0.0002 | I | -0.0066 | CCD+B | 26/12 | Lehký M. | 0.40-m f/5 + CCD G2-1600 + BVRI |
| 56630.28510 | 0.0001 | I | -0.0061 | CCD+I | 25/12 | Lehký M. | 0.40-m f/5 + CCD G2-1600 + BVRI |
| 56630.28523 | 0.0002 | I | -0.0060 | CCD+V | 26/12 | Lehký M. | 0.40-m f/5 + CCD G2-1600 + BVRI |
| 56630.28558 | 0.0001 | I | -0.0057 | CCD+R | 24/11 | Lehký M. | 0.40-m f/5 + CCD G2-1600 + BVRI |
| 56760.38925 | 0.0004 | I | -0.0057 | CCD+Clear | 230/132 | Urbaník M. | 120/420 G1 0300 |
| | | | show all | | | | |
| HS Aur | | | | | | | |
| 56750.44031 | 0.0004 | II | +0.0144 | CCD+Clear | 234/138 | Urbaník M. | ED 80/600,0.5x Reductor, G1 0300 |
| KU Aur | | | | | | | |
| 56608.60433 | 0.0006 | II | +0.0127 | CCD+R | 113/60 | Lehký M. | EQ6 + 0.25-m f/4 + CCD ST7 + R |
| 56700.31283 | 0.0003 | I | +0.0105 | CCD+R | 30/9 | Lehký M. | EQ6 + 0.25-m f/4 + CCD ST7 + R |
| | | | show all | | | | |
| LT Aur | | | | | | | |
| 56695.26908 | 0.0001 | I | -0.0040 | CCD+Clear | 69/34 | Bílek F. | Newton 0.2m f4,4 CCD Atik 314L+ |
| RY Aur | | | | | | | |
| 56717.31465 | 0.0017 | II | ------- | CCD+R | 216/52 | Lehký M. | EQ6 + 0.25-m f/4 + CCD ST7 + R |
| USNO-B1.0 1197-0128756 Aur | | | | | | | |
| 56639.43858 | 0.0004 | I | ------- | CCD+Clear | 147/72 | Esseiva N. | Celestron 11Edge F/D7, QSI516swg, Lodestar autoguider |
| V0410 Aur | | | | | | | |
| 56654.35582 | 0.0004 | II | +0.0336 | CCD+R | 118/24 | Lehký M. | 0.40-m f/5 + CCD G2-1600 + BVRI |
| 56654.35689 | 0.0002 | II | +0.0347 | CCD+B | 120/26 | Lehký M. | 0.40-m f/5 + CCD G2-1600 + BVRI |
| 56654.35719 | 0.0003 | II | +0.0350 | CCD+V | 120/27 | Lehký M. | 0.40-m f/5 + CCD G2-1600 + BVRI |
| 56654.35795 | 0.0003 | II | +0.0357 | CCD+I | 114/26 | Lehký M. | 0.40-m f/5 + CCD G2-1600 + BVRI |
| 56654.53880 | 0.0005 | I | +0.0332 | CCD+V | 120/87 | Lehký M. | 0.40-m f/5 + CCD G2-1600 + BVRI |
| 56654.53978 | 0.0006 | I | +0.0342 | CCD+I | 114/84 | Lehký M. | 0.40-m f/5 + CCD G2-1600 + BVRI |
| 56654.54041 | 0.0004 | I | +0.0348 | CCD+B | 120/88 | Lehký M. | 0.40-m f/5 + CCD G2-1600 + BVRI |
| 56654.54158 | 0.0005 | I | +0.0360 | CCD+R | 118/87 | Lehký M. | 0.40-m f/5 + CCD G2-1600 + BVRI |
| | | | show all | | | | |
| V0495 Aur | | | | | | | |
| 56693.51328 | 0.0002 | I | +0.0121 | CCD+V | 93/47 | Šmelcer L. | Celestron 355/2460 + CCD G2 1600 |
| 56693.51449 | 0.0001 | I | +0.0133 | CCD+R | 93/48 | Šmelcer L. | Celestron 355/2460 + CCD G2 1600 |
| | | | show all | | | | |
| V0534 Aur | | | | | | | |
| 56693.51372 | 0.0002 | I | -0.0278 | CCD+R | 323/255 | Šmelcer L. | Newton 254/1200 + CCD G2 402 |
| 56693.51463 | 0.0006 | I | -0.0269 | CCD+B | 299/246 | Šmelcer L. | Newton 254/1200 + CCD G2 402 |
| | | | show all | | | | |
| V0607 Aur | | | | | | | |
| 56693.33546 | 0.0001 | I | +0.0015 | CCD+Clear | 196/99 | Urbaník M. | ED 80/600,0.5x Reductor, G1 0300 |
| V0608 Aur | | | | | | | |
| 56728.34837 | 0.0001 | I | -0.0020 | CCD+R | 246/135 | Hanžl D. | 0.2-m RL + CCD G2 8300 (DATEL telescope) |
| V0609 Aur | | | | | | | |
| 56713.32671 | 0.0002 | I | +0.0150 | CCD+R | 300/119 | Hanžl D. | 0.2-m RL + CCD G2 8300 (DATEL telescope) |
| 56728.28215 | 0.0006 | II | ------- | CCD+R | 245/49 | Hanžl D. | 0.2-m RL + CCD G2 8300 (DATEL telescope) |
| V0612 Aur | | | | | | | |
| 56712.40546 | 0.0001 | I | +0.0030 | CCD+R | 192/88 | Hanžl D. | 0.2-m RL + CCD G2 8300 (DATEL telescope) |
| V0627 Aur | | | | | | | |
| 56709.26266 | 0.0006 | II | ------- | CCD+R | 197/43 | Hanžl D. | 0.2-m RL + CCD G2 8300 (DATEL telescope) |
| V0640 Aur | | | | | | | |
| 56542.61356 | 0.0001 | I | +0.0040 | CCD+R | 75/49 | Lehký M. | EQ6 + 0.25-m f/4 + CCD ST7 + R |
| 56610.34618 | 0.0002 | II | +0.0051 | CCD+R | 70/42 | Lehký M. | EQ6 + 0.25-m f/4 + CCD ST7 + R |
| 56630.68191 | 0.0003 | II | +0.0050 | CCD+R | 80/40 | Lehký M. | EQ6 + 0.25-m f/4 + CCD ST7 + R |
| | | | show all | | | | |
| V0641 Aur | | | | | | | |
| 56727.37164 | 0.0001 | I | -0.0018 | CCD+R | 259/158 | Hanžl D. | 0.2-m RL + CCD G2 8300 (DATEL telescope) |
| V0644 Aur | | | | | | | |
| 56666.30331 | 0.0002 | I | -0.0020 | CCD+R | 93/43 | Lehký M. | EQ6 + 0.25-m f/4 + CCD ST7 + R |
| 56728.35098 | 0.0002 | II | ------- | CCD+R | 150/81 | Lehký M. | EQ6 + 0.25-m f/4 + CCD ST7 + R |
| V0646 Aur | | | | | | | |
| 56709.55681 | 0.0005 | I | +0.1044 | CCD+Clear | 156/135 | Bílek F. | Newton 0.2m f4,4 CCD Atik 314L+ |
| 56711.52769 | 0.0003 | I | -0.1187 | CCD+Clear | 145/114 | Bílek F. | Newton 0.2m f4,4 CCD Atik 314L+ |
| 56712.40523 | 0.0002 | I | -0.1188 | CCD+Clear | 101/74 | Bílek F. | Newton 0.2m f4,4 CCD Atik 314L+ |
| 56713.28376 | 0.0008 | I | -0.1178 | CCD+Clear | 128/5 | Bílek F. | Newton 0.2m f4,4 CCD Atik 314L+ |
| 56725.34597 | 0.0003 | I | +0.0968 | CCD+Clear | 128/50 | Bílek F. | Newton 0.2m f4,4 CCD Atik 314L+ |
| | | | show all | | | | |
| VSX J062606.6+275559 Aur | | | | | | | |
| 56693.27317 | 0.0012 | I | ------- | CCD+R | 204/18 | Šmelcer L. | Newton 254/1200 + CCD G2 402 |
| 56693.43345 | 0.0009 | II | ------- | CCD+R | 204/138 | Šmelcer L. | Newton 254/1200 + CCD G2 402 |
| 56716.42288 | 0.0009 | I | ------- | CCD+R | 44/19 | Šmelcer L. | Celestron 355/2460 + CCD G2 1600 |
| WW Aur | | | | | | | |
| 56765.31046 | 0.0007 | II | +0.0018 | DSLR | 110/32 | Benáček J. | Canon 600D, Canon 24-105mm f4 |
| ZZ Aur | | | | | | | |
| 56582.53856 | 0.0001 | I | +0.0021 | CCD+V | 325/170 | Magris M. | Schmidt-Newton 0.25m F/4, CCD MX 716 |
| AQ Boo | | | | | | | |
| 56717.57179 | 0.0002 | I | -0.0060 | CCD+R | 67/34 | Lehký M. | EQ6 + 0.25-m f/4 + CCD ST7 + R |
| 56778.36897 | 0.0002 | II | -0.0062 | CCD+R | 89/54 | Lehký M. | EQ6 + 0.25-m f/4 + CCD ST7 + R |
| | | | show all | | | | |
| EF Boo | | | | | | | |
| 56745.33831 | 0.0001 | I | +0.0105 | CCD+R | 540/139 | Lehký M. | EQ6 + 0.25-m f/4 + CCD ST7 + R |
| 56745.54767 | 0.0001 | II | +0.0088 | CCD+R | 540/405 | Lehký M. | EQ6 + 0.25-m f/4 + CCD ST7 + R |
| | | | show all | | | | |
| EW Boo | | | | | | | |
| 56765.51584 | 0.0006 | I | -0.0038 | CCD+B | 47/26 | Lehký M. | 0.40-m f/5 + CCD G2-1600 + BVRI |
| 56765.51609 | 0.0004 | I | -0.0036 | CCD+V | 38/18 | Lehký M. | 0.40-m f/5 + CCD G2-1600 + BVRI |
| 56765.51837 | 0.0003 | I | -0.0013 | CCD+R | 45/24 | Lehký M. | 0.40-m f/5 + CCD G2-1600 + BVRI |
| 56765.51869 | 0.0004 | I | -0.0010 | CCD+I | 52/29 | Lehký M. | 0.40-m f/5 + CCD G2-1600 + BVRI |
| | | | show all | | | | |
| FP Boo | | | | | | | |
| 56862.41814 | 0.0005 | I | -0.1015 | CCD+R | 143/67 | Lehký M. | EQ6 + 0.25-m f/4 + CCD ST7 + R |
| GSC 02016-00444 Boo | | | | | | | |
| 56737.63157 | 0.0016 | I | ------- | CCD+R | 156/119 | Lehký M. | EQ6 + 0.25-m f/4 + CCD ST7 + R |
| 56797.46398 | 0.0016 | I | ------- | CCD+R | 166/108 | Lehký M. | EQ6 + 0.25-m f/4 + CCD ST7 + R |
| GG Boo | | | | | | | |
| 56455.44227 | 0.0008 | II | ------- | CCD+R | 225/107 | Lehký M. | EQ6 + 0.25-m f/4 + CCD ST7 + R |
| 56781.42768 | 0.0003 | I | +0.0001 | CCD+R | 200/139 | Lehký M. | EQ6 + 0.25-m f/4 + CCD ST7 + R |
| GK Boo | | | | | | | |
| 56747.58851 | 0.0001 | II | +0.0058 | CCD+R | 267/219 | Lehký M. | EQ6 + 0.25-m f/4 + CCD ST7 + R |
| 56799.42706 | 0.0001 | I | +0.0064 | CCD+R | 298/196 | Lehký M. | EQ6 + 0.25-m f/4 + CCD ST7 + R |
| 56842.42781 | 0.0002 | I | +0.0077 | CCD+Clear | 159/114 | Urbaník M. | ED 80/600,0.5x Reductor, G1 0300 |
| | | | show all | | | | |
| GM Boo | | | | | | | |
| 56737.66185 | 0.0003 | II | +0.0062 | CCD+R | 175/160 | Lehký M. | EQ6 + 0.25-m f/4 + CCD ST7 + R |
| 56797.42639 | 0.0003 | I | +0.0050 | CCD+R | 175/82 | Lehký M. | EQ6 + 0.25-m f/4 + CCD ST7 + R |
| | | | show all | | | | |
| GN Boo | | | | | | | |
| 56718.60904 | 0.0002 | I | +0.0048 | CCD+R | 165/78 | Lehký M. | EQ6 + 0.25-m f/4 + CCD ST7 + R |
| 56729.61695 | 0.0001 | II | +0.0042 | CCD+R | 164/80 | Lehký M. | EQ6 + 0.25-m f/4 + CCD ST7 + R |
| | | | show all | | | | |
| GR Boo | | | | | | | |
| 56745.44440 | 0.0002 | II | -0.0066 | CCD+R | 93/41 | Lehký M. | EQ6 + 0.25-m f/4 + CCD ST7 + R |
| 56798.36695 | 0.0002 | I | -0.0062 | CCD+R | 78/40 | Lehký M. | EQ6 + 0.25-m f/4 + CCD ST7 + R |
| | | | show all | | | | |
| GS Boo | | | | | | | |
| 56711.60853 | 0.0003 | I | -0.0124 | CCD+V | 20/8 | Lehký M. | 0.40-m f/5 + CCD G2-1600 + BVRI |
| 56711.60875 | 0.0002 | I | -0.0122 | CCD+R | 24/8 | Lehký M. | 0.40-m f/5 + CCD G2-1600 + BVRI |
| 56711.60882 | 0.0003 | I | -0.0121 | CCD+B | 24/10 | Lehký M. | 0.40-m f/5 + CCD G2-1600 + BVRI |
| 56711.60889 | 0.0003 | I | -0.0120 | CCD+I | 24/9 | Lehký M. | 0.40-m f/5 + CCD G2-1600 + BVRI |
| 56745.54204 | 0.0001 | I | -0.0129 | CCD+R | 206/80 | Hanžl D. | 0.2-m RL + CCD G2 8300 (DATEL telescope) |
| | | | show all | | | | |
| GT Boo | | | | | | | |
| 56728.46581 | 0.0005 | II | +0.0036 | CCD+R | 207/31 | Lehký M. | EQ6 + 0.25-m f/4 + CCD ST7 + R |
| GU Boo | | | | | | | |
| 56683.66694 | 0.0001 | I | -0.0055 | CCD+R | 53/35 | Lehký M. | 0.40-m f/5 + CCD G2-1600 + BVRI |
| 56683.66720 | 0.0003 | I | -0.0052 | CCD+V | 52/34 | Lehký M. | 0.40-m f/5 + CCD G2-1600 + BVRI |
| 56701.50702 | 0.0002 | II | -0.0041 | CCD+R | 45/32 | Lehký M. | 0.40-m f/5 + CCD G2-1600 + BVRI |
| 56701.50702 | 0.0002 | II | -0.0041 | CCD+V | 45/32 | Lehký M. | 0.40-m f/5 + CCD G2-1600 + BVRI |
| 56712.50095 | 0.0002 | I | -0.0066 | CCD+V | 44/32 | Lehký M. | 0.40-m f/5 + CCD G2-1600 + BVRI |
| 56712.50200 | 0.0001 | I | -0.0055 | CCD+R | 45/32 | Lehký M. | 0.40-m f/5 + CCD G2-1600 + BVRI |
| 56730.58495 | 0.0001 | I | -0.0056 | CCD+R | 193/95 | Hanžl D. | 0.2-m RL + CCD G2 8300 (DATEL telescope) |
| 56819.53274 | 0.0002 | I | -0.0066 | CCD+R | 197/180 | Šmelcer L. | Newton 254/1200 + CCD G2 402 |
| 56845.43685 | 0.0001 | I | -0.0052 | CCD+R | 238/128 | Šmelcer L. | Newton 254/1200 + CCD G2 402 |
| | | | show all | | | | |
| GV Boo | | | | | | | |
| 56840.42076 | 0.0003 | II | -0.0307 | CCD+R | 143/76 | Lehký M. | EQ6 + 0.25-m f/4 + CCD ST7 + R |
| 56845.39149 | 0.0004 | I | -0.0262 | CCD+R | 100/53 | Lehký M. | EQ6 + 0.25-m f/4 + CCD ST7 + R |
| | | | show all | | | | |
| IL Boo | | | | | | | |
| 56808.43145 | 0.0002 | II | -0.0001 | CCD+R | 77/33 | Mazanec J. | N400, G2 402 |
| 56808.43224 | 0.0002 | II | +0.0006 | CCD+Clear | 80/36 | Mazanec J. | N400, G2 402 |
| 56808.43245 | 0.0003 | II | +0.0009 | CCD+I | 80/36 | Mazanec J. | N400, G2 402 |
| | | | show all | | | | |
| IN Boo | | | | | | | |
| 56816.42330 | 0.0002 | I | +0.0081 | CCD+I | 254/56 | Mazanec J. | N400, G2 402 |
| 56816.42340 | 0.0001 | I | +0.0082 | CCD+R | 257/55 | Mazanec J. | N400, G2 402 |
| 56816.56511 | 0.0006 | II | +0.0069 | CCD+I | 254/232 | Mazanec J. | N400, G2 402 |
| 56816.56604 | 0.0005 | II | +0.0078 | CCD+R | 257/233 | Mazanec J. | N400, G2 402 |
| | | | show all | | | | |
| IX Boo | | | | | | | |
| 56738.34169 | 0.0003 | I | -0.2062 | CCD+Clear | 111/58 | Jacobsen J. | 400 mm f/5 Newton, Starlight SXVR-H16 |
| KK Boo | | | | | | | |
| 56713.50106 | 0.0007 | II | ------- | CCD+R | 122/12 | Hanžl D. | 0.2-m RL + CCD G2 8300 (DATEL telescope) |
| 56713.64020 | 0.0008 | I | +0.0741 | CCD+R | 122/106 | Hanžl D. | 0.2-m RL + CCD G2 8300 (DATEL telescope) |
| 56745.46179 | 0.0008 | I | +0.0748 | CCD+R | 105/83 | Marchi F. | Newton 250 mm F5, FLI Kaf 260, UBVRI |
| 56749.36889 | 0.0011 | I | +0.0741 | CCD+R | 47/21 | Marchi F. | Newton 250 mm F5, FLI Kaf 260, UBVRI |
| 56750.48450 | 0.0005 | I | +0.0732 | CCD+V | 98/44 | Lehký M. | 0.40-m f/5 + CCD G2-1600 + BVRI |
| 56750.48646 | 0.0003 | I | +0.0752 | CCD+R | 110/48 | Lehký M. | 0.40-m f/5 + CCD G2-1600 + BVRI |
| 56750.62301 | 0.0008 | II | ------- | CCD+R | 110/100 | Lehký M. | 0.40-m f/5 + CCD G2-1600 + BVRI |
| 56750.62387 | 0.0011 | II | ------- | CCD+V | 98/94 | Lehký M. | 0.40-m f/5 + CCD G2-1600 + BVRI |
| 56761.37350 | 0.0006 | I | +0.0761 | CCD+R | 43/26 | Marchi F. | Newton 250 mm F5, FLI Kaf 260, UBVRI |
| 56798.49728 | 0.0004 | I | +0.0756 | CCD+R | 161/101 | Hanžl D. | 0.2-m RL + CCD G2 8300 (DATEL telescope) |
| | | | show all | | | | |
| KW Boo | | | | | | | |
| 56717.56083 | 0.0008 | II | -0.0885 | CCD+R | 138/70 | Hanžl D. | 0.2-m RL + CCD G2 8300 (DATEL telescope) |
| 56797.44978 | 0.0006 | II | -0.0609 | CCD+R | 330/150 | Hanžl D. | 0.2-m RL + CCD G2 8300 (DATEL telescope) |
| | | | show all | | | | |
| MQ Boo | | | | | | | |
| 56798.51620 | 0.0001 | I | +0.0071 | CCD+R | 149/57 | Mazanec J. | N400, G2 402 |
| 56798.51644 | 0.0002 | I | +0.0073 | CCD+I | 149/58 | Mazanec J. | N400, G2 402 |
| | | | show all | | | | |
| MT Boo | | | | | | | |
| 56824.42812 | 0.0003 | I | -0.0347 | CCD+Clear | 126/36 | Audejean M. | 0.32-m f/6 Newtonian reflector + CCD |
| MW Boo | | | | | | | |
| 56765.35061 | 0.0004 | I | -0.0199 | CCD+R | 352/73 | Hanžl D. | 0.2-m RL + CCD G2 8300 (DATEL telescope) |
| 56765.53016 | 0.0004 | II | ------- | CCD+R | 352/298 | Hanžl D. | 0.2-m RL + CCD G2 8300 (DATEL telescope) |
| NT Boo | | | | | | | |
| 56726.60232 | 0.0004 | I | +0.0604 | CCD+R | 217/159 | Hanžl D. | 0.2-m RL + CCD G2 8300 (DATEL telescope) |
| NX Boo | | | | | | | |
| 56745.55978 | 0.0005 | II | -0.0036 | CCD+R | 126/70 | Hanžl D. | 0.2-m RL + CCD G2 8300 (DATEL telescope) |
| 56760.37572 | 0.0003 | II | -0.0046 | CCD+Clear | 252/33 | Mazanec J. | N400, G2 402 |
| 56760.37736 | 0.0003 | II | -0.0030 | CCD+R | 255/37 | Mazanec J. | N400, G2 402 |
| 56760.50160 | 0.0001 | I | -0.0039 | CCD+R | 255/214 | Mazanec J. | N400, G2 402 |
| 56760.50161 | 0.0002 | I | -0.0038 | CCD+Clear | 252/212 | Mazanec J. | N400, G2 402 |
| 56764.39841 | 0.0004 | II | -0.0001 | CCD+I | 253/105 | Mazanec J. | N400, G2 402 |
| 56764.40457 | 0.0005 | II | +0.0061 | CCD+R | 238/103 | Mazanec J. | N400, G2 402 |
| 56798.42326 | 0.0002 | I | -0.0036 | CCD+I | 119/69 | Mazanec J. | N400, G2 402 |
| 56798.42361 | 0.0001 | I | -0.0032 | CCD+R | 148/100 | Mazanec J. | N400, G2 402 |
| 56815.37520 | 0.0002 | II | -0.0037 | CCD+R | 168/33 | Mazanec J. | N400, G2 402 |
| 56815.37558 | 0.0002 | II | -0.0033 | CCD+V | 166/31 | Mazanec J. | N400, G2 402 |
| 56815.37590 | 0.0002 | II | -0.0030 | CCD+I | 168/34 | Mazanec J. | N400, G2 402 |
| 56815.50072 | 0.0002 | I | -0.0033 | CCD+I | 168/135 | Mazanec J. | N400, G2 402 |
| 56815.50081 | 0.0002 | I | -0.0032 | CCD+R | 168/134 | Mazanec J. | N400, G2 402 |
| 56815.50135 | 0.0002 | I | -0.0027 | CCD+V | 166/132 | Mazanec J. | N400, G2 402 |
| | | | show all | | | | |
| NY Boo | | | | | | | |
| 56737.64068 | 0.0002 | I | +0.0641 | CCD+R | 208/155 | Hanžl D. | 0.2-m RL + CCD G2 8300 (DATEL telescope) |
| PS Boo | | | | | | | |
| 56725.55369 | 0.0004 | I | +0.0500 | CCD+R | 212/86 | Hanžl D. | 0.2-m RL + CCD G2 8300 (DATEL telescope) |
| PV Boo | | | | | | | |
| 56730.56097 | 0.0003 | II | ------- | CCD+R | 191/64 | Hanžl D. | 0.2-m RL + CCD G2 8300 (DATEL telescope) |
| PY Boo | | | | | | | |
| 56796.44009 | 0.0001 | I | -0.0014 | CCD+Clear | 120/49 | Audejean M. | 0.32-m f/6 Newtonian reflector + CCD |
| 56827.44468 | 0.0001 | II | +0.0008 | CCD+Clear | 130/55 | Audejean M. | 0.32-m f/6 Newtonian reflector + CCD |
| | | | show all | | | | |
| PZ Boo | | | | | | | |
| 56727.50018 | 0.0003 | I | +0.0032 | CCD+I | 175/79 | Mazanec J. | N400, G2 402 |

QV Boo

| | | | | | | | |
|---|---|---|---|---|---|---|---|
| 56799.35039 | 0.0012 | II | ------- | CCD+R | 109/12 | Hanžl D. | 0.2-m RL + CCD G2 8300 (DATEL telescope) |
| 56799.51297 | 0.0010 | I | -0.0202 | CCD+R | 109/83 | Hanžl D. | 0.2-m RL + CCD G2 8300 (DATEL telescope) |

QW Boo

| | | | | | | | |
|---|---|---|---|---|---|---|---|
| 56709.62845 | 0.0002 | I | +0.0211 | CCD+R | 65/37 | Lehký M. | 0.40-m f/5 + CCD G2-1600 + BVRI |
| 56709.62875 | 0.0002 | I | +0.0214 | CCD+I | 67/39 | Lehký M. | 0.40-m f/5 + CCD G2-1600 + BVRI |
| 56709.62895 | 0.0003 | I | +0.0216 | CCD+V | 64/38 | Lehký M. | 0.40-m f/5 + CCD G2-1600 + BVRI |
| 56725.48288 | 0.0003 | II | ------- | CCD+R | 95/22 | Lehký M. | 0.40-m f/5 + CCD G2-1600 + BVRI |
| 56725.48291 | 0.0003 | II | ------- | CCD+I | 96/22 | Lehký M. | 0.40-m f/5 + CCD G2-1600 + BVRI |
| 56725.48330 | 0.0003 | II | ------- | CCD+V | 88/19 | Lehký M. | 0.40-m f/5 + CCD G2-1600 + BVRI |
| 56725.62510 | 0.0003 | I | +0.0199 | CCD+R | 95/74 | Lehký M. | 0.40-m f/5 + CCD G2-1600 + BVRI |
| 56725.62618 | 0.0004 | I | +0.0210 | CCD+V | 88/69 | Lehký M. | 0.40-m f/5 + CCD G2-1600 + BVRI |
| 56725.62631 | 0.0002 | I | +0.0211 | CCD+I | 96/77 | Lehký M. | 0.40-m f/5 + CCD G2-1600 + BVRI |
| 56728.38999 | 0.0007 | II | ------- | CCD+V | 113/8 | Lehký M. | 0.40-m f/5 + CCD G2-1600 + BVRI |
| 56728.39009 | 0.0004 | II | ------- | CCD+I | 111/6 | Lehký M. | 0.40-m f/5 + CCD G2-1600 + BVRI |
| 56728.39268 | 0.0003 | II | ------- | CCD+R | 110/8 | Lehký M. | 0.40-m f/5 + CCD G2-1600 + BVRI |
| 56728.53510 | 0.0002 | I | +0.0212 | CCD+I | 111/59 | Lehký M. | 0.40-m f/5 + CCD G2-1600 + BVRI |
| 56728.53514 | 0.0003 | I | +0.0212 | CCD+R | 110/63 | Lehký M. | 0.40-m f/5 + CCD G2-1600 + BVRI |
| 56728.53519 | 0.0002 | I | +0.0213 | CCD+V | 113/65 | Lehký M. | 0.40-m f/5 + CCD G2-1600 + BVRI |
| 56729.40726 | 0.0002 | I | +0.0208 | CCD+R | 115/17 | Lehký M. | 0.40-m f/5 + CCD G2-1600 + BVRI |
| 56729.40848 | 0.0003 | I | +0.0220 | CCD+I | 112/14 | Lehký M. | 0.40-m f/5 + CCD G2-1600 + BVRI |
| 56729.40852 | 0.0003 | I | +0.0220 | CCD+V | 109/17 | Lehký M. | 0.40-m f/5 + CCD G2-1600 + BVRI |
| 56729.55457 | 0.0002 | II | ------- | CCD+I | 112/67 | Lehký M. | 0.40-m f/5 + CCD G2-1600 + BVRI |
| 56729.55479 | 0.0002 | II | ------- | CCD+V | 109/67 | Lehký M. | 0.40-m f/5 + CCD G2-1600 + BVRI |
| 56729.55509 | 0.0003 | II | ------- | CCD+R | 115/69 | Lehký M. | 0.40-m f/5 + CCD G2-1600 + BVRI |
| 56730.42670 | 0.0003 | II | ------- | CCD+R | 110/19 | Lehký M. | 0.40-m f/5 + CCD G2-1600 + BVRI |
| 56730.42695 | 0.0003 | II | ------- | CCD+I | 112/18 | Lehký M. | 0.40-m f/5 + CCD G2-1600 + BVRI |
| 56730.42722 | 0.0003 | II | ------- | CCD+V | 113/19 | Lehký M. | 0.40-m f/5 + CCD G2-1600 + BVRI |
| 56730.57116 | 0.0002 | I | +0.0212 | CCD+I | 112/73 | Lehký M. | 0.40-m f/5 + CCD G2-1600 + BVRI |
| 56730.57165 | 0.0003 | I | +0.0217 | CCD+V | 113/73 | Lehký M. | 0.40-m f/5 + CCD G2-1600 + BVRI |
| 56730.57176 | 0.0002 | I | +0.0218 | CCD+R | 110/74 | Lehký M. | 0.40-m f/5 + CCD G2-1600 + BVRI |
| 56746.42589 | 0.0003 | II | ------- | CCD+I | 115/30 | Lehký M. | 0.40-m f/5 + CCD G2-1600 + BVRI |
| 56746.42604 | 0.0003 | II | ------- | CCD+R | 118/30 | Lehký M. | 0.40-m f/5 + CCD G2-1600 + BVRI |
| 56746.42659 | 0.0004 | II | ------- | CCD+V | 115/32 | Lehký M. | 0.40-m f/5 + CCD G2-1600 + BVRI |
| 56746.56742 | 0.0003 | I | +0.0196 | CCD+I | 115/82 | Lehký M. | 0.40-m f/5 + CCD G2-1600 + BVRI |
| 56746.56835 | 0.0002 | I | +0.0205 | CCD+V | 115/85 | Lehký M. | 0.40-m f/5 + CCD G2-1600 + BVRI |
| 56746.56940 | 0.0002 | I | +0.0216 | CCD+R | 118/86 | Lehký M. | 0.40-m f/5 + CCD G2-1600 + BVRI |
| 56747.44151 | 0.0003 | I | +0.0211 | CCD+I | 116/36 | Lehký M. | 0.40-m f/5 + CCD G2-1600 + BVRI |
| 56747.44194 | 0.0002 | I | +0.0215 | CCD+V | 106/34 | Lehký M. | 0.40-m f/5 + CCD G2-1600 + BVRI |
| 56747.44214 | 0.0003 | I | +0.0217 | CCD+R | 114/35 | Lehký M. | 0.40-m f/5 + CCD G2-1600 + BVRI |
| 56747.58972 | 0.0005 | II | ------- | CCD+I | 116/92 | Lehký M. | 0.40-m f/5 + CCD G2-1600 + BVRI |
| 56747.58996 | 0.0005 | II | ------- | CCD+V | 106/84 | Lehký M. | 0.40-m f/5 + CCD G2-1600 + BVRI |
| 56747.59057 | 0.0004 | II | ------- | CCD+R | 114/91 | Lehký M. | 0.40-m f/5 + CCD G2-1600 + BVRI |
| 56765.33142 | 0.0005 | II | ------- | CCD+R | 245/15 | Mazanec J. | N400, G2 402 |
| 56765.33218 | 0.0006 | II | ------- | CCD+I | 259/18 | Mazanec J. | N400, G2 402 |
| 56765.47523 | 0.0001 | I | +0.0209 | CCD+R | 245/168 | Mazanec J. | N400, G2 402 |
| 56765.47538 | 0.0001 | I | +0.0210 | CCD+I | 259/183 | Mazanec J. | N400, G2 402 |
| 56778.41996 | 0.0002 | I | -0.1236 | CCD+R | 417/284 | Hanžl D. | 0.2-m RL + CCD G2 8300 (DATEL telescope) |
| 56798.49067 | 0.0003 | II | ------- | CCD+I | 80/52 | Lehký M. | 0.40-m f/5 + CCD G2-1600 + BVRI |
| 56798.49137 | 0.0003 | II | ------- | CCD+V | 85/57 | Lehký M. | 0.40-m f/5 + CCD G2-1600 + BVRI |
| 56798.49171 | 0.0003 | II | ------- | CCD+R | 85/57 | Lehký M. | 0.40-m f/5 + CCD G2-1600 + BVRI |
| 56799.36321 | 0.0005 | II | ------- | CCD+V | 78/6 | Lehký M. | 0.40-m f/5 + CCD G2-1600 + BVRI |
| 56799.36326 | 0.0004 | II | ------- | CCD+R | 77/6 | Lehký M. | 0.40-m f/5 + CCD G2-1600 + BVRI |
| 56799.36326 | 0.0006 | II | ------- | CCD+I | 75/6 | Lehký M. | 0.40-m f/5 + CCD G2-1600 + BVRI |
| 56799.50721 | 0.0003 | I | +0.0210 | CCD+R | 77/56 | Lehký M. | 0.40-m f/5 + CCD G2-1600 + BVRI |
| 56799.50729 | 0.0003 | I | +0.0211 | CCD+V | 78/56 | Lehký M. | 0.40-m f/5 + CCD G2-1600 + BVRI |
| 56799.50735 | 0.0003 | I | +0.0212 | CCD+I | 75/55 | Lehký M. | 0.40-m f/5 + CCD G2-1600 + BVRI |
| 56815.50538 | 0.0001 | I | +0.0214 | CCD+R | 66/53 | Lehký M. | 0.40-m f/5 + CCD G2-1600 + BVRI |
| 56815.50553 | 0.0002 | I | +0.0215 | CCD+I | 66/53 | Lehký M. | 0.40-m f/5 + CCD G2-1600 + BVRI |
| 56815.50567 | 0.0003 | I | +0.0217 | CCD+V | 69/57 | Lehký M. | 0.40-m f/5 + CCD G2-1600 + BVRI |
| 56818.41403 | 0.0001 | I | +0.0213 | CCD+V | 260/82 | Mazanec J. | N400, G2 402 |
| 56818.41406 | 0.0002 | I | +0.0213 | CCD+R | 263/87 | Mazanec J. | N400, G2 402 |
| 56818.41454 | 0.0002 | I | +0.0218 | CCD+I | 256/85 | Mazanec J. | N400, G2 402 |
| 56818.56027 | 0.0003 | II | ------- | CCD+V | 260/244 | Mazanec J. | N400, G2 402 |
| | | | show all | | | | |

QX Boo

| | | | | | | | |
|---|---|---|---|---|---|---|---|
| 56777.39534 | 0.0003 | I | -0.0014 | CCD+Clear | 206/57 | Mazanec J. | N400, G2 402 |
| 56777.39536 | 0.0003 | I | -0.0014 | CCD+I | 222/61 | Mazanec J. | N400, G2 402 |
| 56777.39554 | 0.0003 | I | -0.0012 | CCD+R | 215/62 | Mazanec J. | N400, G2 402 |
| 56777.57394 | 0.0003 | II | -0.0028 | CCD+I | 222/197 | Mazanec J. | N400, G2 402 |
| 56777.57463 | 0.0005 | II | -0.0021 | CCD+Clear | 206/186 | Mazanec J. | N400, G2 402 |
| 56777.57478 | 0.0001 | II | -0.0020 | CCD+R | 215/193 | Mazanec J. | N400, G2 402 |
| 56803.34769 | 0.0005 | I | -0.0028 | CCD+I | 250/19 | Mazanec J. | N400, G2 402 |
| 56803.34817 | 0.0005 | I | -0.0023 | CCD+Clear | 251/19 | Mazanec J. | N400, G2 402 |
| 56803.34898 | 0.0003 | I | -0.0015 | CCD+R | 266/23 | Mazanec J. | N400, G2 402 |
| 56803.52917 | 0.0002 | II | -0.0013 | CCD+I | 250/204 | Mazanec J. | N400, G2 402 |
| 56803.52952 | 0.0001 | II | -0.0009 | CCD+R | 266/219 | Mazanec J. | N400, G2 402 |
| 56803.52956 | 0.0002 | II | -0.0009 | CCD+Clear | 251/206 | Mazanec J. | N400, G2 402 |
| 56817.40714 | 0.0001 | I | -0.0016 | CCD+I | 77/27 | Fatka P. | Newton 400, G2-0402 |
| 56817.40721 | 0.0002 | I | -0.0015 | CCD+V | 77/27 | Fatka P. | Newton 400, G2-0402 |
| 56817.40727 | 0.0002 | I | -0.0014 | CCD+R | 78/27 | Fatka P. | Newton 400, G2-0402 |
| 56823.53408 | 0.0002 | I | -0.0026 | CCD+B | 77/60 | Mazanec J. | N400, G2 402 |
| 56823.53464 | 0.0002 | I | -0.0020 | CCD+R | 77/61 | Mazanec J. | N400, G2 402 |
| 56823.53469 | 0.0001 | I | -0.0020 | CCD+V | 77/61 | Mazanec J. | N400, G2 402 |
| 56823.53486 | 0.0001 | I | -0.0018 | CCD+I | 73/56 | Mazanec J. | N400, G2 402 |
| | | | show all | | | | |
| QY Boo | | | | | | | |
| 56737.45864 | 0.0012 | I | +0.0026 | CCD+R | 84/9 | Vrašťák M. | 0,24m f/5 RL+CCD G2-1600, pointer 80/400+G1-300 |
| 56737.45894 | 0.0016 | I | +0.0029 | CCD+V | 84/9 | Vrašťák M. | 0,24m f/5 RL+CCD G2-1600, pointer 80/400+G1-300 |
| 56737.46009 | 0.0020 | I | +0.0041 | CCD+I | 87/11 | Vrašťák M. | 0,24m f/5 RL+CCD G2-1600, pointer 80/400+G1-300 |
| 56737.63199 | 0.0007 | II | ------- | CCD+I | 87/74 | Vrašťák M. | 0,24m f/5 RL+CCD G2-1600, pointer 80/400+G1-300 |
| 56737.63444 | 0.0012 | II | ------- | CCD+V | 84/72 | Vrašťák M. | 0,24m f/5 RL+CCD G2-1600, pointer 80/400+G1-300 |
| 56737.63567 | 0.0007 | II | ------- | CCD+R | 84/72 | Vrašťák M. | 0,24m f/5 RL+CCD G2-1600, pointer 80/400+G1-300 |
| | | | show all | | | | |
| SU Boo | | | | | | | |
| 56711.60003 | 0.0002 | I | +0.0120 | CCD+R | 101/50 | Lehký M. | EQ6 + 0.25-m f/4 + CCD ST7 + R |
| 56729.54730 | 0.0007 | II | +0.0056 | CCD+R | 309/134 | Hanžl D. | 0.2-m RL + CCD G2 8300 (DATEL telescope) |
| | | | show all | | | | |
| SY Boo | | | | | | | |
| 56712.61780 | 0.0006 | II | -0.0114 | CCD+R | 183/83 | Lehký M. | EQ6 + 0.25-m f/4 + CCD ST7 + R |
| TZ Boo | | | | | | | |
| 56819.41595 | 0.0007 | II | -0.0067 | CCD+Clear | 96/35 | Šmelcer L. | Celestron 280/1765 + CCD G2 4000 |
| 56819.41859 | 0.0010 | II | -0.0041 | CCD+R | 78/19 | Šmelcer L. | Celestron 280/1765 + CCD G2 4000 |
| 56819.41871 | 0.0008 | II | -0.0040 | CCD+V | 50/15 | Šmelcer L. | Celestron 280/1765 + CCD G2 4000 |
| | | | show all | | | | |
| AO Cam | | | | | | | |
| 56692.33494 | 0.0001 | II | -0.0097 | CCD+B | 17/7 | Lehký M. | 0.40-m f/5 + CCD G2-1600 + BVRI |
| 56692.33506 | 0.0002 | II | -0.0095 | CCD+I | 17/7 | Lehký M. | 0.40-m f/5 + CCD G2-1600 + BVRI |
| 56692.33539 | 0.0001 | II | -0.0092 | CCD+V | 18/7 | Lehký M. | 0.40-m f/5 + CCD G2-1600 + BVRI |
| 56692.33586 | 0.0001 | II | -0.0087 | CCD+R | 17/7 | Lehký M. | 0.40-m f/5 + CCD G2-1600 + BVRI |
| 56714.27224 | 0.0004 | I | -0.0109 | CCD+B | 17/7 | Lehký M. | 0.40-m f/5 + CCD G2-1600 + BVRI |
| 56714.27267 | 0.0002 | I | -0.0105 | CCD+R | 17/7 | Lehký M. | 0.40-m f/5 + CCD G2-1600 + BVRI |
| 56714.27283 | 0.0002 | I | -0.0103 | CCD+V | 17/7 | Lehký M. | 0.40-m f/5 + CCD G2-1600 + BVRI |
| 56714.27295 | 0.0002 | I | -0.0102 | CCD+I | 17/7 | Lehký M. | 0.40-m f/5 + CCD G2-1600 + BVRI |
| | | | show all | | | | |
| AW Cam | | | | | | | |
| 56749.38931 | 0.0003 | I | -0.0035 | CCD+Clear | 263/136 | Urbaník M. | ED 80/600,0.5x Reductor, G1 0300 |
| LR Cam | | | | | | | |
| 56746.36676 | 0.0004 | I | +0.0093 | CCD+Clear | 846/251 | Kuchťák B. | Sonnar 180 + G1 |
| 56746.36739 | 0.0003 | I | +0.0099 | CCD+Clear | 854/250 | Kuchťák B. | Sonnar 180 + G1 |
| 56746.58556 | 0.0007 | II | +0.0110 | CCD+Clear | 854/672 | Kuchťák B. | Sonnar 180 + G1 |
| | | | show all | | | | |
| NR Cam | | | | | | | |
| 56747.38747 | 0.0008 | I | -0.0027 | CCD+Clear | 635/213 | Kuchťák B. | Sonnar 180 + G1 |
| 56747.51374 | 0.0006 | II | -0.0045 | CCD+Clear | 635/478 | Kuchťák B. | Sonnar 180 + G1 |
| | | | show all | | | | |

NSVS 01031772 Cam

| | | | | | | | |
|---|---|---|---|---|---|---|---|
| 56684.35000 | 0.0001 | II | ------- | CCD+R | 248/99 | Šmelcer L. | Celestron 355/2460 + CCD G2 1600 |
| 56684.35062 | 0.0002 | II | ------- | CCD+V | 235/100 | Šmelcer L. | Celestron 355/2460 + CCD G2 1600 |
| 56684.53429 | 0.0001 | I | ------- | CCD+V | 235/214 | Šmelcer L. | Celestron 355/2460 + CCD G2 1600 |
| 56684.53435 | 0.0001 | I | ------- | CCD+R | 248/217 | Šmelcer L. | Celestron 355/2460 + CCD G2 1600 |
| 56728.34286 | 0.0001 | I | ------- | CCD+V | 163/25 | Šmelcer L. | Celestron 355/2460 + CCD G2 1600 |
| 56728.34296 | 0.0001 | I | ------- | CCD+R | 145/26 | Šmelcer L. | Celestron 355/2460 + CCD G2 1600 |
| 56728.52726 | 0.0001 | II | ------- | CCD+R | 145/107 | Šmelcer L. | Celestron 355/2460 + CCD G2 1600 |
| 56728.52732 | 0.0002 | II | ------- | CCD+V | 163/126 | Šmelcer L. | Celestron 355/2460 + CCD G2 1600 |
| 56742.33191 | 0.0001 | I | ------- | CCD+V | 112/24 | Šmelcer L. | Celestron 355/2460 + CCD G2 1600 |
| 56742.33220 | 0.0001 | I | ------- | CCD+R | 110/23 | Šmelcer L. | Celestron 355/2460 + CCD G2 1600 |
| 56742.33221 | 0.0002 | I | ------- | CCD+I | 107/22 | Šmelcer L. | Celestron 355/2460 + CCD G2 1600 |
| 56742.51628 | 0.0002 | II | ------- | CCD+I | 107/98 | Šmelcer L. | Celestron 355/2460 + CCD G2 1600 |
| 56742.51634 | 0.0002 | II | ------- | CCD+V | 112/104 | Šmelcer L. | Celestron 355/2460 + CCD G2 1600 |
| 56742.51668 | 0.0002 | II | ------- | CCD+R | 110/102 | Šmelcer L. | Celestron 355/2460 + CCD G2 1600 |
| 56745.46143 | 0.0001 | II | ------- | CCD+R | 34/20 | Šmelcer L. | Celestron 355/2460 + CCD G2 1600 |
| 56745.46157 | 0.0001 | II | ------- | CCD+I | 33/20 | Šmelcer L. | Celestron 355/2460 + CCD G2 1600 |
| 56746.38180 | 0.0002 | I | ------- | CCD+I | 172/75 | Šmelcer L. | Celestron 355/2460 + CCD G2 1600 |
| 56746.38197 | 0.0001 | I | ------- | CCD+R | 183/79 | Šmelcer L. | Celestron 355/2460 + CCD G2 1600 |
| 56749.32668 | 0.0001 | I | ------- | CCD+I | 150/8 | Šmelcer L. | Celestron 355/2460 + CCD G2 1600 |
| 56749.32682 | 0.0001 | I | ------- | CCD+R | 159/7 | Šmelcer L. | Celestron 355/2460 + CCD G2 1600 |
| 56749.51104 | 0.0001 | II | ------- | CCD+R | 159/128 | Šmelcer L. | Celestron 355/2460 + CCD G2 1600 |
| 56749.51141 | 0.0002 | II | ------- | CCD+I | 150/126 | Šmelcer L. | Celestron 355/2460 + CCD G2 1600 |
| 56750.43101 | 0.0001 | I | ------- | CCD+R | 145/87 | Šmelcer L. | Celestron 355/2460 + CCD G2 1600 |
| 56750.43178 | 0.0001 | I | ------- | CCD+I | 137/80 | Šmelcer L. | Celestron 355/2460 + CCD G2 1600 |
| 56751.35159 | 0.0001 | II | ------- | CCD+R | 119/52 | Šmelcer L. | Celestron 355/2460 + CCD G2 1600 |
| 56751.35218 | 0.0002 | II | ------- | CCD+I | 106/48 | Šmelcer L. | Celestron 355/2460 + CCD G2 1600 |
| 56755.40131 | 0.0001 | II | ------- | CCD+I | 104/71 | Šmelcer L. | Celestron 355/2460 + CCD G2 1600 |
| 56755.40148 | 0.0001 | II | ------- | CCD+R | 106/74 | Šmelcer L. | Celestron 355/2460 + CCD G2 1600 |
| 56765.34093 | 0.0001 | II | ------- | CCD+R | 174/34 | Šmelcer L. | Celestron 355/2460 + CCD G2 1600 |
| 56765.34102 | 0.0001 | II | ------- | CCD+I | 169/34 | Šmelcer L. | Celestron 355/2460 + CCD G2 1600 |
| 56765.52518 | 0.0001 | I | ------- | CCD+R | 174/148 | Šmelcer L. | Celestron 355/2460 + CCD G2 1600 |
| 56765.52567 | 0.0001 | I | ------- | CCD+I | 169/148 | Šmelcer L. | Celestron 355/2460 + CCD G2 1600 |
| 56813.38362 | 0.0002 | I | ------- | CCD+R | 112/22 | Šmelcer L. | Celestron 355/2460 + CCD G2 1600 |
| 56813.38426 | 0.0003 | I | ------- | CCD+V | 98/15 | Šmelcer L. | Celestron 355/2460 + CCD G2 1600 |
| 56819.45798 | 0.0001 | II | ------- | CCD+V | 86/66 | Šmelcer L. | Celestron 355/2460 + CCD G2 1600 |
| 56819.45821 | 0.0001 | II | ------- | CCD+R | 97/67 | Šmelcer L. | Celestron 355/2460 + CCD G2 1600 |
| 56824.42753 | 0.0001 | I | ------- | CCD+V | 108/45 | Šmelcer L. | Celestron 355/2460 + CCD G2 1600 |
| 56824.42819 | 0.0001 | I | ------- | CCD+R | 82/47 | Šmelcer L. | Celestron 355/2460 + CCD G2 1600 |
| 56827.37322 | 0.0001 | I | ------- | CCD+R | 99/12 | Šmelcer L. | Celestron 355/2460 + CCD G2 1600 |
| 56827.37345 | 0.0001 | I | ------- | CCD+V | 110/11 | Šmelcer L. | Celestron 355/2460 + CCD G2 1600 |
| 56891.42913 | 0.0001 | I | ------- | CCD+Clear | 84/43 | Šmelcer L. | Celestron 280/1765 + CCD G2 4000 |
| 56891.42932 | 0.0002 | I | ------- | CCD+R | 84/44 | Šmelcer L. | Celestron 280/1765 + CCD G2 4000 |
| 56891.42943 | 0.0002 | I | ------- | CCD+V | 77/40 | Šmelcer L. | Celestron 280/1765 + CCD G2 4000 |
| 56898.42370 | 0.0000 | I | ------- | CCD+R | 378/209 | Šmelcer L. | Celestron 355/2460 + CCD G2 1600 |
| QU Cam | | | | | | | |
| 56730.36676 | 0.0001 | I | -0.0053 | DSLR | 235/121 | S. Jakš, M. Horník | RF Comet finder, Canon 600D |
| QY Cam | | | | | | | |
| 56692.37612 | 0.0004 | II | ------- | CCD+R | 164/84 | Lehký M. | EQ6 + 0.25-m f/4 + CCD ST7 + R |
| V0337 Cam | | | | | | | |
| 56887.48687 | 0.0003 | I | -0.0003 | DSLR | 165/112 | Medulka T. | Refraktor 90/500 + Canon 450D |
| 56898.45921 | 0.0010 | II | ------- | DSLR | 97/43 | Medulka T. | Refraktor 90/500 + Canon 450D |
| V0382 Cam | | | | | | | |
| 56630.56830 | 0.0001 | I | -0.0020 | CCD+R | 312/161 | Lehký M. | EQ6 + 0.25-m f/4 + CCD ST7 + R |
| V0426 Cam | | | | | | | |
| 56729.34202 | 0.0002 | I | -0.0076 | CCD+R | 222/109 | Hanžl D. | 0.2-m RL + CCD G2 8300 (DATEL telescope) |
| V0453 Cam | | | | | | | |
| 56750.30772 | 0.0004 | I | -0.0697 | CCD+R | 250/58 | Hanžl D. | 0.2-m RL + CCD G2 8300 (DATEL telescope) |
| V0456 Cam | | | | | | | |
| 56654.23695 | 0.0005 | I | -0.0787 | CCD+Clear | 255/31 | Bílek F. | Newton 0.2m f4,4 CCD Atik 314L+ |
| V0459 Cam | | | | | | | |
| 56654.29275 | 0.0001 | I | +0.1598 | CCD+R | 479/71 | Lehký M. | EQ6 + 0.25-m f/4 + CCD ST7 + R |
| 56726.51939 | 0.0001 | I | +0.1609 | CCD+R | 824/598 | Lehký M. | EQ6 + 0.25-m f/4 + CCD ST7 + R |
| | | | show all | | | | |
| V0460 Cam | | | | | | | |
| 56718.26793 | 0.0003 | I | -0.0582 | CCD+R | 95/40 | Hanžl D. | 0.2-m RL + CCD G2 8300 (DATEL telescope) |
| V0490 Cam | | | | | | | |
| 56718.34742 | 0.0006 | II | ------- | CCD+R | 136/38 | Hanžl D. | 0.2-m RL + CCD G2 8300 (DATEL telescope) |
| V0495 Cam | | | | | | | |
| 56737.33413 | 0.0003 | II | ------- | CCD+R | 337/114 | Hanžl D. | 0.2-m RL + CCD G2 8300 (DATEL telescope) |
| ASAS J105115-6032.1 Car | | | | | | | |
| 56814.55806 | 0.0008 | I | ------- | CCD+R | 32/25 | M. Mašek, K. Hoňková, J. Juryšek | FRAM, Nikkor 300mm + G4-16000 |

V0573 Car

| | | | | | | | |
|---|---|---|---|---|---|---|---|
| 56814.61858 | 0.0005 | I | ------- | CCD+V | 80/63 | M. Mašek, K. Hoňková, J. Juryšek | FRAM, 0.3m SCT + CCD G2-1600 |
| 56814.61878 | 0.0008 | I | ------- | CCD+R | 76/61 | M. Mašek, K. Hoňková, J. Juryšek | FRAM, 0.3m SCT + CCD G2-1600 |
| 56814.62041 | 0.0006 | I | ------- | CCD+B | 77/64 | M. Mašek, K. Hoňková, J. Juryšek | FRAM, 0.3m SCT + CCD G2-1600 |

NSVS 1622436 Cas

| | | | | | | | |
|---|---|---|---|---|---|---|---|
| 56891.40434 | 0.0003 | II | ------- | DSLR | 119/67 | Pintr P. | R80/600mm + Canon EOS 350D |

AX Cas

| | | | | | | | |
|---|---|---|---|---|---|---|---|
| 56507.50802 | 0.0004 | II | +0.0084 | CCD+R | 224/147 | Lehký M. | EQ6 + 0.25-m f/4 + CCD ST7 + R |
| 56597.26408 | 0.0001 | I | +0.0084 | CCD+R | 168/56 | Lehký M. | EQ6 + 0.25-m f/4 + CCD ST7 + R |
| 56897.44967 | 0.0001 | I | +0.0072 | CCD+R | 270/178 | Mazanec J. | N400, G2 402 |
| 56897.45250 | 0.0002 | I | +0.0100 | CCD+I | 265/181 | Mazanec J. | N400, G2 402 |
| | | | show all | | | | |

CW Cas

| | | | | | | | |
|---|---|---|---|---|---|---|---|
| 56854.51151 | 0.0004 | II | -0.0158 | CCD+Clear | 330/88 | Magris M. | Schmidt-Newton 0.25m F/4, CCD MX 716 |

DO Cas

| | | | | | | | |
|---|---|---|---|---|---|---|---|
| 56579.31829 | 0.0003 | I | +0.0017 | CCD+V | 1031/543 | Magris M. | Schmidt-Newton 0.25m F/4, CCD Atik Titan |

EG Cas

| | | | | | | | |
|---|---|---|---|---|---|---|---|
| 56609.25934 | 0.0002 | I | -0.0007 | CCD+R | 166/86 | Lehký M. | EQ6 + 0.25-m f/4 + CCD ST7 + R |

GJ 3236 Cas §

| | | | | | | | |
|---|---|---|---|---|---|---|---|
| 56729.46380 | 0.0002 | I | ------- | CCD+R | 117/94 | Šmelcer L. | Newton 254/1200 + CCD G2 402 |
| 56729.46494 | 0.0005 | I | ------- | CCD+V | 103/80 | Šmelcer L. | Newton 254/1200 + CCD G2 402 |
| 56738.33535 | 0.0005 | II | ------- | CCD+R | 110/46 | Šmelcer L. | Newton 254/1200 + CCD G2 402 |
| 56739.49040 | 0.0003 | I | ------- | CCD+R | 251/226 | Šmelcer L. | Newton 254/1200 + CCD G2 402 |
| 56746.43162 | 0.0002 | I | ------- | CCD+R | 324/234 | Šmelcer L. | Newton 254/1200 + CCD G2 402 |
| 56751.44525 | 0.0007 | II | ------- | CCD+R | 202/195 | Šmelcer L. | Newton 254/1200 + CCD G2 402 |
| 56755.30181 | 0.0002 | II | ------- | CCD+R | 181/35 | Šmelcer L. | Newton 254/1200 + CCD G2 402 |
| 56765.32787 | 0.0003 | II | ------- | CCD+R | 131/18 | Šmelcer L. | Newton 254/1200 + CCD G2 402 |
| 56775.35383 | 0.0004 | II | ------- | CCD+R | 116/58 | Šmelcer L. | Newton 254/1200 + CCD G2 402 |
| 56778.43913 | 0.0003 | II | ------- | CCD+I | 239/180 | Šmelcer L. | Celestron 280/1765 + CCD G2 4000 |
| 56778.43980 | 0.0001 | II | ------- | CCD+I | 236/177 | Šmelcer L. | Celestron 355/2460 + CCD G2 1600 |
| 56786.53641 | 0.0002 | I | ------- | CCD+R | 319/292 | Šmelcer L. | Newton 254/1200 + CCD G2 402 |
| 56786.53760 | 0.0001 | I | ------- | CCD+I | 306/269 | Šmelcer L. | Celestron 355/2460 + CCD G2 1600 |
| 56790.39374 | 0.0002 | I | ------- | CCD+R | 202/82 | Šmelcer L. | Newton 254/1200 + CCD G2 402 |
| 56835.51130 | 0.0008 | II | ------- | CCD+V | 157/130 | Šmelcer L. | Celestron 280/1765 + CCD G2 4000 |
| 56835.51144 | 0.0001 | II | ------- | CCD+R | 199/164 | Šmelcer L. | Celestron 355/2460 + CCD G2 1600 |
| 56845.53635 | 0.0006 | II | ------- | CCD+R | 231/224 | Šmelcer L. | Celestron 355/2460 + CCD G2 1600 |
| 56857.49291 | 0.0001 | I | ------- | CCD+R | 237/174 | Šmelcer L. | Celestron 355/2460 + CCD G2 1600 |
| 56857.49329 | 0.0001 | I | ------- | CCD+R | 262/202 | Šmelcer L. | Newton 254/1200 + CCD G2 402 |
| 56887.57170 | 0.0003 | I | ------- | CCD+V | 104/90 | Šmelcer L. | Celestron 355/2460 + CCD G2 1600 |
| 56887.57263 | 0.0001 | I | ------- | CCD+R | 344/308 | Šmelcer L. | Newton 254/1200 + CCD G2 402 |
| 56887.57267 | 0.0002 | I | ------- | CCD+R | 114/99 | Šmelcer L. | Celestron 355/2460 + CCD G2 1600 |
| 56891.42838 | 0.0004 | I | ------- | CCD+V | 197/115 | Šmelcer L. | Celestron 355/2460 + CCD G2 1600 |
| 56891.42874 | 0.0002 | I | ------- | CCD+R | 315/181 | Šmelcer L. | Newton 254/1200 + CCD G2 402 |
| 56894.51395 | 0.0001 | I | ------- | CCD+R | 223/189 | Šmelcer L. | Newton 254/1200 + CCD G2 402 |
| 56894.51497 | 0.0004 | I | ------- | CCD+V | 147/118 | Šmelcer L. | Celestron 355/2460 + CCD G2 1600 |
| 56898.37016 | 0.0001 | I | ------- | CCD+R | 369/118 | Šmelcer L. | Newton 254/1200 + CCD G2 402 |
| MW Cas | | | | | | | |
| 56665.26515 | 0.0005 | II | ------- | CCD+V | 52/43 | Šmelcer L. | Celestron 355/2460 + CCD G2 1600 |
| 56665.26525 | 0.0013 | II | ------- | CCD+R | 50/43 | Šmelcer L. | Celestron 355/2460 + CCD G2 1600 |
| 56670.31351 | 0.0002 | I | +0.1945 | CCD+I | 88/46 | Šmelcer L. | Celestron 355/2460 + CCD G2 1600 |
| 56670.31371 | 0.0002 | I | +0.1947 | CCD+R | 82/40 | Šmelcer L. | Celestron 355/2460 + CCD G2 1600 |
| 56670.31383 | 0.0002 | I | +0.1948 | CCD+V | 83/49 | Šmelcer L. | Celestron 355/2460 + CCD G2 1600 |
| 56897.53654 | 0.0003 | II | ------- | CCD+R | 90/33 | Mazanec J. | N400, G2 402 |
| 56897.53665 | 0.0002 | II | ------- | CCD+I | 185/68 | Mazanec J. | N400, G2 402 |
| | | | show all | | | | |
| NZ Cas | | | | | | | |
| 56543.48192 | 0.0010 | I | +0.0042 | CCD+R | 148/116 | Hanžl D. | 0.2-m RL |
| OQ Cas | | | | | | | |
| 56596.32777 | 0.0002 | II | -0.0129 | CCD+R | 205/53 | Lehký M. | 0.40-m f/5 + CCD G2-1600 + BVRI |
| 56596.32818 | 0.0003 | II | -0.0125 | CCD+I | 210/51 | Lehký M. | 0.40-m f/5 + CCD G2-1600 + BVRI |
| 56643.63002 | 0.0008 | II | -0.0124 | CCD+R | 195/164 | Lehký M. | 0.40-m f/5 + CCD G2-1600 + BVRI |
| 56643.63043 | 0.0009 | II | -0.0120 | CCD+I | 187/156 | Lehký M. | 0.40-m f/5 + CCD G2-1600 + BVRI |
| | | | show all | | | | |
| QQ Cas | | | | | | | |
| 56587.28995 | 0.0006 | II | +0.0093 | CCD+Clear | 184/102 | Urbaník M. | ED 80/600,0.5x Reductor, G1 0300 |
| TV Cas | | | | | | | |
| 56837.44839 | 0.0002 | I | -0.0072 | DSLR | 120/55 | Medulka T. | Refraktor 90/500 + Canon 450D |
| V0366 Cas | | | | | | | |
| 56865.48644 | 0.0003 | II | +0.0150 | CCD+R | 64/35 | Lehký M. | 0.40-m f/5 + CCD G2-1600 + BVRI |
| 56865.48662 | 0.0003 | II | +0.0152 | CCD+V | 63/36 | Lehký M. | 0.40-m f/5 + CCD G2-1600 + BVRI |
| 56865.48677 | 0.0003 | II | +0.0153 | CCD+I | 66/36 | Lehký M. | 0.40-m f/5 + CCD G2-1600 + BVRI |
| | | | show all | | | | |
| V0381 Cas | | | | | | | |
| 56895.35298 | 0.0003 | I | +0.0181 | CCD+R | 78/50 | Šmelcer L. | Celestron 280/1765 + CCD G2 4000 |
| 56895.35655 | 0.0005 | I | +0.0217 | CCD+Clear | 71/50 | Šmelcer L. | Celestron 280/1765 + CCD G2 4000 |
| 56895.35761 | 0.0006 | I | +0.0228 | CCD+V | 64/48 | Šmelcer L. | Celestron 280/1765 + CCD G2 4000 |
| | | | show all | | | | |
| V0541 Cas | | | | | | | |
| 56559.31155 | 0.0011 | I | -0.0023 | CCD+V | 62/38 | Magris M. | Schmidt-Newton 0.25m F/4, CCD MX 716 |
| V0651 Cas | | | | | | | |
| 56597.27424 | 0.0004 | II | +0.0004 | CCD+V | 123/69 | Magris M. | Schmidt-Newton 0.25m F/4, Atik Titan |
| V0793 Cas | | | | | | | |
| 56869.39852 | 0.0009 | I | -0.0832 | DSLR | 130/33 | Medulka T. | Refraktor 90/500 + Canon 450D |
| V0959 Cas | | | | | | | |
| 56542.34299 | 0.0003 | I | -0.0038 | CCD+R | 180/73 | Lehký M. | EQ6 + 0.25-m f/4 + CCD ST7 + R |
| V0961 Cas | | | | | | | |
| 56706.35946 | 0.0015 | I | -0.0064 | CCD+R | 186/93 | Hanžl D. | 0.2-m RL + CCD G2 8300 (DATEL telescope) |
| V1031 Cas | | | | | | | |
| 56481.52237 | 0.0002 | II | ------- | CCD+R | 110/70 | Lehký M. | EQ6 + 0.25-m f/4 + CCD ST7 + R |
| 56630.31023 | 0.0004 | I | +0.0059 | CCD+R | 79/52 | Lehký M. | EQ6 + 0.25-m f/4 + CCD ST7 + R |
| V1053 Cas | | | | | | | |
| 56543.39327 | 0.0022 | II | +0.0166 | CCD+R | 108/25 | Lehký M. | 0.40-m f/5 + CCD G2-1600 + BVRI |
| 56543.39859 | 0.0018 | II | +0.0219 | CCD+I | 108/25 | Lehký M. | 0.40-m f/5 + CCD G2-1600 + BVRI |
| 56596.31693 | 0.0014 | I | +0.0174 | CCD+R | 159/36 | Lehký M. | 0.40-m f/5 + CCD G2-1600 + BVRI |
| 56596.32023 | 0.0011 | I | +0.0207 | CCD+I | 153/34 | Lehký M. | 0.40-m f/5 + CCD G2-1600 + BVRI |
| 56597.28949 | 0.0022 | II | +0.0156 | CCD+I | 123/29 | Lehký M. | 0.40-m f/5 + CCD G2-1600 + BVRI |
| 56597.29072 | 0.0023 | II | +0.0168 | CCD+R | 124/36 | Lehký M. | 0.40-m f/5 + CCD G2-1600 + BVRI |
| 56643.39124 | 0.0029 | II | +0.0125 | CCD+R | 141/60 | Lehký M. | 0.40-m f/5 + CCD G2-1600 + BVRI |
| 56643.39890 | 0.0032 | II | +0.0202 | CCD+I | 142/55 | Lehký M. | 0.40-m f/5 + CCD G2-1600 + BVRI |
| | | | show all | | | | |
| V1060 Cas | | | | | | | |
| 56506.45664 | 0.0001 | I | -0.0003 | CCD+R | 649/303 | Lehký M. | EQ6 + 0.25-m f/4 + CCD ST7 + R |
| 56861.48537 | 0.0002 | II | -0.0034 | CCD+R | 241/164 | Lehký M. | EQ6 + 0.25-m f/4 + CCD ST7 + R |
| | | | show all | | | | |
| V1067 Cas | | | | | | | |
| 56670.58414 | 0.0002 | I | -0.1229 | CCD+R | 148/93 | Hanžl D. | 0.2-m RL + CCD G2 8300 (DATEL telescope) |
| 56683.39587 | 0.0001 | I | -0.1216 | CCD+R | 241/110 | Hanžl D. | 0.2-m RL + CCD G2 8300 (DATEL telescope) |
| 56692.48757 | 0.0001 | I | -0.1212 | CCD+R | 190/56 | Hanžl D. | 0.2-m RL + CCD G2 8300 (DATEL telescope) |
| 56693.31388 | 0.0001 | I | -0.1213 | CCD+R | 143/118 | Hanžl D. | 0.2-m RL + CCD G2 8300 (DATEL telescope) |
| | | | show all | | | | |
| V1070 Cas | | | | | | | |
| 56843.46648 | 0.0006 | I | +0.0080 | DSLR | 120/75 | Medulka T. | Refraktor 90/500 + Canon 450D |
| 56852.41169 | 0.0010 | I | +0.0042 | DSLR | 73/38 | Medulka T. | Refraktor 90/500 + Canon 450D |
| | | | show all | | | | |
| V1106 Cas | | | | | | | |
| 56522.58086 | 0.0035 | II | ------- | CCD+R | 67/54 | Lehký M. | EQ6 + 0.25-m f/4 + CCD ST7 + R |
| 56864.46931 | 0.0008 | II | ------- | CCD+R | 149/55 | Lehký M. | EQ6 + 0.25-m f/4 + CCD ST7 + R |
| V1107 Cas | | | | | | | |
| 56507.45692 | 0.0002 | II | -0.0016 | CCD+R | 218/105 | Lehký M. | EQ6 + 0.25-m f/4 + CCD ST7 + R |
| 56597.27328 | 0.0002 | I | -0.0007 | CCD+R | 168/65 | Lehký M. | 0.40-m f/5 + CCD G2-1600 + BVRI |
| 56725.36888 | 0.0004 | II | +0.0010 | CCD+R | 128/97 | Lehký M. | EQ6 + 0.25-m f/4 + CCD ST7 + R |
| | | | show all | | | | |
| V1170 Cas | | | | | | | |
| 56245.39582 | 0.0020 | II | ------- | CCD+R | 543/181 | Lehký M. | EQ6 + 0.25-m f/4 + CCD ST7 + R |
| 56629.31922 | 0.0016 | I | +0.1256 | CCD+R | 513/148 | Lehký M. | EQ6 + 0.25-m f/4 + CCD ST7 + R |
| ZZ Cas | | | | | | | |
| 56670.31843 | 0.0003 | I | +0.0085 | CCD+R | 105/57 | Šmelcer L. | Celestron 280/1765 + CCD ST7 |
| 56670.32514 | 0.0004 | I | +0.0152 | CCD+V | 112/57 | Šmelcer L. | Celestron 280/1765 + CCD ST7 |
| | | | show all | | | | |
| ASAS J114951-6101.1 Cen | | | | | | | |
| 56680.72522 | 0.0019 | I | ------- | CCD+I | 61/32 | M. Mašek, K. Hoňková, J. Juryšek | FRAM, Nikkor 300mm + G4-16000 |
| 56680.72868 | 0.0017 | I | ------- | CCD+R | 71/39 | M. Mašek, K. Hoňková, J. Juryšek | FRAM, Nikkor 300mm + G4-16000 |
| 56680.72939 | 0.0005 | I | ------- | CCD+B | 72/51 | M. Mašek, K. Hoňková, J. Juryšek | FRAM, Nikkor 300mm + G4-16000 |
| ASAS J115108-6355.1 Cen | | | | | | | |
| 56774.66078 | 0.0016 | I | ------- | CCD+I | 42/10 | M. Mašek, K. Hoňková, J. Juryšek | FRAM, Nikkor 300mm + G4-16000 |
| 56774.66840 | 0.0016 | I | ------- | CCD+R | 33/8 | M. Mašek, K. Hoňková, J. Juryšek | FRAM, Nikkor 300mm + G4-16000 |
| BF Cen | | | | | | | |
| 56774.67403 | 0.0011 | I | +0.0012 | CCD+R | 40/12 | M. Mašek, K. Hoňková, J. Juryšek | FRAM, Nikkor 300mm + G4-16000 |
| 56774.67568 | 0.0005 | I | +0.0029 | CCD+B | 55/24 | M. Mašek, K. Hoňková, J. Juryšek | FRAM, Nikkor 300mm + G4-16000 |
| 56774.67703 | 0.0011 | I | +0.0042 | CCD+I | 48/16 | M. Mašek, K. Hoňková, J. Juryšek | FRAM, Nikkor 300mm + G4-16000 |
| 56774.67720 | 0.0016 | I | +0.0044 | CCD+V | 28/13 | M. Mašek, K. Hoňková, J. Juryšek | FRAM, Nikkor 300mm + G4-16000 |
| | | | show all | | | | |
| BH Cen | | | | | | | |
| 56766.80416 | 0.0005 | I | +0.0124 | CCD+R | 77/60 | M. Mašek, K. Hoňková, J. Juryšek | FRAM, Nikkor 300mm + G4-16000 |
| 56766.80510 | 0.0004 | I | +0.0134 | CCD+B | 43/29 | M. Mašek, K. Hoňková, J. Juryšek | FRAM, Nikkor 300mm + G4-16000 |
| 56766.80674 | 0.0008 | I | +0.0150 | CCD+I | 77/62 | M. Mašek, K. Hoňková, J. Juryšek | FRAM, Nikkor 300mm + G4-16000 |
| 56774.71989 | 0.0005 | I | +0.0123 | CCD+I | 44/21 | M. Mašek, K. Hoňková, J. Juryšek | FRAM, Nikkor 300mm + G4-16000 |
| 56774.72037 | 0.0004 | I | +0.0128 | CCD+R | 36/16 | M. Mašek, K. Hoňková, J. Juryšek | FRAM, Nikkor 300mm + G4-16000 |
| 56774.72136 | 0.0004 | I | +0.0138 | CCD+B | 52/29 | M. Mašek, K. Hoňková, J. Juryšek | FRAM, Nikkor 300mm + G4-16000 |
| | | | show all | | | | |
| GSC 08976-03898 Cen | | | | | | | |
| 56680.76064 | 0.0005 | I | ------- | CCD+R | 169/122 | M. Mašek, K. Hoňková, J. Juryšek | FRAM, 0.3m SCT + CCD G2-1600 |
| 56680.76134 | 0.0008 | I | ------- | CCD+V | 172/134 | M. Mašek, K. Hoňková, J. Juryšek | FRAM, 0.3m SCT + CCD G2-1600 |
| 56680.76199 | 0.0012 | I | ------- | CCD+I | 146/116 | M. Mašek, K. Hoňková, J. Juryšek | FRAM, 0.3m SCT + CCD G2-1600 |
| 56680.76286 | 0.0006 | I | ------- | CCD+B | 168/122 | M. Mašek, K. Hoňková, J. Juryšek | FRAM, 0.3m SCT + CCD G2-1600 |
| GSC 08977-00474 Cen | | | | | | | |
| 56680.69842 | 0.0011 | I | ------- | CCD+I | 59/27 | M. Mašek, K. Hoňková, J. Juryšek | FRAM, Nikkor 300mm + G4-16000 |
| 56680.70134 | 0.0012 | I | ------- | CCD+B | 75/46 | M. Mašek, K. Hoňková, J. Juryšek | FRAM, Nikkor 300mm + G4-16000 |
| 56680.70355 | 0.0009 | I | ------- | CCD+R | 71/34 | M. Mašek, K. Hoňková, J. Juryšek | FRAM, Nikkor 300mm + G4-16000 |
| 56766.54213 | 0.0008 | I | ------- | CCD+I | 78/14 | M. Mašek, K. Hoňková, J. Juryšek | FRAM, Nikkor 300mm + G4-16000 |
| 56766.54661 | 0.0005 | I | ------- | CCD+R | 77/14 | M. Mašek, K. Hoňková, J. Juryšek | FRAM, Nikkor 300mm + G4-16000 |
| 56766.70898 | 0.0010 | II | ------- | CCD+R | 77/43 | M. Mašek, K. Hoňková, J. Juryšek | FRAM, Nikkor 300mm + G4-16000 |
| 56766.71105 | 0.0006 | II | ------- | CCD+I | 78/45 | M. Mašek, K. Hoňková, J. Juryšek | FRAM, Nikkor 300mm + G4-16000 |
| GSC 08977-01661 Cen | | | | | | | |
| 56680.80657 | 0.0007 | I | ------- | CCD+B | 74/60 | M. Mašek, K. Hoňková, J. Juryšek | FRAM, Nikkor 300mm + G4-16000 |
| 56680.80903 | 0.0022 | I | ------- | CCD+I | 62/51 | M. Mašek, K. Hoňková, J. Juryšek | FRAM, Nikkor 300mm + G4-16000 |
| SV Cen | | | | | | | |
| 56680.80258 | 0.0005 | I | -0.2896 | CCD+B | 73/61 | M. Mašek, K. Hoňková, J. Juryšek | FRAM, Nikkor 300mm + G4-16000 |
| 56680.80367 | 0.0011 | I | -0.2885 | CCD+R | 68/57 | M. Mašek, K. Hoňková, J. Juryšek | FRAM, Nikkor 300mm + G4-16000 |
| 56680.80546 | 0.0007 | I | -0.2867 | CCD+I | 61/49 | M. Mašek, K. Hoňková, J. Juryšek | FRAM, Nikkor 300mm + G4-16000 |
| 56768.64488 | 0.0006 | I | -0.3033 | CCD+R | 44/28 | M. Mašek, K. Hoňková, J. Juryšek | FRAM, Nikkor 300mm + G4-16000 |
| 56768.64496 | 0.0004 | I | -0.3032 | CCD+I | 48/32 | M. Mašek, K. Hoňková, J. Juryšek | FRAM, Nikkor 300mm + G4-16000 |
| 56768.64684 | 0.0004 | I | -0.3013 | CCD+B | 46/30 | M. Mašek, K. Hoňková, J. Juryšek | FRAM, Nikkor 300mm + G4-16000 |
| | | | show all | | | | |
| V0690 Cen | | | | | | | |
| 56680.65620 | 0.0018 | I | +0.0054 | CCD+B | 64/33 | M. Mašek, K. Hoňková, J. Juryšek | FRAM, Nikkor 300mm + G4-16000 |
| V0692 Cen | | | | | | | |
| 56766.62861 | 0.0024 | I | +0.6656 | CCD+R | 76/28 | M. Mašek, K. Hoňková, J. Juryšek | FRAM, Nikkor 300mm + G4-16000 |
| 56766.64778 | 0.0017 | I | +0.6848 | CCD+I | 71/31 | M. Mašek, K. Hoňková, J. Juryšek | FRAM, Nikkor 300mm + G4-16000 |
| | | | show all | | | | |
| V0916 Cen | | | | | | | |
| 56766.74946 | 0.0009 | I | -0.0045 | CCD+V | 233/161 | M. Mašek, K. Hoňková, J. Juryšek | FRAM, 0.3m SCT + CCD G2-1600 |
| 56766.74964 | 0.0023 | I | -0.0043 | CCD+I | 233/166 | M. Mašek, K. Hoňková, J. Juryšek | FRAM, 0.3m SCT + CCD G2-1600 |
| 56766.75262 | 0.0013 | I | -0.0013 | CCD+R | 208/138 | M. Mašek, K. Hoňková, J. Juryšek | FRAM, 0.3m SCT + CCD G2-1600 |
| 56766.75887 | 0.0017 | I | +0.0049 | CCD+B | 193/156 | M. Mašek, K. Hoňková, J. Juryšek | FRAM, 0.3m SCT + CCD G2-1600 |
| | | | show all | | | | |
| V1104 Cen | | | | | | | |
| 56766.66466 | 0.0016 | I | +0.0027 | CCD+I | 75/36 | M. Mašek, K. Hoňková, J. Juryšek | FRAM, Nikkor 300mm + G4-16000 |
| 56766.66607 | 0.0027 | I | +0.0041 | CCD+B | 39/10 | M. Mašek, K. Hoňková, J. Juryšek | FRAM, Nikkor 300mm + G4-16000 |
| 56766.67430 | 0.0013 | I | +0.0123 | CCD+R | 76/36 | M. Mašek, K. Hoňková, J. Juryšek | FRAM, Nikkor 300mm + G4-16000 |
| | | | show all | | | | |
| V1237 Cen | | | | | | | |
| 56766.69375 | 0.0007 | I | -0.0054 | CCD+R | 47/12 | M. Mašek, K. Hoňková, J. Juryšek | FRAM, Nikkor 300mm + G4-16000 |
| 56766.69544 | 0.0007 | I | -0.0037 | CCD+I | 47/13 | M. Mašek, K. Hoňková, J. Juryšek | FRAM, Nikkor 300mm + G4-16000 |
| 56774.71310 | 0.0009 | II | ------- | CCD+I | 42/18 | M. Mašek, K. Hoňková, J. Juryšek | FRAM, Nikkor 300mm + G4-16000 |
| 56774.71440 | 0.0008 | II | ------- | CCD+R | 34/15 | M. Mašek, K. Hoňková, J. Juryšek | FRAM, Nikkor 300mm + G4-16000 |
| 56774.71504 | 0.0010 | II | ------- | CCD+B | 40/18 | M. Mašek, K. Hoňková, J. Juryšek | FRAM, Nikkor 300mm + G4-16000 |
| 56774.71730 | 0.0022 | II | ------- | CCD+V | 29/12 | M. Mašek, K. Hoňková, J. Juryšek | FRAM, Nikkor 300mm + G4-16000 |
| | | | show all | | | | |
| V0358 Cep | | | | | | | |
| 56569.28563 | 0.0004 | II | +0.0021 | CCD+R | 265/47 | Lehký M. | EQ6 + 0.25-m f/4 + CCD ST7 + R |
| 56569.51930 | 0.0002 | I | -0.0011 | CCD+R | 265/242 | Lehký M. | EQ6 + 0.25-m f/4 + CCD ST7 + R |
| 56854.40005 | 0.0007 | II | +0.0004 | CCD+R | 162/46 | Lehký M. | EQ6 + 0.25-m f/4 + CCD ST7 + R |
| | | | show all | | | | |
| V0757 Cep | | | | | | | |
| 56726.39764 | 0.0003 | I | +0.0228 | CCD+R | 123/39 | Šmelcer L. | Celestron 355/2460 + CCD G2 1600 |
| 56726.39781 | 0.0005 | I | +0.0229 | CCD+V | 122/39 | Šmelcer L. | Celestron 355/2460 + CCD G2 1600 |
| | | | show all | | | | |
| V0800 Cep | | | | | | | |
| 56569.46337 | 0.0002 | I | +0.2024 | CCD+R | 268/202 | Lehký M. | EQ6 + 0.25-m f/4 + CCD ST7 + R |
| 56854.48539 | 0.0002 | I | +0.2124 | CCD+R | 163/100 | Lehký M. | EQ6 + 0.25-m f/4 + CCD ST7 + R |
| | | | show all | | | | |
| V0806 Cep | | | | | | | |
| 56683.24265 | 0.0002 | II | ------- | CCD+R | 142/62 | Lehký M. | EQ6 + 0.25-m f/4 + CCD ST7 + R |
| V0816 Cep | | | | | | | |
| 56894.44137 | 0.0004 | I | -0.0315 | CCD+R | 295/177 | Hanžl D. | 0.2-m RL + CCD G2 8300 (DATEL telescope) |
| V0843 Cep | | | | | | | |
| 56865.49065 | 0.0004 | I | +0.0087 | CCD+R | 260/141 | Hanžl D. | 0.2-m RL + CCD G2 8300 (DATEL telescope) |
| V0845 Cep | | | | | | | |
| 56865.41348 | 0.0007 | I | -0.0535 | CCD+R | 260/42 | Hanžl D. | 0.2-m RL + CCD G2 8300 (DATEL telescope) |
| V0934 Cep | | | | | | | |
| 56878.51982 | 0.0003 | I | +0.0055 | CCD+R | 75/43 | Hanžl D. | 0.2-m RL + CCD G2 8300 (DATEL telescope) |
| V0951 Cep | | | | | | | |
| 56858.46503 | 0.0001 | I | -0.0520 | CCD+Clear | 145/78 | Bílek F. | Newton 0.2m f4,4 CCD Atik 314L+ |
| EE Cet | | | | | | | |
| 56565.55444 | 0.0004 | I | +0.0239 | CCD+Clear | 102/63 | Mašek M. | R70/700mm, 0,5x reducer + CCD Meade DSI |
| AM CMi | | | | | | | |
| 56718.35083 | 0.0004 | I | +0.0138 | CCD+Clear | 230/146 | Urbaník M. | ED 80/600,0.5x Reductor, G1 0300 |
| TT CMi | | | | | | | |
| 56608.60121 | 0.0005 | I | +0.0007 | CCD+V | 41/19 | Lehký M. | 0.40-m f/5 + CCD G2-1600 + BVRI |
| 56608.60140 | 0.0005 | I | +0.0009 | CCD+I | 41/20 | Lehký M. | 0.40-m f/5 + CCD G2-1600 + BVRI |
| 56608.60256 | 0.0004 | I | +0.0020 | CCD+R | 40/18 | Lehký M. | 0.40-m f/5 + CCD G2-1600 + BVRI |
| | | | show all | | | | |
| AE Cnc | | | | | | | |
| 56718.35936 | 0.0003 | I | -0.0118 | CCD+V | 56/18 | Lehký M. | 0.40-m f/5 + CCD G2-1600 + BVRI |
| 56718.35985 | 0.0003 | I | -0.0113 | CCD+R | 60/20 | Lehký M. | 0.40-m f/5 + CCD G2-1600 + BVRI |
| 56718.35997 | 0.0003 | I | -0.0112 | CCD+I | 60/21 | Lehký M. | 0.40-m f/5 + CCD G2-1600 + BVRI |
| | | | show all | | | | |
| HN Cnc | | | | | | | |
| 56716.28472 | 0.0006 | I | -0.0104 | CCD+R | 72/35 | Lehký M. | EQ6 + 0.25-m f/4 + CCD ST7 + R |
| 56737.30111 | 0.0002 | II | -0.0105 | CCD+R | 94/38 | Lehký M. | EQ6 + 0.25-m f/4 + CCD ST7 + R |
| | | | show all | | | | |
| IN Cnc | | | | | | | |
| 56716.43515 | 0.0003 | I | -0.0087 | CCD+R | 99/57 | Lehký M. | EQ6 + 0.25-m f/4 + CCD ST7 + R |
| 56765.32738 | 0.0004 | II | -0.0088 | CCD+R | 69/30 | Lehký M. | EQ6 + 0.25-m f/4 + CCD ST7 + R |
| | | | show all | | | | |
| IO Cnc | | | | | | | |
| 56747.42578 | 0.0009 | I | -0.0110 | CCD+R | 155/132 | Lehký M. | EQ6 + 0.25-m f/4 + CCD ST7 + R |
| 56765.33536 | 0.0003 | II | -0.0078 | CCD+R | 67/36 | Lehký M. | EQ6 + 0.25-m f/4 + CCD ST7 + R |
| | | | show all | | | | |
| IT Cnc | | | | | | | |
| 56727.34074 | 0.0004 | II | -0.1068 | CCD+R | 280/38 | Lehký M. | EQ6 + 0.25-m f/4 + CCD ST7 + R |
| 56727.51994 | 0.0005 | I | -0.1095 | CCD+R | 280/222 | Lehký M. | EQ6 + 0.25-m f/4 + CCD ST7 + R |
| | | | show all | | | | |
| IU Cnc | | | | | | | |
| 56643.59685 | 0.0006 | I | -0.0094 | CCD+R | 164/23 | Lehký M. | EQ6 + 0.25-m f/4 + CCD ST7 + R |
| KY Cnc | | | | | | | |
| 56752.38673 | 0.0003 | I | +0.0095 | CCD+Clear | 256/136 | Urbaník M. | ED 80/600,0.5x Reductor, G1 0300 |
| NSVS 10123419 Cnc | | | | | | | |
| 56727.45521 | 0.0004 | II | ------- | CCD+R | 285/160 | Lehký M. | EQ6 + 0.25-m f/4 + CCD ST7 + R |
| TX Cnc | | | | | | | |
| 56584.63541 | 0.0010 | I | +0.0009 | CCD+V | 470/232 | Magris M. | Schmidt-Newton 0.25m F/4, CCD MX 716 |
| 56602.63167 | 0.0004 | I | +0.0017 | CCD+Clear | 653/449 | Magris M. | Schmidt-Newton 0.25m F/4, CCD MX 716 |
| | | | show all | | | | |
| WW Cnc | | | | | | | |
| 56714.28845 | 0.0001 | I | +0.0043 | CCD+R | 75/34 | Lehký M. | EQ6 + 0.25-m f/4 + CCD ST7 + R |
| 56719.31045 | 0.0003 | II | +0.0046 | CCD+R | 159/80 | Lehký M. | EQ6 + 0.25-m f/4 + CCD ST7 + R |
| | | | show all | | | | |
| RU Col | | | | | | | |
| 56640.73754 | 0.0000 | I | +0.0915 | CCD+Clear | 682/590 | Mina, F.; Artola, R.; Zalazar, J. | 0.36-m f/5.10 Schmidt-Cassegrain + CCD |
| CC Com | | | | | | | |
| 56746.53978 | 0.0001 | II | +0.0008 | CCD+Clear | 384/349 | Lomoz F. | Newton 254/1016+G2-8300 |
| EK Com | | | | | | | |
| 56712.43592 | 0.0002 | I | +0.0051 | CCD+V | 87/26 | Šmelcer L. | Celestron 355/2460 + CCD G2 1600 |
| 56712.43644 | 0.0001 | I | +0.0056 | CCD+R | 104/24 | Šmelcer L. | Celestron 355/2460 + CCD G2 1600 |
| 56712.43645 | 0.0003 | I | +0.0057 | CCD+I | 91/25 | Šmelcer L. | Celestron 355/2460 + CCD G2 1600 |
| 56712.56873 | 0.0003 | II | +0.0049 | CCD+R | 104/96 | Šmelcer L. | Celestron 355/2460 + CCD G2 1600 |
| | | | show all | | | | |
| KK Com | | | | | | | |
| 56725.53007 | 0.0011 | I | +0.0063 | CCD+Clear | 101/32 | Hladík B. | RF 8/200, ATIK 320E, CG-4 |
| LL Com | | | | | | | |
| 56481.38801 | 0.0006 | I | +0.0050 | CCD+R | 36/25 | Lehký M. | EQ6 + 0.25-m f/4 + CCD ST7 + R |
| 56706.61945 | 0.0003 | II | +0.0094 | CCD+R | 97/48 | Lehký M. | EQ6 + 0.25-m f/4 + CCD ST7 + R |
| 56712.51894 | 0.0002 | I | +0.0082 | CCD+R | 49/25 | Lehký M. | EQ6 + 0.25-m f/4 + CCD ST7 + R |
| | | | show all | | | | |
| LQ Com | | | | | | | |
| 56723.46870 | 0.0004 | II | +0.0015 | CCD+R | 117/41 | Lehký M. | EQ6 + 0.25-m f/4 + CCD ST7 + R |
| 56786.44875 | 0.0002 | I | -0.0002 | CCD+Clear | 136/53 | Hájek P., Lehký M. | 0.20-m f/4 reflector + CCD ST6 |
| | | | show all | | | | |
| LT Com | | | | | | | |
| 56460.39462 | 0.0004 | I | +0.0077 | CCD+R | 104/60 | Lehký M. | EQ6 + 0.25-m f/4 + CCD ST7 + R |
| 56729.48023 | 0.0003 | II | +0.0074 | CCD+R | 186/114 | Lehký M. | EQ6 + 0.25-m f/4 + CCD ST7 + R |
| 56737.41687 | 0.0002 | I | +0.0071 | CCD+R | 126/69 | Lehký M. | EQ6 + 0.25-m f/4 + CCD ST7 + R |
| | | | show all | | | | |
| MM Com | | | | | | | |
| 56765.44522 | 0.0001 | II | +0.0139 | CCD+R | 74/45 | Lehký M. | EQ6 + 0.25-m f/4 + CCD ST7 + R |
| MR Com | | | | | | | |
| 56701.50241 | 0.0002 | II | -0.0020 | CCD+R | 142/51 | Lehký M. | EQ6 + 0.25-m f/4 + CCD ST7 + R |
| AY CrB | | | | | | | |
| 56797.48777 | 0.0003 | I | -0.0316 | CCD+Clear | 143/24 | Čaloud J. | Newton 270/2150 G2 - 1600 |
| BD CrB | | | | | | | |
| 56816.46357 | 0.0006 | I | +0.0093 | CCD+R | 148/79 | Hanžl D. | 0.2-m RL + CCD G2 8300 (DATEL telescope) |
| BX CrB | | | | | | | |
| 56738.58846 | 0.0010 | II | ------- | CCD+I | 78/65 | Vrašťák M. | 0,24m f/5 RL+CCD G2-1600, pointer 80/400+G1-300 |
| 56738.59133 | 0.0011 | II | ------- | CCD+R | 78/65 | Vrašťák M. | 0,24m f/5 RL+CCD G2-1600, pointer 80/400+G1-300 |
| 56738.59201 | 0.0008 | II | ------- | CCD+V | 78/65 | Vrašťák M. | 0,24m f/5 RL+CCD G2-1600, pointer 80/400+G1-300 |
| UX CrB | | | | | | | |
| 56778.50392 | 0.0004 | I | -0.2631 | CCD+R | 100/72 | Lehký M. | EQ6 + 0.25-m f/4 + CCD ST7 + R |
| DE CVn | | | | | | | |
| 56746.38398 | 0.0006 | I | -0.0062 | CCD+R | 274/139 | Hanžl D. | 0.2-m RL + CCD G2 8300 (DATEL telescope) |
| EF CVn | | | | | | | |
| 56725.47827 | 0.0003 | I | +0.0047 | CCD+R | 202/41 | Lehký M. | EQ6 + 0.25-m f/4 + CCD ST7 + R |
| 56725.61429 | 0.0003 | II | +0.0047 | CCD+R | 202/151 | Lehký M. | EQ6 + 0.25-m f/4 + CCD ST7 + R |
| | | | show all | | | | |
| EG CVn | | | | | | | |
| 56670.67907 | 0.0003 | I | +0.0122 | CCD+R | 95/49 | Lehký M. | EQ6 + 0.25-m f/4 + CCD ST7 + R |
| EL CVn | | | | | | | |
| 56730.41520 | 0.0002 | I | -0.0014 | CCD+R | 342/184 | Lehký M. | EQ6 + 0.25-m f/4 + CCD ST7 + R |
| EO CVn | | | | | | | |
| 56643.66545 | 0.0003 | I | -0.0088 | CCD+R | 128/80 | Hanžl D. | 0.2-m RL + CCD G2 8300 (DATEL telescope) |
| 56747.38255 | 0.0003 | I | -0.0051 | CCD+R | 278/132 | Hanžl D. | 0.2-m RL + CCD G2 8300 (DATEL telescope) |
| | | | show all | | | | |
| FN CVn | | | | | | | |
| 56726.52613 | 0.0003 | I | +0.0455 | CCD+R | 103/43 | Lehký M. | 0.40-m f/5 + CCD G2-1600 + BVRI |
| 56726.52636 | 0.0003 | I | +0.0458 | CCD+V | 105/48 | Lehký M. | 0.40-m f/5 + CCD G2-1600 + BVRI |
| 56726.52716 | 0.0002 | I | +0.0465 | CCD+I | 101/46 | Lehký M. | 0.40-m f/5 + CCD G2-1600 + BVRI |
| | | | show all | | | | |
| FU CVn | | | | | | | |
| 56777.48813 | 0.0005 | I | -0.0099 | DSLR | 156/74 | Walter F. | RF Comet finder 20/137, Canon 350 D |
| FY CVn | | | | | | | |
| 56706.60926 | 0.0004 | I | +0.0591 | CCD+R | 220/186 | Hanžl D. | 0.2-m RL + CCD G2 8300 (DATEL telescope) |
| 56709.50454 | 0.0003 | II | ------- | CCD+R | 384/126 | Hanžl D. | 0.2-m RL + CCD G2 8300 (DATEL telescope) |
| FZ CVn | | | | | | | |
| 56706.61456 | 0.0004 | II | -0.0016 | CCD+R | 209/190 | Hanžl D. | 0.2-m RL + CCD G2 8300 (DATEL telescope) |
| 56709.48121 | 0.0003 | II | -0.0016 | CCD+R | 382/96 | Hanžl D. | 0.2-m RL + CCD G2 8300 (DATEL telescope) |
| 56709.65677 | 0.0003 | I | -0.0054 | CCD+R | 382/321 | Hanžl D. | 0.2-m RL + CCD G2 8300 (DATEL telescope) |
| | | | show all | | | | |
| GSC 03557-01334 Cyg | | | | | | | |
| 56588.34826 | 0.0009 | I | ------- | CCD+V | 127/69 | Šmelcer L. | Newton 254/1200 + CCD G2 402 |
| 56588.34968 | 0.0009 | I | ------- | CCD+R | 115/58 | Šmelcer L. | Newton 254/1200 + CCD G2 402 |
| GO Cyg | | | | | | | |
| 56854.41176 | 0.0002 | I | -0.0043 | DSLR | 195/83 | Medulka T. | Refraktor 90/500 + Canon 450D |
| OO Cyg | | | | | | | |
| 56842.43107 | 0.0002 | I | -0.0524 | CCD+R | 120/40 | Hanžl D. | 0.2-m RL + CCD G2 8300 (DATEL telescope) |
| V0456 Cyg | | | | | | | |
| 56845.35898 | 0.0006 | I | -0.0016 | DSLR | 154/22 | Walter F., Divišová L. | RL Finder 20/137 Canon 350D |
| V0508 Cyg | | | | | | | |
| 56461.38754 | 0.0006 | I | +0.1178 | CCD+V | 28/13 | Lehký M. | 0.40-m f/5 + CCD G2-1600 + BVRI |
| 56461.38971 | 0.0003 | I | +0.1200 | CCD+I | 29/15 | Lehký M. | 0.40-m f/5 + CCD G2-1600 + BVRI |
| 56461.39025 | 0.0003 | I | +0.1205 | CCD+R | 32/15 | Lehký M. | 0.40-m f/5 + CCD G2-1600 + BVRI |
| 56482.44059 | 0.0004 | I | +0.1221 | CCD+V | 81/35 | Lehký M. | 0.40-m f/5 + CCD G2-1600 + BVRI |
| 56482.44064 | 0.0003 | I | +0.1222 | CCD+R | 85/35 | Lehký M. | 0.40-m f/5 + CCD G2-1600 + BVRI |
| 56482.44068 | 0.0002 | I | +0.1222 | CCD+I | 78/30 | Lehký M. | 0.40-m f/5 + CCD G2-1600 + BVRI |
| 56491.40639 | 0.0002 | II | +0.1225 | CCD+R | 82/18 | Lehký M. | 0.40-m f/5 + CCD G2-1600 + BVRI |
| 56491.40705 | 0.0002 | II | +0.1231 | CCD+I | 83/18 | Lehký M. | 0.40-m f/5 + CCD G2-1600 + BVRI |
| 56491.40721 | 0.0005 | II | +0.1233 | CCD+V | 82/20 | Lehký M. | 0.40-m f/5 + CCD G2-1600 + BVRI |
| 56496.47246 | 0.0004 | I | +0.1215 | CCD+V | 95/50 | Lehký M. | 0.40-m f/5 + CCD G2-1600 + BVRI |
| 56496.47292 | 0.0002 | I | +0.1219 | CCD+I | 96/48 | Lehký M. | 0.40-m f/5 + CCD G2-1600 + BVRI |
| 56496.47342 | 0.0002 | I | +0.1224 | CCD+R | 99/51 | Lehký M. | 0.40-m f/5 + CCD G2-1600 + BVRI |
| 56500.37268 | 0.0004 | I | +0.1238 | CCD+R | 91/14 | Lehký M. | 0.40-m f/5 + CCD G2-1600 + BVRI |
| 56500.37282 | 0.0005 | I | +0.1239 | CCD+I | 89/14 | Lehký M. | 0.40-m f/5 + CCD G2-1600 + BVRI |
| 56500.37309 | 0.0006 | I | +0.1242 | CCD+V | 85/15 | Lehký M. | 0.40-m f/5 + CCD G2-1600 + BVRI |
| 56501.54112 | 0.0003 | II | +0.1226 | CCD+V | 96/76 | Lehký M. | 0.40-m f/5 + CCD G2-1600 + BVRI |
| 56501.54146 | 0.0003 | II | +0.1230 | CCD+I | 97/75 | Lehký M. | 0.40-m f/5 + CCD G2-1600 + BVRI |
| 56501.54191 | 0.0003 | II | +0.1234 | CCD+R | 102/81 | Lehký M. | 0.40-m f/5 + CCD G2-1600 + BVRI |
| 56542.47342 | 0.0002 | I | +0.1270 | CCD+I | 114/68 | Lehký M. | 0.40-m f/5 + CCD G2-1600 + BVRI |
| 56542.47354 | 0.0004 | I | +0.1271 | CCD+V | 113/69 | Lehký M. | 0.40-m f/5 + CCD G2-1600 + BVRI |
| 56542.47392 | 0.0002 | I | +0.1275 | CCD+R | 118/76 | Lehký M. | 0.40-m f/5 + CCD G2-1600 + BVRI |
| 56569.37166 | 0.0002 | II | +0.1293 | CCD+R | 78/26 | Lehký M. | 0.40-m f/5 + CCD G2-1600 + BVRI |
| 56569.37199 | 0.0002 | II | +0.1297 | CCD+I | 74/25 | Lehký M. | 0.40-m f/5 + CCD G2-1600 + BVRI |
| 56569.37213 | 0.0003 | II | +0.1298 | CCD+V | 76/27 | Lehký M. | 0.40-m f/5 + CCD G2-1600 + BVRI |
| 56878.50740 | 0.0002 | I | +0.1602 | CCD+R | 101/64 | Lehký M. | 0.40-m f/5 + CCD G2-1600 + BVRI |
| 56878.50755 | 0.0002 | I | +0.1604 | CCD+I | 104/65 | Lehký M. | 0.40-m f/5 + CCD G2-1600 + BVRI |
| 56878.50779 | 0.0004 | I | +0.1606 | CCD+V | 98/60 | Lehký M. | 0.40-m f/5 + CCD G2-1600 + BVRI |
| | | | show all | | | | |
| V0749 Cyg | | | | | | | |
| 56821.41963 | 0.0007 | II | -0.0003 | CCD+Clear | 278/84 | Trnka J. | 200/1 000, ST-9E |
| | | | | | | | |
| V0807 Cyg | | | | | | | |
| 56481.46326 | 0.0006 | II | -0.0251 | CCD+V | 62/22 | Lehký M. | 0.40-m f/5 + CCD G2-1600 + BVRI |
| 56481.46470 | 0.0004 | II | -0.0237 | CCD+R | 62/20 | Lehký M. | 0.40-m f/5 + CCD G2-1600 + BVRI |
| 56483.46196 | 0.0003 | I | -0.0227 | CCD+R | 70/34 | Lehký M. | 0.40-m f/5 + CCD G2-1600 + BVRI |
| 56483.46232 | 0.0005 | I | -0.0224 | CCD+V | 66/32 | Lehký M. | 0.40-m f/5 + CCD G2-1600 + BVRI |
| 56495.44171 | 0.0004 | I | -0.0244 | CCD+V | 90/36 | Lehký M. | 0.40-m f/5 + CCD G2-1600 + BVRI |
| 56495.44188 | 0.0003 | I | -0.0243 | CCD+R | 98/39 | Lehký M. | 0.40-m f/5 + CCD G2-1600 + BVRI |
| 56519.40665 | 0.0002 | I | -0.0224 | CCD+R | 77/31 | Lehký M. | 0.40-m f/5 + CCD G2-1600 + BVRI |
| 56519.40736 | 0.0003 | I | -0.0217 | CCD+V | 102/27 | Lehký M. | 0.40-m f/5 + CCD G2-1600 + BVRI |
| 56521.40229 | 0.0004 | II | -0.0243 | CCD+R | 112/30 | Lehký M. | 0.40-m f/5 + CCD G2-1600 + BVRI |
| 56521.40341 | 0.0004 | II | -0.0231 | CCD+V | 107/31 | Lehký M. | 0.40-m f/5 + CCD G2-1600 + BVRI |
| 56541.37281 | 0.0007 | II | -0.0228 | CCD+V | 92/23 | Lehký M. | 0.40-m f/5 + CCD G2-1600 + BVRI |
| 56541.37284 | 0.0004 | II | -0.0228 | CCD+R | 96/25 | Lehký M. | 0.40-m f/5 + CCD G2-1600 + BVRI |
| | | | show all | | | | |
| V0859 Cyg | | | | | | | |
| 56462.45639 | 0.0002 | II | +0.0081 | CCD+R | 167/84 | Lehký M. | EQ6 + 0.25-m f/4 + CCD ST7 + R |
| 56482.50491 | 0.0002 | I | +0.0091 | CCD+R | 136/68 | Lehký M. | EQ6 + 0.25-m f/4 + CCD ST7 + R |
| 56865.43523 | 0.0003 | II | +0.0103 | CCD+R | 82/42 | Lehký M. | EQ6 + 0.25-m f/4 + CCD ST7 + R |
| | | | show all | | | | |
| V0887 Cyg | | | | | | | |
| 56824.45253 | 0.0002 | I | -0.0089 | CCD+Clear | 267/108 | Trnka J. | 200/1 000, ST-9E |
| | | | | | | | |
| V0889 Cyg | | | | | | | |
| 56835.40924 | 0.0008 | I | -0.0121 | CCD+Clear | 244/100 | Urbaník M. | ED 80/600,0.5x Reductor, G1 0300 |
| | | | | | | | |
| V0995 Cyg | | | | | | | |
| 56588.28527 | 0.0001 | I | +0.0326 | CCD+V | 147/35 | Šmelcer L. | Newton 254/1200 + CCD G2 402 |
| 56588.28550 | 0.0001 | I | +0.0328 | CCD+R | 152/41 | Šmelcer L. | Newton 254/1200 + CCD G2 402 |
| | | | show all | | | | |
| V1073 Cyg | | | | | | | |
| 56574.32099 | 0.0001 | I | -0.0041 | CCD+V | 429/177 | Šmelcer L. | Newton 254/1200 + CCD G2 402 |
| 56574.32138 | 0.0002 | I | -0.0037 | CCD+R | 427/181 | Šmelcer L. | Newton 254/1200 + CCD G2 402 |
| 56574.32157 | 0.0001 | I | -0.0036 | CCD+B | 413/178 | Šmelcer L. | Newton 254/1200 + CCD G2 402 |
| | | | show all | | | | |
| V1141 Cyg | | | | | | | |
| 56872.52045 | 0.0007 | II | +0.0022 | DSLR | 52/33 | Mašek M. | N150/750mm + Canon EOS 1000D |
| V1898 Cyg | | | | | | | |
| 56596.43551 | 0.0006 | I | +0.0111 | CCD+V | 101/49 | Šmelcer L. | Celestron 355/2460 + CCD G2 1600 |
| 56596.43745 | 0.0007 | I | +0.0131 | CCD+I | 81/44 | Šmelcer L. | Celestron 355/2460 + CCD G2 1600 |
| 56596.44065 | 0.0010 | I | +0.0163 | CCD+R | 98/42 | Šmelcer L. | Celestron 355/2460 + CCD G2 1600 |
| | | | show all | | | | |
| V2197 Cyg | | | | | | | |
| 56878.47356 | 0.0018 | II | -0.0007 | CCD+Clear | 115/58 | Hladík B. | RF F200, ATIK 320E, CG-4 |
| V2278 Cyg | | | | | | | |
| 56845.52989 | 0.0004 | I | +0.0583 | CCD+I | 119/67 | Mazanec J. | N400, G2 402 |
| 56845.53197 | 0.0004 | I | +0.0604 | CCD+R | 128/81 | Mazanec J. | N400, G2 402 |
| 56855.48051 | 0.0010 | II | +0.0589 | CCD+I | 348/157 | Mazanec J. | N400, G2 402 |
| 56855.48875 | 0.0006 | II | +0.0672 | CCD+R | 317/171 | Mazanec J. | N400, G2 402 |
| 56863.44549 | 0.0004 | II | +0.0639 | CCD+I | 336/165 | Mazanec J. | N400, G2 402 |
| 56863.44905 | 0.0006 | II | +0.0674 | CCD+R | 294/158 | Mazanec J. | N400, G2 402 |
| 56878.47610 | 0.0005 | I | +0.0789 | CCD+I | 171/46 | Mazanec J. | N400, G2 402 |
| 56878.47630 | 0.0003 | I | +0.0791 | CCD+R | 193/55 | Mazanec J. | N400, G2 402 |
| | | | show all | | | | |
| V2477 Cyg | | | | | | | |
| 56819.41424 | 0.0001 | II | +0.0024 | CCD+Clear | 203/125 | Urbaník M. | ED 80/600,0.5x Reductor, G1 0300 |
| 56833.41987 | 0.0003 | II | +0.0018 | CCD+Clear | 185/114 | Urbaník M. | ED 80/600,0.5x Reductor, G1 0300 |
| | | | show all | | | | |
| V2544 Cyg | | | | | | | |
| 56857.42497 | 0.0002 | I | +0.0083 | CCD+Clear | 64/28 | Šmelcer L. | Celestron 280/1765 + CCD G2 4000 |
| 56857.42499 | 0.0004 | I | +0.0084 | CCD+R | 65/28 | Šmelcer L. | Celestron 280/1765 + CCD G2 4000 |
| 56857.42555 | 0.0003 | I | +0.0089 | CCD+V | 65/31 | Šmelcer L. | Celestron 280/1765 + CCD G2 4000 |
| | | | show all | | | | |
| V2626 Cyg | | | | | | | |
| 56864.45095 | 0.0006 | I | -0.0326 | CCD+R | 113/25 | Hanžl D. | 0.2-m RL + CCD G2 8300 (DATEL telescope) |
| WZ Cyg | | | | | | | |
| 56878.48217 | 0.0003 | I | -0.0153 | CCD+Clear | 121/72 | Hladík B. | RF F200, ATIK 320E, CG-4 |
| BG Del | | | | | | | |
| 56818.44547 | 0.0003 | I | -0.0145 | CCD+Clear | 180/72 | Trnka J. | 200/1 000, ST-9E |
| OZ Del | | | | | | | |
| 56844.57395 | 0.0004 | I | -0.0612 | CCD+V | 66/34 | Campos F. | N-200 f/4.7, NEQ6; Atik 314L+, BVRcIc |
| AV Dor | | | | | | | |
| 56883.88416 | 0.0004 | I | +0.0003 | CCD+R | 55/40 | M. Mašek, K. Hoňková, J. Juryšek | FRAM, 0.3m SCT + CCD G2-1600 |
| 56883.88485 | 0.0004 | I | +0.0010 | CCD+V | 55/42 | M. Mašek, K. Hoňková, J. Juryšek | FRAM, 0.3m SCT + CCD G2-1600 |
| 56883.88494 | 0.0004 | I | +0.0011 | CCD+I | 56/40 | M. Mašek, K. Hoňková, J. Juryšek | FRAM, 0.3m SCT + CCD G2-1600 |
| 56883.88965 | 0.0005 | I | +0.0058 | CCD+B | 43/29 | M. Mašek, K. Hoňková, J. Juryšek | FRAM, 0.3m SCT + CCD G2-1600 |
| | | | show all | | | | |
| AK Dra | | | | | | | |
| 56861.53945 | 0.0003 | I | +0.0340 | CCD+V | 60/47 | Šmelcer L. | Celestron 280/1765 + CCD G2 4000 |
| 56861.54015 | 0.0001 | I | +0.0347 | CCD+Clear | 59/47 | Šmelcer L. | Celestron 280/1765 + CCD G2 4000 |
| 56861.54029 | 0.0004 | I | +0.0348 | CCD+R | 66/50 | Šmelcer L. | Celestron 280/1765 + CCD G2 4000 |
| | | | show all | | | | |
| AU Dra | | | | | | | |
| 56851.47159 | 0.0005 | I | -0.0085 | CCD+V | 57/45 | Lehký M. | 0.40-m f/5 + CCD G2-1600 + BVRI |
| 56851.47291 | 0.0004 | I | -0.0072 | CCD+I | 59/48 | Lehký M. | 0.40-m f/5 + CCD G2-1600 + BVRI |
| 56851.47304 | 0.0004 | I | -0.0071 | CCD+R | 60/48 | Lehký M. | 0.40-m f/5 + CCD G2-1600 + BVRI |
| | | | show all | | | | |
| BX Dra | | | | | | | |
| 56711.46151 | 0.0002 | I | +0.0133 | CCD+R | 144/68 | Lehký M. | EQ6 + 0.25-m f/4 + CCD ST7 + R |
| 56746.49265 | 0.0002 | II | +0.0133 | CCD+R | 142/61 | Lehký M. | EQ6 + 0.25-m f/4 + CCD ST7 + R |
| | | | show all | | | | |
| CM Dra | | | | | | | |
| 56894.42075 | 0.0002 | I | +0.0033 | CCD+R | 103/45 | Šmelcer L. | Celestron 280/1765 + CCD G2 4000 |
| 56894.42085 | 0.0001 | I | +0.0034 | CCD+Clear | 108/51 | Šmelcer L. | Celestron 280/1765 + CCD G2 4000 |
| | | | show all | | | | |
| CV Dra | | | | | | | |
| 56821.40911 | 0.0002 | I | +0.0085 | CCD+V | 85/37 | Lehký M. | 0.40-m f/5 + CCD G2-1600 + BVRI |
| 56821.40960 | 0.0002 | I | +0.0090 | CCD+R | 83/37 | Lehký M. | 0.40-m f/5 + CCD G2-1600 + BVRI |
| 56821.41070 | 0.0003 | I | +0.0101 | CCD+B | 78/35 | Lehký M. | 0.40-m f/5 + CCD G2-1600 + BVRI |
| 56821.41075 | 0.0002 | I | +0.0102 | CCD+I | 79/35 | Lehký M. | 0.40-m f/5 + CCD G2-1600 + BVRI |
| 56862.47764 | 0.0003 | II | +0.0089 | CCD+I | 161/106 | Lehký M. | 0.40-m f/5 + CCD G2-1600 + BVRI |
| 56862.47833 | 0.0003 | II | +0.0096 | CCD+R | 158/104 | Lehký M. | 0.40-m f/5 + CCD G2-1600 + BVRI |
| 56862.47841 | 0.0003 | II | +0.0097 | CCD+V | 161/106 | Lehký M. | 0.40-m f/5 + CCD G2-1600 + BVRI |
| 56862.47883 | 0.0003 | II | +0.0101 | CCD+B | 160/104 | Lehký M. | 0.40-m f/5 + CCD G2-1600 + BVRI |
| | | | show all | | | | |
| EF Dra | | | | | | | |
| 56476.46846 | 0.0005 | I | +0.0104 | CCD+B | 25/11 | Lehký M. | 0.40-m f/5 + CCD G2-1600 + BVRI |
| 56476.46858 | 0.0005 | I | +0.0105 | CCD+I | 27/14 | Lehký M. | 0.40-m f/5 + CCD G2-1600 + BVRI |
| 56476.46874 | 0.0003 | I | +0.0106 | CCD+V | 26/12 | Lehký M. | 0.40-m f/5 + CCD G2-1600 + BVRI |
| 56476.47028 | 0.0008 | I | +0.0122 | CCD+R | 25/11 | Lehký M. | 0.40-m f/5 + CCD G2-1600 + BVRI |
| 56765.45203 | 0.0016 | II | +0.0182 | CCD+I | 40/29 | Šmelcer L. | Celestron 280/1765 + CCD G2 4000 |
| 56765.45248 | 0.0010 | II | +0.0186 | CCD+R | 43/34 | Šmelcer L. | Celestron 280/1765 + CCD G2 4000 |
| 56765.45289 | 0.0003 | II | +0.0190 | CCD+B | 65/44 | Šmelcer L. | Celestron 280/1765 + CCD G2 4000 |
| 56765.45362 | 0.0006 | II | +0.0198 | CCD+V | 37/28 | Šmelcer L. | Celestron 280/1765 + CCD G2 4000 |
| 56765.45368 | 0.0007 | II | +0.0198 | CCD+Clear | 44/28 | Šmelcer L. | Celestron 280/1765 + CCD G2 4000 |
| | | | show all | | | | |
| GM Dra | | | | | | | |
| 56845.43360 | 0.0004 | I | -0.0057 | DSLR | 124/100 | Medulka T. | Refraktor 90/500 + Canon 450D |
| HZ Dra | | | | | | | |
| 56460.43956 | 0.0008 | I | -0.0101 | CCD+V | 90/43 | Lehký M. | CG4 + 0.096-m f/4.4 + CCD ST5C + BVRI |
| 56460.43960 | 0.0012 | I | -0.0100 | CCD+I | 73/37 | Lehký M. | CG4 + 0.096-m f/4.4 + CCD ST5C + BVRI |
| 56460.44042 | 0.0022 | I | -0.0092 | CCD+R | 78/35 | Lehký M. | CG4 + 0.096-m f/4.4 + CCD ST5C + BVRI |
| 56865.46387 | 0.0006 | I | -0.0037 | CCD+Clear | 195/118 | Audejean M. | Tele F = 150 mm + CCD Kaf 1603 |
| | | | show all | | | | |
| MZ Dra | | | | | | | |
| 56712.59623 | 0.0001 | I | +0.2072 | CCD+R | 261/110 | Hanžl D. | 0.2-m RL + CCD G2 8300 (DATEL telescope) |
| 56728.56220 | 0.0001 | I | +0.2067 | CCD+R | 303/139 | Hanžl D. | 0.2-m RL + CCD G2 8300 (DATEL telescope) |
| | | | show all | | | | |
| NN Dra | | | | | | | |
| 56683.49434 | 0.0003 | I | -0.0028 | CCD+R | 223/51 | Lehký M. | EQ6 + 0.25-m f/4 + CCD ST7 + R |
| 56683.68717 | 0.0002 | II | -0.0010 | CCD+R | 223/207 | Lehký M. | EQ6 + 0.25-m f/4 + CCD ST7 + R |
| | | | show all | | | | |
| NSVS 01286630 Dra | | | | | | | |
| 56786.52032 | 0.0003 | II | ------- | CCD+R | 77/67 | Šmelcer L. | Celestron 280/1765 + CCD G2 4000 |
| 56786.52058 | 0.0002 | II | ------- | CCD+Clear | 90/73 | Šmelcer L. | Celestron 280/1765 + CCD G2 4000 |
| 56786.52100 | 0.0005 | II | ------- | CCD+I | 84/74 | Šmelcer L. | Celestron 280/1765 + CCD G2 4000 |
| 56790.35887 | 0.0003 | II | ------- | CCD+V | 55/6 | Šmelcer L. | Celestron 280/1765 + CCD G2 4000 |
| 56790.35944 | 0.0001 | II | ------- | CCD+V | 88/11 | Šmelcer L. | Celestron 355/2460 + CCD G2 1600 |
| 56790.35957 | 0.0001 | II | ------- | CCD+Clear | 48/6 | Šmelcer L. | Celestron 280/1765 + CCD G2 4000 |
| 56790.35959 | 0.0001 | II | ------- | CCD+R | 76/11 | Šmelcer L. | Celestron 355/2460 + CCD G2 1600 |
| 56790.35971 | 0.0001 | II | ------- | CCD+R | 57/5 | Šmelcer L. | Celestron 280/1765 + CCD G2 4000 |
| 56790.35975 | 0.0002 | II | ------- | CCD+I | 74/11 | Šmelcer L. | Celestron 355/2460 + CCD G2 1600 |
| 56790.36015 | 0.0004 | II | ------- | CCD+I | 50/5 | Šmelcer L. | Celestron 280/1765 + CCD G2 4000 |
| 56813.39417 | 0.0003 | II | ------- | CCD+V | 54/5 | Šmelcer L. | Celestron 280/1765 + CCD G2 4000 |
| 56813.39534 | 0.0002 | II | ------- | CCD+Clear | 54/6 | Šmelcer L. | Celestron 280/1765 + CCD G2 4000 |
| 56813.39553 | 0.0003 | II | ------- | CCD+R | 60/6 | Šmelcer L. | Celestron 280/1765 + CCD G2 4000 |
| 56818.38590 | 0.0001 | I | ------- | CCD+Clear | 64/12 | Šmelcer L. | Celestron 280/1765 + CCD G2 4000 |
| 56818.38623 | 0.0002 | I | ------- | CCD+V | 83/14 | Šmelcer L. | Celestron 280/1765 + CCD G2 4000 |
| 56818.38657 | 0.0002 | I | ------- | CCD+R | 82/13 | Šmelcer L. | Celestron 280/1765 + CCD G2 4000 |
| 56824.52839 | 0.0001 | II | ------- | CCD+R | 149/145 | Šmelcer L. | Celestron 280/1765 + CCD G2 4000 |
| 56825.48870 | 0.0002 | I | ------- | CCD+R | 140/111 | Šmelcer L. | Celestron 280/1765 + CCD G2 4000 |
| 56827.40837 | 0.0001 | I | ------- | CCD+R | 123/55 | Šmelcer L. | Celestron 280/1765 + CCD G2 4000 |
| 56832.39941 | 0.0001 | I | ------- | CCD+R | 125/32 | Šmelcer L. | Celestron 280/1765 + CCD G2 4000 |
| NSVS 2690221 Dra | | | | | | | |
| 56683.57820 | 0.0003 | II | ------- | CCD+R | 222/116 | Lehký M. | EQ6 + 0.25-m f/4 + CCD ST7 + R |
| QU Dra | | | | | | | |
| 56815.45666 | 0.0003 | I | -0.0006 | CCD+R | 149/76 | Hanžl D. | 0.2-m RL + CCD G2 8300 (DATEL telescope) |
| V0341 Dra | | | | | | | |
| 56826.41478 | 0.0003 | I | +0.0056 | CCD+Clear | 272/174 | Urbaník M. | ED 80/600,0.5x Reductor, G1 0300 |
| V0349 Dra | | | | | | | |
| 56572.35362 | 0.0001 | I | -0.0119 | CCD+Clear | 260/145 | Jacobsen J. | EQ6, 100 mm refractor, Atik 314 |
| V0372 Dra | | | | | | | |
| 56790.37363 | 0.0002 | I | +0.0097 | CCD+Clear | 267/94 | Urbaník M. | ED 80/600,0.5x Reductor, G1 0300 |
| V0374 Dra | | | | | | | |
| 56728.40979 | 0.0015 | I | +0.0292 | CCD+Clear | 135/104 | Jacobsen J. | EQ6, 80 mm refractor, SX-M9 |
| VSX J201516.1+645805 Dra | | | | | | | |
| 56657.25041 | 0.0005 | I | ------- | CCD+Clear | 204/59 | Trnka J. | 200/1 000, ST-9E |
| Z Dra | | | | | | | |
| 56703.51507 | 0.0001 | I | +0.0181 | CCD+B | 65/49 | Šmelcer L. | Celestron 355/2460 + CCD G2 1600 |
| 56703.51526 | 0.0001 | I | +0.0183 | CCD+I | 72/53 | Šmelcer L. | Celestron 355/2460 + CCD G2 1600 |
| 56703.51541 | 0.0001 | I | +0.0184 | CCD+V | 74/55 | Šmelcer L. | Celestron 355/2460 + CCD G2 1600 |
| 56703.51554 | 0.0001 | I | +0.0186 | CCD+R | 74/55 | Šmelcer L. | Celestron 355/2460 + CCD G2 1600 |
| 56771.38829 | 0.0001 | I | +0.0198 | CCD+R | 36/22 | Šmelcer L. | Celestron 280/1765 + CCD G2 4000 |
| 56771.38833 | 0.0002 | I | +0.0198 | CCD+Clear | 36/23 | Šmelcer L. | Celestron 280/1765 + CCD G2 4000 |
| 56771.38848 | 0.0001 | I | +0.0200 | CCD+V | 38/23 | Šmelcer L. | Celestron 280/1765 + CCD G2 4000 |
| 56771.38860 | 0.0002 | I | +0.0201 | CCD+I | 36/22 | Šmelcer L. | Celestron 280/1765 + CCD G2 4000 |
| 56771.38869 | 0.0002 | I | +0.0202 | CCD+B | 42/24 | Šmelcer L. | Celestron 280/1765 + CCD G2 4000 |
| | | | show all | | | | |
| ASAS J030827-1609.1 Eri | | | | | | | |
| 56626.63248 | 0.0010 | I | ------- | CCD+R | 43/27 | M. Mašek, K. Hoňková, J. Juryšek | FRAM, Nikkor 300mm + G4-16000 |
| 56631.63163 | 0.0024 | I | ------- | CCD+R | 53/33 | M. Mašek, K. Hoňková, J. Juryšek | FRAM, Nikkor 300mm + G4-16000 |
| GH Eri | | | | | | | |
| 56896.76742 | 0.0002 | I | +0.0133 | CCD+I | 27/15 | M. Mašek, K. Hoňková, J. Juryšek | FRAM, 0.3m SCT + CCD G2-1600 |
| 56896.76753 | 0.0001 | I | +0.0134 | CCD+B | 28/16 | M. Mašek, K. Hoňková, J. Juryšek | FRAM, 0.3m SCT + CCD G2-1600 |
| 56896.76754 | 0.0001 | I | +0.0134 | CCD+V | 28/17 | M. Mašek, K. Hoňková, J. Juryšek | FRAM, 0.3m SCT + CCD G2-1600 |
| 56896.76771 | 0.0001 | I | +0.0136 | CCD+R | 27/16 | M. Mašek, K. Hoňková, J. Juryšek | FRAM, 0.3m SCT + CCD G2-1600 |
| | | | show all | | | | |
| HX Eri | | | | | | | |
| 56629.56533 | 0.0009 | I | -0.0204 | CCD+R | 56/8 | M. Mašek, K. Hoňková, J. Juryšek | FRAM, Nikkor 300mm + G4-16000 |
| 56629.56771 | 0.0007 | I | -0.0180 | CCD+V | 46/7 | M. Mašek, K. Hoňková, J. Juryšek | FRAM, Nikkor 300mm + G4-16000 |
| | | | show all | | | | |
| FG Gem | | | | | | | |
| 56725.36078 | 0.0002 | I | +0.0058 | CCD+I | 75/36 | Šmelcer L. | Celestron 355/2460 + CCD G2 1600 |
| 56725.36083 | 0.0002 | I | +0.0058 | CCD+R | 87/41 | Šmelcer L. | Celestron 355/2460 + CCD G2 1600 |
| 56725.36093 | 0.0002 | I | +0.0059 | CCD+V | 76/35 | Šmelcer L. | Celestron 355/2460 + CCD G2 1600 |
| | | | show all | | | | |
| IV Gem | | | | | | | |
| 56745.37189 | 0.0014 | II | +0.0121 | CCD+R | 229/118 | Hanžl D. | 0.2-m RL + CCD G2 8300 (DATEL telescope) |
| KV Gem | | | | | | | |
| 56583.55772 | 0.0001 | I | -0.0087 | CCD+V | 157/59 | Magris M. | Schmidt-Newton 0.25m F/4, CCD Atik Titan |
| 56694.34248 | 0.0001 | I | -0.0073 | CCD+R | 234/134 | Šmelcer L. | Newton 254/1200 + CCD G2 402 |
| 56703.30462 | 0.0005 | I | -0.0082 | CCD+V | 223/12 | Šmelcer L. | Newton 254/1200 + CCD G2 402 |
| 56703.30470 | 0.0002 | I | -0.0081 | CCD+R | 211/9 | Šmelcer L. | Newton 254/1200 + CCD G2 402 |
| 56703.48443 | 0.0003 | II | -0.0076 | CCD+V | 223/180 | Šmelcer L. | Newton 254/1200 + CCD G2 402 |
| 56703.48451 | 0.0002 | II | -0.0075 | CCD+R | 211/163 | Šmelcer L. | Newton 254/1200 + CCD G2 402 |
| 56709.39953 | 0.0001 | I | -0.0081 | CCD+V | 231/170 | Šmelcer L. | Newton 254/1200 + CCD G2 402 |
| 56709.39970 | 0.0001 | I | -0.0080 | CCD+R | 238/169 | Šmelcer L. | Newton 254/1200 + CCD G2 402 |
| 56729.29865 | 0.0002 | II | -0.0070 | CCD+I | 69/16 | Šmelcer L. | Celestron 355/2460 + CCD G2 1600 |
| 56729.29897 | 0.0002 | II | -0.0066 | CCD+R | 74/18 | Šmelcer L. | Celestron 355/2460 + CCD G2 1600 |
| 56729.29908 | 0.0002 | II | -0.0065 | CCD+V | 77/15 | Šmelcer L. | Celestron 355/2460 + CCD G2 1600 |
| 56739.33762 | 0.0004 | II | -0.0066 | CCD+I | 62/28 | Šmelcer L. | Celestron 355/2460 + CCD G2 1600 |
| 56739.33898 | 0.0004 | II | -0.0052 | CCD+R | 59/27 | Šmelcer L. | Celestron 355/2460 + CCD G2 1600 |
| 56739.33911 | 0.0004 | II | -0.0051 | CCD+V | 49/21 | Šmelcer L. | Celestron 355/2460 + CCD G2 1600 |
| | | | show all | | | | |
| QW Gem | | | | | | | |
| 56629.53780 | 0.0001 | II | -0.0039 | CCD+B | 98/36 | Lehký M. | 0.40-m f/5 + CCD G2-1600 + BVRI |
| 56629.53799 | 0.0001 | II | -0.0037 | CCD+I | 97/34 | Lehký M. | 0.40-m f/5 + CCD G2-1600 + BVRI |
| 56629.53808 | 0.0001 | II | -0.0036 | CCD+V | 97/34 | Lehký M. | 0.40-m f/5 + CCD G2-1600 + BVRI |
| 56629.53865 | 0.0001 | II | -0.0030 | CCD+R | 95/35 | Lehký M. | 0.40-m f/5 + CCD G2-1600 + BVRI |
| 56630.43298 | 0.0002 | I | -0.0041 | CCD+I | 85/37 | Lehký M. | 0.40-m f/5 + CCD G2-1600 + BVRI |
| 56630.43354 | 0.0002 | I | -0.0035 | CCD+R | 88/40 | Lehký M. | 0.40-m f/5 + CCD G2-1600 + BVRI |
| 56630.43381 | 0.0002 | I | -0.0032 | CCD+V | 87/40 | Lehký M. | 0.40-m f/5 + CCD G2-1600 + BVRI |
| 56630.43398 | 0.0002 | I | -0.0031 | CCD+B | 79/37 | Lehký M. | 0.40-m f/5 + CCD G2-1600 + BVRI |
| | | | show all | | | | |
| USNO-B1.0 1135-0102876 Gem | | | | | | | |
| 56629.41524 | 0.0008 | I | ------- | CCD+V | 43/34 | Šmelcer L. | Celestron 355/2460 + CCD G2 1600 |
| 56629.41563 | 0.0009 | I | ------- | CCD+I | 43/34 | Šmelcer L. | Celestron 355/2460 + CCD G2 1600 |
| 56629.41907 | 0.0006 | I | ------- | CCD+R | 93/64 | Šmelcer L. | Celestron 355/2460 + CCD G2 1600 |
| V0380 Gem | | | | | | | |
| 56725.34725 | 0.0001 | II | -0.0108 | CCD+I | 297/101 | Mazanec J. | N400, G2 402 |
| 56727.36697 | 0.0001 | II | -0.0107 | CCD+I | 303/139 | Mazanec J. | N400, G2 402 |
| 56727.36739 | 0.0001 | II | -0.0103 | CCD+Clear | 299/141 | Mazanec J. | N400, G2 402 |
| 56729.38624 | 0.0001 | II | -0.0111 | CCD+Clear | 656/111 | Lomoz F. | Newton 300/1200+ST2000XM |
| 56729.38861 | 0.0002 | II | -0.0088 | CCD+I | 305/204 | Mazanec J. | N400, G2 402 |
| 56729.38861 | 0.0001 | II | -0.0088 | CCD+R | 320/216 | Mazanec J. | N400, G2 402 |
| 56746.38546 | 0.0002 | I | -0.0111 | CCD+R | 227/167 | Mazanec J. | N400, G2 402 |
| 56746.38593 | 0.0003 | I | -0.0106 | CCD+I | 202/170 | Mazanec J. | N400, G2 402 |
| | | | show all | | | | |
| V0382 Gem | | | | | | | |
| 56712.33356 | 0.0001 | I | -0.0204 | CCD+R | 313/144 | Lehký M. | EQ6 + 0.25-m f/4 + CCD ST7 + R |
| 56725.34664 | 0.0002 | I | -0.0240 | CCD+Clear | 179/92 | Mašek M. | N150/600mm + CCD Meade DSI |
| 56746.31959 | 0.0003 | II | ------- | CCD+R | 156/58 | Lehký M. | EQ6 + 0.25-m f/4 + CCD ST7 + R |
| | | | show all | | | | |
| V0388 Gem | | | | | | | |
| 56643.50663 | 0.0001 | I | -0.0039 | CCD+R | 103/46 | Šmelcer L. | Newton 254/1200 + CCD G2 402 |
| 56643.50674 | 0.0001 | I | -0.0038 | CCD+V | 91/44 | Šmelcer L. | Newton 254/1200 + CCD G2 402 |
| | | | show all | | | | |
| V0404 Gem | | | | | | | |
| 56694.35704 | 0.0005 | I | +0.0135 | CCD+R | 188/113 | Šmelcer L. | Newton 254/1200 + CCD G2 402 |
| 56703.42026 | 0.0003 | I | +0.0104 | CCD+V | 223/128 | Šmelcer L. | Newton 254/1200 + CCD G2 402 |
| 56703.42104 | 0.0001 | I | +0.0112 | CCD+R | 187/93 | Šmelcer L. | Newton 254/1200 + CCD G2 402 |
| 56709.34889 | 0.0001 | I | +0.0111 | CCD+R | 240/120 | Šmelcer L. | Newton 254/1200 + CCD G2 402 |
| 56709.34893 | 0.0002 | I | +0.0111 | CCD+V | 199/96 | Šmelcer L. | Newton 254/1200 + CCD G2 402 |
| 56729.39682 | 0.0003 | II | +0.0085 | CCD+R | 66/53 | Šmelcer L. | Celestron 355/2460 + CCD G2 1600 |
| 56729.39751 | 0.0003 | II | +0.0092 | CCD+V | 76/55 | Šmelcer L. | Celestron 355/2460 + CCD G2 1600 |
| 56729.39759 | 0.0005 | II | +0.0093 | CCD+I | 64/54 | Šmelcer L. | Celestron 355/2460 + CCD G2 1600 |
| 56739.33851 | 0.0005 | I | +0.0121 | CCD+R | 59/26 | Šmelcer L. | Celestron 355/2460 + CCD G2 1600 |
| 56739.33876 | 0.0003 | I | +0.0123 | CCD+I | 54/28 | Šmelcer L. | Celestron 355/2460 + CCD G2 1600 |
| 56739.33890 | 0.0003 | I | +0.0124 | CCD+V | 51/23 | Šmelcer L. | Celestron 355/2460 + CCD G2 1600 |
| | | | show all | | | | |
| V0405 Gem | | | | | | | |
| 56694.33895 | 0.0011 | II | -0.0043 | CCD+R | 176/100 | Šmelcer L. | Newton 254/1200 + CCD G2 402 |
| 56703.33741 | 0.0006 | I | -0.0168 | CCD+R | 171/28 | Šmelcer L. | Newton 254/1200 + CCD G2 402 |
| 56703.34061 | 0.0007 | I | -0.0136 | CCD+V | 204/51 | Šmelcer L. | Newton 254/1200 + CCD G2 402 |
| 56709.34236 | 0.0005 | I | -0.0193 | CCD+V | 166/92 | Šmelcer L. | Newton 254/1200 + CCD G2 402 |
| 56709.34350 | 0.0001 | I | -0.0181 | CCD+R | 193/103 | Šmelcer L. | Newton 254/1200 + CCD G2 402 |
| 56739.37875 | 0.0015 | I | -0.0201 | CCD+I | 42/33 | Šmelcer L. | Celestron 355/2460 + CCD G2 1600 |
| 56739.38373 | 0.0032 | I | -0.0151 | CCD+V | 42/35 | Šmelcer L. | Celestron 355/2460 + CCD G2 1600 |
| 56739.38417 | 0.0014 | I | -0.0147 | CCD+R | 47/37 | Šmelcer L. | Celestron 355/2460 + CCD G2 1600 |
| | | | show all | | | | |
| V0428 Gem | | | | | | | |
| 56680.42042 | 0.0001 | I | -0.0131 | CCD+R | 122/69 | F. Scaggiante, D. Zardin | Newton 410/1710 CCD Audine KAF402ME |
| WW Gem | | | | | | | |
| 56629.39562 | 0.0003 | II | +0.0080 | CCD+V | 95/47 | Šmelcer L. | Celestron 355/2460 + CCD G2 1600 |
| 56629.39571 | 0.0004 | II | +0.0081 | CCD+R | 92/53 | Šmelcer L. | Celestron 355/2460 + CCD G2 1600 |
| 56629.39769 | 0.0004 | II | +0.0101 | CCD+I | 100/55 | Šmelcer L. | Celestron 355/2460 + CCD G2 1600 |
| | | | show all | | | | |
| Y Gru | | | | | | | |
| 56806.81879 | 0.0004 | I | -0.0457 | CCD+Clear | 62/9 | C. Quiñones, L. Tapia | Schmidt-Cassegrain 14\" + RF + CCD |
| FW Her | | | | | | | |
| 56712.60162 | 0.0003 | I | +0.0166 | CCD+R | 67/27 | Lehký M. | 0.40-m f/5 + CCD G2-1600 + BVRI |
| 56712.60236 | 0.0003 | I | +0.0174 | CCD+V | 67/28 | Lehký M. | 0.40-m f/5 + CCD G2-1600 + BVRI |
| 56842.50347 | 0.0010 | II | +0.0113 | CCD+V | 83/57 | Lehký M. | 0.40-m f/5 + CCD G2-1600 + BVRI |
| 56842.50560 | 0.0008 | II | +0.0134 | CCD+R | 87/59 | Lehký M. | 0.40-m f/5 + CCD G2-1600 + BVRI |
| | | | show all | | | | |
| GSC 02594-00971 Her | | | | | | | |
| 56802.49606 | 0.0005 | I | ------- | CCD+Clear | 247/156 | Trnka J. | 200/1 000, ST-9E |
| IX Her | | | | | | | |
| 56845.51928 | 0.0004 | I | +0.1430 | CCD+I | 76/55 | Lehký M. | 0.40-m f/5 + CCD G2-1600 + BVRI |
| 56845.51947 | 0.0003 | I | +0.1432 | CCD+R | 87/64 | Lehký M. | 0.40-m f/5 + CCD G2-1600 + BVRI |
| 56845.51994 | 0.0005 | I | +0.1437 | CCD+V | 80/57 | Lehký M. | 0.40-m f/5 + CCD G2-1600 + BVRI |
| | | | show all | | | | |
| ROTSE1 J170438.01+330348.3 Her | | | | | | | |
| 56802.41172 | 0.0004 | I | ------- | CCD+Clear | 246/48 | Trnka J. | 200/1 000, ST-9E |
| ROTSE1 J173413.59+440118.5 Her | | | | | | | |
| 56461.47572 | 0.0004 | I | ------- | CCD+R | 155/101 | Lehký M. | EQ6 + 0.25-m f/4 + CCD ST7 + R |
| 56818.37485 | 0.0007 | I | ------- | CCD+R | 154/25 | Lehký M. | EQ6 + 0.25-m f/4 + CCD ST7 + R |
| 56818.49138 | 0.0008 | II | ------- | CCD+R | 154/124 | Lehký M. | EQ6 + 0.25-m f/4 + CCD ST7 + R |
| 56878.46130 | 0.0014 | I | ------- | CCD+R | 155/107 | Lehký M. | EQ6 + 0.25-m f/4 + CCD ST7 + R |
| V0381 Her | | | | | | | |
| 56476.40565 | 0.0001 | I | +0.0086 | CCD+R | 85/50 | Lehký M. | EQ6 + 0.25-m f/4 + CCD ST7 + R |
| V0829 Her | | | | | | | |
| 56781.55277 | 0.0003 | I | +0.0070 | CCD+R | 134/88 | Lehký M. | EQ6 + 0.25-m f/4 + CCD ST7 + R |
| 56815.39767 | 0.0004 | II | +0.0056 | CCD+R | 108/67 | Lehký M. | EQ6 + 0.25-m f/4 + CCD ST7 + R |
| | | | show all | | | | |
| V1002 Her | | | | | | | |
| 56831.52103 | 0.0015 | I | -0.0084 | CCD+Clear | 162/98 | Hladík B. | RF F200, ATIK 320E, CG-4 |
| V1031 Her | | | | | | | |
| 56750.55300 | 0.0003 | I | +0.0056 | CCD+R | 192/104 | Hanžl D. | 0.2-m RL + CCD G2 8300 (DATEL telescope) |
| V1035 Her | | | | | | | |
| 56839.41498 | 0.0002 | I | +0.0041 | DSLR | 63/22 | Walter F. | RF Comet finder 20/137, Canon 350 D |
| V1036 Her | | | | | | | |
| 56456.43938 | 0.0001 | II | +0.0073 | CCD+R | 283/118 | Lehký M. | EQ6 + 0.25-m f/4 + CCD ST7 + R |
| 56750.59094 | 0.0002 | II | +0.0099 | CCD+R | 202/124 | Lehký M. | EQ6 + 0.25-m f/4 + CCD ST7 + R |
| 56851.44594 | 0.0002 | I | +0.0083 | CCD+R | 170/114 | Lehký M. | EQ6 + 0.25-m f/4 + CCD ST7 + R |
| | | | show all | | | | |
| V1041 Her | | | | | | | |
| 56455.41629 | 0.0007 | I | -0.0058 | CCD+B | 32/17 | Lehký M. | 0.40-m f/5 + CCD G2-1600 + BVRI |
| 56455.41707 | 0.0003 | I | -0.0050 | CCD+R | 31/17 | Lehký M. | 0.40-m f/5 + CCD G2-1600 + BVRI |
| 56455.41733 | 0.0004 | I | -0.0047 | CCD+V | 28/16 | Lehký M. | 0.40-m f/5 + CCD G2-1600 + BVRI |
| 56455.41737 | 0.0005 | I | -0.0047 | CCD+I | 31/15 | Lehký M. | 0.40-m f/5 + CCD G2-1600 + BVRI |
| 56816.38819 | 0.0012 | I | -0.0104 | CCD+B | 20/10 | Lehký M. | 0.40-m f/5 + CCD G2-1600 + BVRI |
| 56816.38893 | 0.0004 | I | -0.0096 | CCD+R | 21/9 | Lehký M. | 0.40-m f/5 + CCD G2-1600 + BVRI |
| 56816.38958 | 0.0005 | I | -0.0090 | CCD+I | 19/8 | Lehký M. | 0.40-m f/5 + CCD G2-1600 + BVRI |
| 56816.38975 | 0.0005 | I | -0.0088 | CCD+V | 19/9 | Lehký M. | 0.40-m f/5 + CCD G2-1600 + BVRI |
| | | | show all | | | | |
| V1051 Her | | | | | | | |
| 56856.50280 | 0.0008 | II | +0.1756 | CCD+R | 162/122 | Lehký M. | EQ6 + 0.25-m f/4 + CCD ST7 + R |
| V1055 Her | | | | | | | |
| 56463.42651 | 0.0002 | II | -0.0032 | CCD+R | 126/63 | Lehký M. | EQ6 + 0.25-m f/4 + CCD ST7 + R |
| 56816.37008 | 0.0003 | II | -0.0023 | CCD+R | 52/27 | Lehký M. | EQ6 + 0.25-m f/4 + CCD ST7 + R |
| 56842.39018 | 0.0002 | I | -0.0035 | CCD+R | 78/36 | Lehký M. | EQ6 + 0.25-m f/4 + CCD ST7 + R |
| | | | show all | | | | |
| V1062 Her | | | | | | | |
| 56461.39992 | 0.0002 | I | +0.0136 | CCD+R | 166/35 | Lehký M. | EQ6 + 0.25-m f/4 + CCD ST7 + R |
| 56461.52362 | 0.0002 | II | +0.0116 | CCD+R | 166/136 | Lehký M. | EQ6 + 0.25-m f/4 + CCD ST7 + R |
| 56818.44229 | 0.0004 | I | +0.0183 | CCD+R | 145/83 | Lehký M. | EQ6 + 0.25-m f/4 + CCD ST7 + R |
| 56878.40555 | 0.0008 | II | +0.0143 | CCD+R | 164/70 | Lehký M. | EQ6 + 0.25-m f/4 + CCD ST7 + R |
| | | | show all | | | | |
| V1064 Her | | | | | | | |
| 56808.56536 | 0.0003 | II | +0.0277 | CCD+I | 171/114 | Mazanec J. | N400, G2 402 |
| 56808.56864 | 0.0002 | II | +0.0310 | CCD+R | 153/109 | Mazanec J. | N400, G2 402 |
| 56835.50603 | 0.0003 | I | +0.0291 | CCD+R | 119/82 | Mazanec J. | N400, G2 402 |
| 56835.50613 | 0.0003 | I | +0.0292 | CCD+I | 119/84 | Mazanec J. | N400, G2 402 |
| 56835.50649 | 0.0003 | I | +0.0295 | CCD+V | 119/82 | Mazanec J. | N400, G2 402 |
| | | | show all | | | | |
| V1070 Her | | | | | | | |
| 56818.45363 | 0.0005 | I | +0.5991 | CCD+R | 144/73 | Hanžl D. | 0.2-m RL + CCD G2 8300 (DATEL telescope) |
| V1091 Her | | | | | | | |
| 56805.31182 | 0.0002 | II | -0.0065 | CCD+Clear | 202/88 | Öğmen Y. | 14"LX200R + SBIG ST-8XME |
| V1097 Her | | | | | | | |
| 56460.51905 | 0.0001 | I | +0.0100 | CCD+R | 96/73 | Lehký M. | EQ6 + 0.25-m f/4 + CCD ST7 + R |
| 56522.40565 | 0.0001 | II | +0.0114 | CCD+R | 179/122 | Lehký M. | EQ6 + 0.25-m f/4 + CCD ST7 + R |
| 56777.52651 | 0.0001 | II | +0.0134 | CCD+R | 228/127 | Lehký M. | EQ6 + 0.25-m f/4 + CCD ST7 + R |
| 56817.39838 | 0.0001 | I | +0.0117 | CCD+R | 130/77 | Lehký M. | EQ6 + 0.25-m f/4 + CCD ST7 + R |
| | | | show all | | | | |
| V1100 Her | | | | | | | |
| 56475.42699 | 0.0001 | I | +0.0025 | CCD+R | 174/103 | Lehký M. | EQ6 + 0.25-m f/4 + CCD ST7 + R |
| 56727.64345 | 0.0001 | I | -0.0014 | CCD+R | 144/86 | Lehký M. | EQ6 + 0.25-m f/4 + CCD ST7 + R |
| 56821.49056 | 0.0002 | II | -0.0002 | CCD+R | 174/118 | Lehký M. | EQ6 + 0.25-m f/4 + CCD ST7 + R |
| | | | show all | | | | |
| V1106 Her | | | | | | | |
| 56489.36100 | 0.0008 | II | -0.0001 | CCD+R | 281/13 | Lehký M. | EQ6 + 0.25-m f/4 + CCD ST7 + R |
| 56489.48862 | 0.0002 | I | +0.0005 | CCD+R | 281/178 | Lehký M. | EQ6 + 0.25-m f/4 + CCD ST7 + R |
| 56490.38082 | 0.0005 | II | +0.0013 | CCD+R | 280/29 | Lehký M. | EQ6 + 0.25-m f/4 + CCD ST7 + R |
| 56490.50607 | 0.0002 | I | -0.0005 | CCD+R | 280/184 | Lehký M. | EQ6 + 0.25-m f/4 + CCD ST7 + R |
| 56491.39717 | 0.0003 | II | -0.0008 | CCD+R | 173/58 | Lehký M. | EQ6 + 0.25-m f/4 + CCD ST7 + R |
| 56496.36343 | 0.0003 | I | +0.0011 | CCD+R | 242/32 | Lehký M. | EQ6 + 0.25-m f/4 + CCD ST7 + R |
| 56496.49011 | 0.0003 | II | +0.0002 | CCD+R | 242/194 | Lehký M. | EQ6 + 0.25-m f/4 + CCD ST7 + R |
| 56497.38200 | 0.0002 | I | +0.0013 | CCD+R | 294/52 | Lehký M. | EQ6 + 0.25-m f/4 + CCD ST7 + R |
| 56497.50816 | 0.0002 | II | -0.0002 | CCD+R | 294/217 | Lehký M. | EQ6 + 0.25-m f/4 + CCD ST7 + R |
| 56499.54301 | 0.0003 | II | -0.0021 | CCD+R | 189/149 | Lehký M. | EQ6 + 0.25-m f/4 + CCD ST7 + R |
| 56500.43621 | 0.0002 | I | +0.0003 | CCD+R | 306/135 | Lehký M. | EQ6 + 0.25-m f/4 + CCD ST7 + R |
| 56501.45628 | 0.0002 | I | +0.0019 | CCD+R | 294/146 | Lehký M. | EQ6 + 0.25-m f/4 + CCD ST7 + R |
| 56502.34864 | 0.0004 | II | +0.0029 | CCD+R | 171/22 | Lehký M. | EQ6 + 0.25-m f/4 + CCD ST7 + R |
| | | | show all | | | | |
| V1119 Her | | | | | | | |
| 56892.39786 | 0.0005 | I | +0.0166 | CCD+Clear | 219/129 | Urbaník M. | ED 80/600,0.5x Reductor, G1 0300 |
| V1140 Her | | | | | | | |
| 56510.49025 | 0.0003 | I | +0.0172 | CCD+Clear | 224/118 | Trnka J. | 200/1 000, ST-9E |
| 56786.51176 | 0.0005 | I | +0.0188 | CCD+I | 173/90 | Mazanec J. | N400, G2 402 |
| 56786.51190 | 0.0002 | I | +0.0189 | CCD+R | 158/81 | Mazanec J. | N400, G2 402 |
| | | | show all | | | | |
| V1147 Her | | | | | | | |
| 56781.55579 | 0.0003 | I | +0.0171 | CCD+R | 219/141 | Hanžl D. | 0.2-m RL + CCD G2 8300 (DATEL telescope) |
| V1159 Her | | | | | | | |
| 778.54261 | 0.0001 | II | ------- | CCD+R | 219/0 | Mazanec J. | N400, G2 402 |
| 56778.40146 | 0.0003 | I | -0.0087 | CCD+I | 209/10 | Mazanec J. | N400, G2 402 |
| 56778.40154 | 0.0002 | I | -0.0086 | CCD+R | 219/31 | Mazanec J. | N400, G2 402 |
| 56778.54581 | 0.0002 | II | ------- | CCD+I | 209/169 | Mazanec J. | N400, G2 402 |
| 56817.48115 | 0.0002 | I | -0.0086 | CCD+R | 150/85 | Hanžl D. | 0.2-m RL + CCD G2 8300 (DATEL telescope) |
| | | | show all | | | | |
| V1173 Her | | | | | | | |
| 56728.32034 | 0.0004 | II | ------- | CCD+Clear | 262/47 | Jacobsen J. | LXD75, 66 mm refractor, SX-M7 |
| 56728.45459 | 0.0001 | I | +0.0988 | CCD+Clear | 262/220 | Jacobsen J. | LXD75, 66 mm refractor, SX-M7 |
| V1183 Her | | | | | | | |
| 56821.44454 | 0.0003 | I | -0.0200 | CCD+R | 145/62 | Hanžl D. | 0.2-m RL + CCD G2 8300 (DATEL telescope) |
| V1198 Her | | | | | | | |
| 56814.48222 | 0.0003 | II | ------- | DSLR | 216/103 | Walter F. | RF Comet finder 20/137, Canon 350 D |
| 56818.48265 | 0.0003 | II | ------- | DSLR | 233/158 | Walter F. | RF Comet finder 20/137, Canon 350 D |
| 56819.39079 | 0.0003 | I | +0.0910 | DSLR | 230/46 | Walter F. | RF Comet finder 20/137, Canon 350 D |
| 56819.56751 | 0.0011 | II | ------- | DSLR | 230/215 | Walter F. | RF Comet finder 20/137, Canon 350 D |
| 56824.48030 | 0.0002 | I | +0.0898 | DSLR | 600/304 | Walter F. | RF Comet finder 20/137, Canon 350 D |
| | | | show all | | | | |
| V1216 Her | | | | | | | |
| 56827.39668 | 0.0002 | II | ------- | CCD+R | 143/31 | Hanžl D. | 0.2-m RL + CCD G2 8300 (DATEL telescope) |
| 56827.54739 | 0.0003 | I | +0.0787 | CCD+R | 143/126 | Hanžl D. | 0.2-m RL + CCD G2 8300 (DATEL telescope) |
| V1233 Her | | | | | | | |
| 56821.51955 | 0.0004 | I | +0.0741 | DSLR | 286/206 | Walter F. | RF Comet finder 20/137, Canon 350 D |
| 56826.48906 | 0.0004 | I | +0.0740 | DSLR | 171/106 | Walter F. | RF Comet finder 20/137, Canon 350 D |
| 56842.48384 | 0.0004 | I | -0.0823 | DSLR | 232/142 | Walter F. | RF Comet finder 20/137, Canon 350 D |
| 56878.51310 | 0.0007 | I | -0.0826 | DSLR | 145/96 | Walter F. | RF Comet finder 20/137, Canon 350 D |
| | | | show all | | | | |
| V0948 Her | | | | | | | |
| 56815.45736 | 0.0005 | I | -0.0188 | CCD+Clear | 144/53 | Hladík B. | RF 200/6, ATIK 320E, CG-4 |
| ASAS J102420-1833.4 Hya | | | | | | | |
| 56765.53978 | 0.0019 | I | ------- | CCD+R | 92/22 | M. Mašek, K. Hoňková, J. Juryšek | FRAM, Nikkor 300mm + G4-16000 |
| ASAS J102528-1911.0 Hya | | | | | | | |
| 56765.49156 | 0.0006 | I | ------- | CCD+R | 61/7 | M. Mašek, K. Hoňková, J. Juryšek | FRAM, Nikkor 300mm + G4-16000 |
| 56765.62112 | 0.0011 | II | ------- | CCD+R | 61/41 | M. Mašek, K. Hoňková, J. Juryšek | FRAM, Nikkor 300mm + G4-16000 |
| ASAS J102612-2000.7 Hya | | | | | | | |
| 56765.56319 | 0.0013 | I | ------- | CCD+R | 87/30 | M. Mašek, K. Hoňková, J. Juryšek | FRAM, Nikkor 300mm + G4-16000 |
| EU Hya | | | | | | | |
| 56711.46521 | 0.0003 | I | -0.0089 | CCD+V | 39/18 | Lehký M. | 0.40-m f/5 + CCD G2-1600 + BVRI |
| 56711.46528 | 0.0005 | I | -0.0088 | CCD+I | 37/17 | Lehký M. | 0.40-m f/5 + CCD G2-1600 + BVRI |
| 56711.46560 | 0.0007 | I | -0.0085 | CCD+B | 31/14 | Lehký M. | 0.40-m f/5 + CCD G2-1600 + BVRI |
| 56711.46620 | 0.0003 | I | -0.0079 | CCD+R | 37/17 | Lehký M. | 0.40-m f/5 + CCD G2-1600 + BVRI |
| | | | show all | | | | |
| V0470 Hya | | | | | | | |
| 56726.35068 | 0.0006 | I | +0.0117 | CCD+Clear | 48/30 | Mašek M. | N150/600mm + CCD Orion SS |
| AA Hyi | | | | | | | |
| 56886.85233 | 0.0001 | I | +0.0133 | CCD+R | 77/38 | M. Mašek, K. Hoňková, J. Juryšek | FRAM, 0.3m SCT + CCD G2-1600 |
| 56886.85235 | 0.0001 | I | +0.0134 | CCD+V | 76/38 | M. Mašek, K. Hoňková, J. Juryšek | FRAM, 0.3m SCT + CCD G2-1600 |
| | | | show all | | | | |
| ASAS J204836-4609.7 Ind | | | | | | | |
| 56803.78157 | 0.0018 | I | ------- | CCD+B | 38/20 | M. Mašek, K. Hoňková, J. Juryšek | FRAM, Nikkor 300mm + G4-16000 |
| 56803.78883 | 0.0012 | I | ------- | CCD+R | 31/14 | M. Mašek, K. Hoňková, J. Juryšek | FRAM, Nikkor 300mm + G4-16000 |
| 56803.79219 | 0.0014 | I | ------- | CCD+V | 32/18 | M. Mašek, K. Hoňková, J. Juryšek | FRAM, Nikkor 300mm + G4-16000 |
| BS Ind | | | | | | | |
| 56847.81901 | 0.0008 | I | -0.0496 | CCD+B | 66/49 | M. Mašek, K. Hoňková, J. Juryšek | FRAM, Nikkor 300mm + G4-16000 |
| 56847.82051 | 0.0005 | I | -0.0481 | CCD+R | 56/43 | M. Mašek, K. Hoňková, J. Juryšek | FRAM, Nikkor 300mm + G4-16000 |
| 56847.82122 | 0.0004 | I | -0.0474 | CCD+I | 52/38 | M. Mašek, K. Hoňková, J. Juryšek | FRAM, Nikkor 300mm + G4-16000 |
| | | | show all | | | | |
| RY Ind | | | | | | | |
| 56785.77134 | 0.0003 | I | +0.0034 | CCD+Clear | 79/29 | C. Quiñones | Schmidt-Cassegrain 14\" + RF + CCD |
| SU Ind | | | | | | | |
| 56803.75487 | 0.0004 | II | +0.0040 | CCD+R | 35/14 | M. Mašek, K. Hoňková, J. Juryšek | FRAM, 0.3m SCT + CCD G2-1600 |
| CzeV161 Lac | § | | | | | | |
| 56499.41071 | 0.0008 | I | ------- | CCD+I | 96/28 | Lehký M. | 0.40-m f/5 + CCD G2-1600 + BVRI |
| 56499.41091 | 0.0008 | I | ------- | CCD+R | 107/30 | Lehký M. | 0.40-m f/5 + CCD G2-1600 + BVRI |
| 56499.54044 | 0.0005 | II | ------- | CCD+R | 107/86 | Lehký M. | 0.40-m f/5 + CCD G2-1600 + BVRI |
| 56499.54225 | 0.0012 | II | ------- | CCD+I | 96/79 | Lehký M. | 0.40-m f/5 + CCD G2-1600 + BVRI |
| CzeV475 Lac | | | | | | | |
| 56574.27009 | 0.0005 | II | ------- | CCD+V | 117/14 | Šmelcer L. | Celestron 355/2460 + CCD G2 1600 |
| 56574.27025 | 0.0001 | II | ------- | CCD+R | 103/10 | Šmelcer L. | Celestron 355/2460 + CCD G2 1600 |
| EM Lac | | | | | | | |
| 56569.53896 | 0.0002 | I | +0.0105 | CCD+R | 36/25 | Šmelcer L. | Celestron 355/2460 + CCD G2 1600 |
| 56569.53978 | 0.0004 | I | +0.0113 | CCD+V | 28/22 | Šmelcer L. | Celestron 355/2460 + CCD G2 1600 |
| 56574.40230 | 0.0002 | II | +0.0092 | CCD+V | 128/83 | Šmelcer L. | Celestron 355/2460 + CCD G2 1600 |
| 56574.40283 | 0.0001 | II | +0.0098 | CCD+R | 124/82 | Šmelcer L. | Celestron 355/2460 + CCD G2 1600 |
| 56597.36178 | 0.0001 | II | +0.0098 | CCD+V | 100/52 | Šmelcer L. | Newton 254/1200 + CCD G2 402 |
| 56597.36189 | 0.0001 | II | +0.0099 | CCD+R | 99/52 | Šmelcer L. | Celestron 355/2460 + CCD G2 1600 |
| | | | show all | | | | |
| LU Lac | | | | | | | |
| 56897.33381 | 0.0003 | II | -0.0115 | CCD+R | 93/31 | Hanžl D. | 0.2-m RL + CCD G2 8300 (DATEL telescope) |
| 56897.48367 | 0.0004 | I | -0.0114 | CCD+R | 93/79 | Hanžl D. | 0.2-m RL + CCD G2 8300 (DATEL telescope) |
| | | | show all | | | | |
| PP Lac | | | | | | | |
| 56857.42980 | 0.0001 | II | +0.0060 | CCD+V | 56/28 | Magris M. | Schmidt-Newton 0.25m F/4, CCD MX 716 |
| SW Lac | | | | | | | |
| 56836.44200 | 0.0005 | II | +0.0223 | DSLR | 74/45 | Medulka T. | Refraktor 90/500 + Canon 450D |
| 56864.50492 | 0.0001 | I | +0.0220 | CCD+R | 155/80 | CCD group in Upice | Newton 150/750 + G2-402 |
| | | | show all | | | | |
| V0430 Lac | | | | | | | |
| 56499.53933 | 0.0003 | I | +0.0333 | CCD+R | 103/85 | Lehký M. | 0.40-m f/5 + CCD G2-1600 + BVRI |
| 56499.53988 | 0.0003 | I | +0.0339 | CCD+I | 100/82 | Lehký M. | 0.40-m f/5 + CCD G2-1600 + BVRI |
| | | | show all | | | | |
| VX Lac | | | | | | | |
| 56563.30226 | 0.0001 | I | +0.0007 | CCD+V | 90/51 | Šmelcer L. | Newton 254/1200 + CCD G2 402 |
| 56563.30249 | 0.0001 | I | +0.0009 | CCD+R | 93/53 | Šmelcer L. | Newton 254/1200 + CCD G2 402 |
| 56650.33611 | 0.0000 | I | +0.0002 | CCD+R | 66/38 | Šmelcer L. | Celestron 355/2460 + CCD G2 1600 |
| 56650.33660 | 0.0001 | I | +0.0007 | CCD+V | 65/38 | Šmelcer L. | Celestron 355/2460 + CCD G2 1600 |
| | | | show all | | | | |
| AM Leo | | | | | | | |
| 56725.48289 | 0.0002 | I | +0.0076 | DSLR | 178/92 | D. Müller, F. Walter | RF Comet finder 20/137, Canon 350 D |
| CE Leo | | | | | | | |
| 56778.28838 | 0.0001 | I | -0.0063 | CCD+Clear | 141/49 | Öğmen Y. | 14"LX200R + SBIG ST-8XME |
| FK Leo | | | | | | | |
| 56719.41142 | 0.0004 | I | -0.0289 | CCD+Clear | 199/72 | Hladík B. | RF 8/200, ATIK 320E, CG-4 |
| GSC 01965-00735 Leo | | | | | | | |
| 56726.40870 | 0.0012 | I | +0.0242 | CCD+Clear | 180/99 | Hladík B. | RF 8/200, ATIK 320E, CG-4 |
| GV Leo | | | | | | | |
| 56630.58922 | 0.0006 | I | -0.0222 | CCD+B | 43/27 | Lehký M. | 0.40-m f/5 + CCD G2-1600 + BVRI |
| 56630.58975 | 0.0004 | I | -0.0217 | CCD+R | 45/27 | Lehký M. | 0.40-m f/5 + CCD G2-1600 + BVRI |
| 56630.58996 | 0.0004 | I | -0.0215 | CCD+V | 43/26 | Lehký M. | 0.40-m f/5 + CCD G2-1600 + BVRI |
| 56630.59068 | 0.0002 | I | -0.0207 | CCD+I | 46/28 | Lehký M. | 0.40-m f/5 + CCD G2-1600 + BVRI |
| 56716.34676 | 0.0014 | II | -0.0186 | CCD+I | 29/9 | Lehký M. | 0.40-m f/5 + CCD G2-1600 + BVRI |
| 56716.35005 | 0.0014 | II | -0.0153 | CCD+B | 25/8 | Lehký M. | 0.40-m f/5 + CCD G2-1600 + BVRI |
| 56716.35027 | 0.0025 | II | -0.0151 | CCD+V | 29/10 | Lehký M. | 0.40-m f/5 + CCD G2-1600 + BVRI |
| 56716.35125 | 0.0011 | II | -0.0141 | CCD+R | 30/10 | Lehký M. | 0.40-m f/5 + CCD G2-1600 + BVRI |
| | | | show all | | | | |
| GW Leo | | | | | | | |
| 56670.68030 | 0.0004 | I | -0.0319 | CCD+R | 45/25 | Lehký M. | 0.40-m f/5 + CCD G2-1600 + BVRI |
| 56670.68130 | 0.0010 | I | -0.0309 | CCD+V | 31/16 | Lehký M. | 0.40-m f/5 + CCD G2-1600 + BVRI |
| 56745.43584 | 0.0008 | II | ------- | CCD+R | 51/18 | Lehký M. | 0.40-m f/5 + CCD G2-1600 + BVRI |
| 56745.43858 | 0.0005 | II | ------- | CCD+V | 55/22 | Lehký M. | 0.40-m f/5 + CCD G2-1600 + BVRI |
| | | | show all | | | | |
| UV Leo | | | | | | | |
| 56731.34753 | 0.0003 | I | +0.0031 | CCD+Clear | 493/238 | Urbaník M. | ED 80/600,0.5x Reductor, G1 0300 |
| 56755.35352 | 0.0001 | I | +0.0056 | CCD+Clear | 309/172 | Urbaník M. | ED 80/600,0.5x Reductor, G1 0300 |
| | | | show all | | | | |
| UZ Leo | | | | | | | |
| 56716.45467 | 0.0002 | I | +0.0122 | CCD+R | 197/111 | Šmelcer L. | Newton 254/1200 + CCD G2 402 |
| 56716.45512 | 0.0003 | I | +0.0126 | CCD+B | 165/97 | Šmelcer L. | Newton 254/1200 + CCD G2 402 |
| 56726.34433 | 0.0003 | I | +0.0130 | CCD+Clear | 296/180 | Urbaník M. | ED 80/600,0.5x Reductor, G1 0300 |
| | | | show all | | | | |
| XY Leo | | | | | | | |
| 56692.60551 | 0.0001 | I | +0.0265 | DSLR | 449/276 | Walter F. | RF Comet finder 20/137, Canon 350 D |
| 56745.44831 | 0.0005 | I | +0.0262 | CCD+Clear | 609/568 | Kuchťák B. | Sonnar 180 + G1 |
| 56778.40350 | 0.0002 | I | +0.0254 | CCD+Clear | 409/226 | Urbaník M. | ED 80/600,0.5x Reductor, G1 0300 |
| | | | show all | | | | |
| Y Leo | | | | | | | |
| 56738.34966 | 0.0001 | I | -0.0229 | CCD+V | 33/13 | Vrašťák M. | 0,24m f/5 RL+CCD G2-1600, pointer 80/400+G1-300 |
| 56738.34979 | 0.0001 | I | -0.0228 | CCD+R | 34/14 | Vrašťák M. | 0,24m f/5 RL+CCD G2-1600, pointer 80/400+G1-300 |
| 56738.34995 | 0.0001 | I | -0.0226 | CCD+I | 34/14 | Vrašťák M. | 0,24m f/5 RL+CCD G2-1600, pointer 80/400+G1-300 |
| | | | show all | | | | |
| IR Lib | | | | | | | |
| 56803.59237 | 0.0006 | I | +0.0654 | CCD+R | 28/20 | M. Mašek, K. Hoňková, J. Juryšek | FRAM, 0.3m SCT + CCD G2-1600 |
| T LMi | | | | | | | |
| 56670.48715 | 0.0002 | I | +0.0042 | CCD+Clear | 248/90 | Zíbar M. | Newton 254/903, G2-0402 (VSES), AG(MMys) on SW 1 |
| BH Lyn | | | | | | | |
| 56588.52120 | 0.0002 | I | +0.0403 | CCD+R | 117/87 | Šmelcer L. | Celestron 355/2460 + CCD G2 1600 |
| 56654.45607 | 0.0002 | I | +0.0401 | CCD+Clear | 40/32 | Bílek F. | Newton 0.2m f4,4 CCD Atik 314L+ |
| 56684.54020 | 0.0002 | I | +0.0404 | CCD+R | 106/51 | Šmelcer L. | Newton 254/1200 + CCD G2 402 |
| 56714.31253 | 0.0002 | I | +0.0406 | CCD+R | 222/53 | Hanžl D. | 0.2-m RL + CCD G2 8300 (DATEL telescope) |
| | | | show all | | | | |
| CF Lyn | | | | | | | |
| 56750.38858 | 0.0001 | I | +0.0102 | CCD+R | 624/372 | Lehký M. | EQ6 + 0.25-m f/4 + CCD ST7 + R |
| CzeV242 Lyn | § | | | | | | |
| 56709.45746 | 0.0003 | I | ------- | CCD+R | 36/14 | Šmelcer L. | Celestron 355/2460 + CCD G2 1600 |
| 56709.45820 | 0.0004 | I | ------- | CCD+V | 31/12 | Šmelcer L. | Celestron 355/2460 + CCD G2 1600 |
| 56709.45881 | 0.0004 | I | ------- | CCD+I | 34/14 | Šmelcer L. | Celestron 355/2460 + CCD G2 1600 |
| 56718.55262 | 0.0005 | I | ------- | CCD+V | 96/65 | Šmelcer L. | Newton 254/1200 + CCD G2 402 |
| 56718.55291 | 0.0004 | I | ------- | CCD+R | 92/61 | Šmelcer L. | Newton 254/1200 + CCD G2 402 |
| EK Lyn | | | | | | | |
| 56701.41948 | 0.0001 | I | -0.0253 | CCD+Clear | 219/99 | Walter F. | RL "MARK" 40/406, SBIG ST10XME |
| EL Lyn | | | | | | | |
| 56718.55446 | 0.0002 | I | -0.0060 | CCD+V | 112/76 | Šmelcer L. | Newton 254/1200 + CCD G2 402 |
| 56718.55473 | 0.0001 | I | -0.0057 | CCD+R | 116/78 | Šmelcer L. | Newton 254/1200 + CCD G2 402 |
| | | | show all | | | | |
| FN Lyn | | | | | | | |
| 56718.30263 | 0.0003 | I | -0.0156 | CCD+R | 132/60 | Lehký M. | EQ6 + 0.25-m f/4 + CCD ST7 + R |
| 56765.38268 | 0.0007 | I | -0.0158 | CCD+Clear | 264/83 | Urbaník M. | 120/420 G1 0300 |
| | | | show all | | | | |
| RZ Lyn | | | | | | | |
| 56683.36183 | 0.0002 | I | -0.0116 | CCD+R | 134/58 | Lehký M. | EQ6 + 0.25-m f/4 + CCD ST7 + R |
| SW Lyn | | | | | | | |
| 56692.48826 | 0.0001 | I | +0.0014 | CCD+B | 141/81 | Šmelcer L. | Newton 254/1200 + CCD G2 402 |
| 56692.48835 | 0.0001 | I | +0.0015 | CCD+R | 139/79 | Šmelcer L. | Newton 254/1200 + CCD G2 402 |
| 56725.33495 | 0.0001 | I | +0.0007 | CCD+R | 79/39 | Lehký M. | 0.40-m f/5 + CCD G2-1600 + BVRI |
| 56725.33518 | 0.0001 | I | +0.0010 | CCD+B | 84/41 | Lehký M. | 0.40-m f/5 + CCD G2-1600 + BVRI |
| 56725.33525 | 0.0001 | I | +0.0010 | CCD+V | 85/41 | Lehký M. | 0.40-m f/5 + CCD G2-1600 + BVRI |
| 56725.33545 | 0.0001 | I | +0.0012 | CCD+I | 79/38 | Lehký M. | 0.40-m f/5 + CCD G2-1600 + BVRI |
| | | | show all | | | | |
| UV Lyn | | | | | | | |
| 56725.37454 | 0.0003 | II | +0.0119 | CCD+R | 109/66 | Šmelcer L. | Newton 254/1200 + CCD G2 402 |
| 56725.37488 | 0.0002 | II | +0.0122 | CCD+B | 108/70 | Šmelcer L. | Newton 254/1200 + CCD G2 402 |
| 56725.37500 | 0.0002 | II | +0.0123 | CCD+V | 110/71 | Šmelcer L. | Newton 254/1200 + CCD G2 402 |
| 56729.31684 | 0.0001 | I | +0.0124 | CCD+R | 395/199 | Lehký M. | EQ6 + 0.25-m f/4 + CCD ST7 + R |
| | | | show all | | | | |
| IP Lyr | | | | | | | |
| 56745.60529 | 0.0007 | II | -0.0018 | CCD+R | 62/43 | Lehký M. | 0.40-m f/5 + CCD G2-1600 + BVRI |
| 56745.60588 | 0.0006 | II | -0.0012 | CCD+V | 61/43 | Lehký M. | 0.40-m f/5 + CCD G2-1600 + BVRI |
| | | | show all | | | | |
| IW Lyr | | | | | | | |
| 56456.40717 | 0.0001 | I | +0.0087 | CCD+R | 76/22 | Lehký M. | 0.40-m f/5 + CCD G2-1600 + BVRI |
| 56456.40724 | 0.0003 | I | +0.0088 | CCD+V | 73/19 | Lehký M. | 0.40-m f/5 + CCD G2-1600 + BVRI |
| 56456.40734 | 0.0001 | I | +0.0089 | CCD+I | 75/21 | Lehký M. | 0.40-m f/5 + CCD G2-1600 + BVRI |
| 56816.54167 | 0.0003 | I | +0.0106 | CCD+V | 40/32 | Lehký M. | 0.40-m f/5 + CCD G2-1600 + BVRI |
| 56816.54197 | 0.0006 | I | +0.0109 | CCD+I | 39/31 | Lehký M. | 0.40-m f/5 + CCD G2-1600 + BVRI |
| 56816.54203 | 0.0003 | I | +0.0110 | CCD+R | 41/33 | Lehký M. | 0.40-m f/5 + CCD G2-1600 + BVRI |
| | | | show all | | | | |
| OT Lyr | | | | | | | |
| 56588.27725 | 0.0009 | I | ------- | CCD+R | 165/94 | Šmelcer L. | Celestron 355/2460 + CCD G2 1600 |
| PV Lyr | | | | | | | |
| 56817.48505 | 0.0005 | I | +0.0085 | CCD+V | 73/50 | Lehký M. | 0.40-m f/5 + CCD G2-1600 + BVRI |
| 56817.48527 | 0.0004 | I | +0.0087 | CCD+R | 80/54 | Lehký M. | 0.40-m f/5 + CCD G2-1600 + BVRI |
| 56817.48659 | 0.0006 | I | +0.0100 | CCD+I | 78/51 | Lehký M. | 0.40-m f/5 + CCD G2-1600 + BVRI |
| | | | show all | | | | |
| V0376 Lyr | | | | | | | |
| 56856.44341 | 0.0031 | II | ------- | CCD+V | 53/29 | Lehký M. | 0.40-m f/5 + CCD G2-1600 + BVRI |
| 56856.44685 | 0.0017 | II | ------- | CCD+I | 48/27 | Lehký M. | 0.40-m f/5 + CCD G2-1600 + BVRI |
| 56856.44985 | 0.0018 | II | ------- | CCD+R | 43/25 | Lehký M. | 0.40-m f/5 + CCD G2-1600 + BVRI |
| V0406 Lyr | | | | | | | |
| 56746.59051 | 0.0002 | II | +0.0141 | CCD+I | 51/27 | Vrašťák M. | 0,24m f/5 RL+CCD G2-1600, pointer 80/400+G1-300 |
| 56746.59155 | 0.0002 | II | +0.0152 | CCD+V | 49/27 | Vrašťák M. | 0,24m f/5 RL+CCD G2-1600, pointer 80/400+G1-300 |
| | | | show all | | | | |
| V0573 Lyr | | | | | | | |
| 56815.50353 | 0.0003 | II | +0.0621 | CCD+R | 135/76 | Lehký M. | EQ6 + 0.25-m f/4 + CCD ST7 + R |
| V0574 Lyr | | | | | | | |
| 56462.44824 | 0.0002 | I | +0.0070 | CCD+I | 62/30 | Lehký M. | 0.40-m f/5 + CCD G2-1600 + BVRI |
| 56462.44835 | 0.0002 | I | +0.0071 | CCD+R | 63/30 | Lehký M. | 0.40-m f/5 + CCD G2-1600 + BVRI |
| 56462.44872 | 0.0002 | I | +0.0074 | CCD+V | 58/25 | Lehký M. | 0.40-m f/5 + CCD G2-1600 + BVRI |
| 56590.27364 | 0.0005 | I | +0.0091 | CCD+R | 26/15 | Šmelcer L. | Celestron 280/1765 + CCD ST7 |
| 56590.27459 | 0.0006 | I | +0.0101 | CCD+V | 24/17 | Šmelcer L. | Celestron 280/1765 + CCD ST7 |
| 56596.28107 | 0.0002 | I | +0.0078 | CCD+R | 85/51 | Šmelcer L. | Celestron 355/2460 + CCD G2 1600 |
| 56596.28158 | 0.0001 | I | +0.0083 | CCD+V | 80/48 | Šmelcer L. | Celestron 355/2460 + CCD G2 1600 |
| 56819.42744 | 0.0008 | I | +0.0098 | CCD+I | 78/28 | Lehký M. | 0.40-m f/5 + CCD G2-1600 + BVRI |
| 56819.42761 | 0.0004 | I | +0.0100 | CCD+R | 83/31 | Lehký M. | 0.40-m f/5 + CCD G2-1600 + BVRI |
| 56819.42862 | 0.0011 | I | +0.0110 | CCD+V | 74/24 | Lehký M. | 0.40-m f/5 + CCD G2-1600 + BVRI |
| 56819.56235 | 0.0003 | II | +0.0081 | CCD+R | 83/75 | Lehký M. | 0.40-m f/5 + CCD G2-1600 + BVRI |
| 56819.56321 | 0.0004 | II | +0.0090 | CCD+V | 74/67 | Lehký M. | 0.40-m f/5 + CCD G2-1600 + BVRI |
| 56819.56328 | 0.0006 | II | +0.0091 | CCD+I | 78/71 | Lehký M. | 0.40-m f/5 + CCD G2-1600 + BVRI |
| | | | show all | | | | |
| V0577 Lyr | | | | | | | |
| 56585.22259 | 0.0008 | I | +0.1476 | CCD+V | 121/8 | Šmelcer L. | Newton 254/1200 + CCD G2 402 |
| 56585.22419 | 0.0009 | I | +0.1492 | CCD+R | 130/5 | Šmelcer L. | Newton 254/1200 + CCD G2 402 |
| 56585.41656 | 0.0004 | II | +0.1466 | CCD+V | 121/107 | Šmelcer L. | Newton 254/1200 + CCD G2 402 |
| 56585.41886 | 0.0005 | II | +0.1489 | CCD+R | 130/108 | Šmelcer L. | Newton 254/1200 + CCD G2 402 |
| | | | show all | | | | |
| V0580 Lyr | | | | | | | |
| 56460.38582 | 0.0002 | II | -0.0095 | CCD+R | 35/14 | Lehký M. | 0.40-m f/5 + CCD G2-1600 + BVRI |
| 56460.38620 | 0.0003 | II | -0.0091 | CCD+V | 33/14 | Lehký M. | 0.40-m f/5 + CCD G2-1600 + BVRI |
| 56460.38624 | 0.0002 | II | -0.0091 | CCD+I | 34/14 | Lehký M. | 0.40-m f/5 + CCD G2-1600 + BVRI |
| 56706.65783 | 0.0003 | II | -0.0106 | CCD+R | 37/19 | Lehký M. | 0.40-m f/5 + CCD G2-1600 + BVRI |
| 56706.65785 | 0.0003 | II | -0.0106 | CCD+V | 39/21 | Lehký M. | 0.40-m f/5 + CCD G2-1600 + BVRI |
| 56706.65823 | 0.0003 | II | -0.0102 | CCD+I | 34/18 | Lehký M. | 0.40-m f/5 + CCD G2-1600 + BVRI |
| 56861.44458 | 0.0003 | I | -0.0133 | CCD+I | 40/7 | Lehký M. | 0.40-m f/5 + CCD G2-1600 + BVRI |
| 56861.44459 | 0.0003 | I | -0.0133 | CCD+R | 38/6 | Lehký M. | 0.40-m f/5 + CCD G2-1600 + BVRI |
| 56861.44461 | 0.0004 | I | -0.0133 | CCD+V | 40/6 | Lehký M. | 0.40-m f/5 + CCD G2-1600 + BVRI |
| | | | show all | | | | |
| V0656 Lyr | | | | | | | |
| 56857.49627 | 0.0002 | I | +0.0272 | CCD+I | 236/126 | Mazanec J. | N400, G2 402 |
| 56857.49629 | 0.0001 | I | +0.0273 | CCD+R | 234/125 | Mazanec J. | N400, G2 402 |
| | | | show all | | | | |
| AU Mon | | | | | | | |
| 56692.67397 | 0.0012 | I | +0.6740 | CCD+V | 130/60 | M. Mašek, K. Hoňková, J. Juryšek | FRAM, 0.3m SCT + CCD G2-1600 |
| 56692.67525 | 0.0007 | I | +0.6753 | CCD+B | 128/57 | M. Mašek, K. Hoňková, J. Juryšek | FRAM, 0.3m SCT + CCD G2-1600 |
| 56692.67689 | 0.0013 | I | +0.6769 | CCD+R | 122/63 | M. Mašek, K. Hoňková, J. Juryšek | FRAM, 0.3m SCT + CCD G2-1600 |
| 56692.67695 | 0.0036 | I | +0.6770 | CCD+I | 112/62 | M. Mašek, K. Hoňková, J. Juryšek | FRAM, 0.3m SCT + CCD G2-1600 |
| | | | show all | | | | |
| GSC 04833-01209 Mon | | | | | | | |
| 56690.62587 | 0.0005 | I | +0.0027 | CCD+R | 47/21 | M. Mašek, K. Hoňková, J. Juryšek | FRAM, 0.3m SCT + CCD G2-1600 |
| 56690.63034 | 0.0006 | I | +0.0071 | CCD+V | 46/23 | M. Mašek, K. Hoňková, J. Juryšek | FRAM, 0.3m SCT + CCD G2-1600 |
| 56690.63056 | 0.0007 | I | +0.0074 | CCD+I | 35/18 | M. Mašek, K. Hoňková, J. Juryšek | FRAM, 0.3m SCT + CCD G2-1600 |
| | | | show all | | | | |
| IZ Mon | | | | | | | |
| 56582.61910 | 0.0004 | I | -0.0096 | CCD+V | 130/46 | Magris M. | Schmidt-Newton 0.25m F/4, CCD Atik Titan |
| 56654.35991 | 0.0004 | I | -0.0108 | CCD+V | 65/18 | Šmelcer L. | Newton 254/1200 + CCD G2 402 |
| 56654.36037 | 0.0006 | I | -0.0104 | CCD+R | 63/17 | Šmelcer L. | Newton 254/1200 + CCD G2 402 |
| 56718.30407 | 0.0002 | I | -0.0107 | CCD+R | 223/50 | Šmelcer L. | Newton 254/1200 + CCD G2 402 |
| | | | show all | | | | |
| KR Mon | | | | | | | |
| 56719.40324 | 0.0008 | I | -0.0022 | CCD+R | 64/43 | Lehký M. | 0.40-m f/5 + CCD G2-1600 + BVRI |
| 56719.40452 | 0.0006 | I | -0.0009 | CCD+I | 59/40 | Lehký M. | 0.40-m f/5 + CCD G2-1600 + BVRI |
| 56719.40469 | 0.0006 | I | -0.0008 | CCD+V | 56/36 | Lehký M. | 0.40-m f/5 + CCD G2-1600 + BVRI |
| 56719.40663 | 0.0006 | I | +0.0012 | CCD+B | 57/39 | Lehký M. | 0.40-m f/5 + CCD G2-1600 + BVRI |
| | | | show all | | | | |
| V0453 Mon | | | | | | | |
| 56692.71870 | 0.0008 | II | -0.0021 | CCD+I | 82/45 | M. Mašek, K. Hoňková, J. Juryšek | FRAM, Nikkor 300mm + G4-16000 |
| 56692.71942 | 0.0005 | II | -0.0014 | CCD+V | 79/41 | M. Mašek, K. Hoňková, J. Juryšek | FRAM, Nikkor 300mm + G4-16000 |
| 56692.71989 | 0.0006 | II | -0.0009 | CCD+B | 63/34 | M. Mašek, K. Hoňková, J. Juryšek | FRAM, Nikkor 300mm + G4-16000 |
| 56692.71989 | 0.0007 | II | -0.0009 | CCD+R | 107/65 | M. Mašek, K. Hoňková, J. Juryšek | FRAM, Nikkor 300mm + G4-16000 |
| | | | show all | | | | |
| V0498 Mon | | | | | | | |
| 56575.79392 | 0.0006 | II | -0.0298 | CCD+V | 106/43 | M. Mašek, K. Hoňková, J. Juryšek | FRAM, 0.3m SCT + CCD G2-1600 |
| 56575.79644 | 0.0004 | II | -0.0273 | CCD+R | 111/40 | M. Mašek, K. Hoňková, J. Juryšek | FRAM, 0.3m SCT + CCD G2-1600 |
| | | | show all | | | | |
| V0532 Mon | | | | | | | |
| 56711.27223 | 0.0008 | I | -0.0195 | CCD+V | 16/9 | Lehký M. | 0.40-m f/5 + CCD G2-1600 + BVRI |
| 56711.27387 | 0.0012 | I | -0.0179 | CCD+I | 18/8 | Lehký M. | 0.40-m f/5 + CCD G2-1600 + BVRI |
| 56711.27531 | 0.0008 | I | -0.0165 | CCD+R | 18/8 | Lehký M. | 0.40-m f/5 + CCD G2-1600 + BVRI |
| 56713.37161 | 0.0003 | II | -0.0221 | CCD+V | 29/16 | Lehký M. | 0.40-m f/5 + CCD G2-1600 + BVRI |
| 56713.37293 | 0.0003 | II | -0.0208 | CCD+R | 30/16 | Lehký M. | 0.40-m f/5 + CCD G2-1600 + BVRI |
| 56713.37362 | 0.0003 | II | -0.0201 | CCD+I | 29/15 | Lehký M. | 0.40-m f/5 + CCD G2-1600 + BVRI |
| | | | show all | | | | |
| V0730 Mon | | | | | | | |
| 56643.46497 | 0.0011 | II | +0.0524 | CCD+Clear | 99/42 | Mašek M. | R70/700mm, 0,5x reducer + CCD Meade DSI |
| V0843 Mon | | | | | | | |
| 56713.38956 | 0.0013 | I | +0.0038 | DSLR | 46/33 | Mašek M. | N150/600mm + Canon EOS 1000D |
| V0882 Mon | | | | | | | |
| 56703.35070 | 0.0001 | I | +0.0155 | CCD+Clear | 105/57 | Mašek M. | N150/600mm + CCD Meade DSI |
| V0922 Mon | | | | | | | |
| 56713.37021 | 0.0002 | I | -0.0067 | DSLR | 203/131 | Mašek M. | N150/600mm + Canon EOS 1000D |
| ASAS J175135-0622.4 Oph | | | | | | | |
| 56498.66346 | 0.0014 | I | ------- | CCD+V | 52/19 | Mašek M. | FRAM, Nikkor 300mm + G4-16000 |
| 56498.66455 | 0.0016 | I | ------- | CCD+R | 70/35 | Mašek M. | FRAM, Nikkor 300mm + G4-16000 |
| 56498.66784 | 0.0024 | I | ------- | CCD+B | 56/23 | Mašek M. | FRAM, Nikkor 300mm + G4-16000 |
| 56498.67124 | 0.0037 | I | ------- | CCD+I | 68/40 | Mašek M. | FRAM, Nikkor 300mm + G4-16000 |
| V0456 Oph | | | | | | | |
| 56843.41097 | 0.0001 | II | -0.0037 | CCD+Clear | 233/102 | Urbaník M. | ED 80/600,0.5x Reductor, G1 0300 |
| V2203 Oph | | | | | | | |
| 56461.52587 | 0.0003 | II | -0.0061 | CCD+R | 34/19 | Lehký M. | 0.40-m f/5 + CCD G2-1600 + BVRI |
| 56461.52602 | 0.0002 | II | -0.0060 | CCD+V | 34/20 | Lehký M. | 0.40-m f/5 + CCD G2-1600 + BVRI |
| 56461.52619 | 0.0002 | II | -0.0058 | CCD+I | 34/19 | Lehký M. | 0.40-m f/5 + CCD G2-1600 + BVRI |
| 56461.52634 | 0.0005 | II | -0.0057 | CCD+B | 34/20 | Lehký M. | 0.40-m f/5 + CCD G2-1600 + BVRI |
| | | | show all | | | | |
| V2377 Oph | | | | | | | |
| 56853.48004 | 0.0012 | II | +0.0126 | DSLR | 117/82 | Medulka T. | Refraktor 90/500 + Canon 450D |
| V2425 Oph | | | | | | | |
| 56798.44392 | 0.0002 | II | -0.0089 | CCD+Clear | 260/97 | M. Mašek, R. Uhlář | N150/750mm + SBIG ST-7 |
| V2713 Oph | | | | | | | |
| 56819.45855 | 0.0002 | I | -0.0087 | CCD+Clear | 108/54 | M. Mašek, J. Filip | N150/750mm + SBIG ST-7 |
| V0647 Ori | | | | | | | |
| 56602.53570 | 0.0002 | I | +0.0061 | CCD+Clear | 207/89 | Magris M. | Schmidt-Newton 0.25m F/4, CCD Atik Titan |
| V2757 Ori | | | | | | | |
| 56695.25778 | 0.0009 | I | +0.0126 | CCD+R | 118/33 | F. Scaggiante, D. Zardin | Newton 410/1710 ccd Audine KAF402ME |
| V2767 Ori | | | | | | | |
| 56629.53048 | 0.0008 | I | +0.0276 | CCD+Clear | 99/54 | Mazanec J. | N400, G2 402 |
| IV Pav | | | | | | | |
| 56805.66986 | 0.0014 | I | ------- | CCD+B | 50/9 | M. Mašek, K. Hoňková, J. Juryšek | FRAM, Nikkor 300mm + G4-16000 |
| 56805.67232 | 0.0012 | I | ------- | CCD+R | 61/9 | M. Mašek, K. Hoňková, J. Juryšek | FRAM, Nikkor 300mm + G4-16000 |
| 56805.67309 | 0.0010 | I | ------- | CCD+I | 61/9 | M. Mašek, K. Hoňková, J. Juryšek | FRAM, Nikkor 300mm + G4-16000 |
| KZ Pav | | | | | | | |
| 56805.83677 | 0.0005 | I | -0.0020 | CCD+I | 62/49 | M. Mašek, K. Hoňková, J. Juryšek | FRAM, Nikkor 300mm + G4-16000 |
| 56805.83682 | 0.0004 | I | -0.0019 | CCD+R | 62/49 | M. Mašek, K. Hoňková, J. Juryšek | FRAM, Nikkor 300mm + G4-16000 |
| 56805.83741 | 0.0005 | I | -0.0013 | CCD+V | 42/32 | M. Mašek, K. Hoňková, J. Juryšek | FRAM, Nikkor 300mm + G4-16000 |
| | | | show all | | | | |
| MW Pav | | | | | | | |
| 56805.76312 | 0.0008 | II | +0.0576 | CCD+I | 55/29 | M. Mašek, K. Hoňková, J. Juryšek | FRAM, Nikkor 300mm + G4-16000 |
| 56805.76430 | 0.0012 | II | +0.0588 | CCD+B | 56/29 | M. Mašek, K. Hoňková, J. Juryšek | FRAM, Nikkor 300mm + G4-16000 |
| 56805.76528 | 0.0014 | II | +0.0598 | CCD+V | 35/18 | M. Mašek, K. Hoňková, J. Juryšek | FRAM, Nikkor 300mm + G4-16000 |
| | | | show all | | | | |
| BB Peg | | | | | | | |
| 56573.42731 | 0.0001 | I | -0.0057 | CCD+Clear | 204/60 | Urbaník M. | ED 80/600,0.5x Reductor, G1 0300 |
| BK Peg | | | | | | | |
| 56594.38555 | 0.0016 | II | -0.0065 | CCD+Clear | 53/24 | Mazanec J. | N400, G2 402 |
| BX Peg | | | | | | | |
| 56888.37706 | 0.0001 | I | +0.0021 | CCD+V | 87/45 | Magris M. | Schmidt-Newton 0.25m F/4, CCD MX 716 |
| GH Peg | | | | | | | |
| 56845.54345 | 0.0003 | I | +0.0064 | CCD+Clear | 98/45 | Hladík B. | RF F200, ATIK 320E, CG-4 |
| GP Peg | | | | | | | |
| 56878.40689 | 0.0002 | I | +0.0012 | CCD+Clear | 181/88 | Urbaník M. | ED 80/600,0.5x Reductor, G1 0300 |
| KW Peg | | | | | | | |
| 56888.35764 | 0.0002 | I | +0.0036 | CCD+V | 87/22 | Magris M. | Schmidt-Newton 0.25m F/4, CCD MX 716 |
| U Peg | | | | | | | |
| 56897.56284 | 0.0004 | II | -0.0066 | DSLR | 76/37 | Mašek M. | Sonnar 180mm + Canon EOS 1000D |
| V0365 Peg | | | | | | | |
| 56871.50597 | 0.0023 | I | -0.0078 | DSLR | 78/57 | Medulka T. | Refraktor 90/500 + Canon 450D |
| V0407 Peg | | | | | | | |
| 56567.34763 | 0.0003 | I | -0.0114 | DSLR | 344/135 | Walter F. | RF Comet finder 20/137, Canon 350 D |
| DK Per | | | | | | | |
| 56609.22956 | 0.0001 | I | -0.0067 | CCD+R | 35/15 | Lehký M. | 0.40-m f/5 + CCD G2-1600 + BVRI |
| 56609.22974 | 0.0001 | I | -0.0065 | CCD+I | 33/13 | Lehký M. | 0.40-m f/5 + CCD G2-1600 + BVRI |
| 56609.22975 | 0.0002 | I | -0.0065 | CCD+V | 37/16 | Lehký M. | 0.40-m f/5 + CCD G2-1600 + BVRI |
| 56693.27621 | 0.0003 | II | -0.0046 | CCD+I | 46/27 | Lehký M. | 0.40-m f/5 + CCD G2-1600 + BVRI |
| 56693.27681 | 0.0003 | II | -0.0040 | CCD+R | 43/25 | Lehký M. | 0.40-m f/5 + CCD G2-1600 + BVRI |
| 56693.27759 | 0.0008 | II | -0.0033 | CCD+V | 48/27 | Lehký M. | 0.40-m f/5 + CCD G2-1600 + BVRI |
| | | | show all | | | | |
| NSVS 1750812 Per | | | | | | | |
| 56180.30401 | 0.0003 | I | ------- | CCD+I | 88/5 | Lehký M. | 0.40-m f/5 + CCD G2-1600 + BVRI |
| 56180.30498 | 0.0006 | I | ------- | CCD+R | 87/6 | Lehký M. | 0.40-m f/5 + CCD G2-1600 + BVRI |
| 56180.49063 | 0.0003 | II | ------- | CCD+R | 87/52 | Lehký M. | 0.40-m f/5 + CCD G2-1600 + BVRI |
| 56180.49185 | 0.0008 | II | ------- | CCD+I | 88/52 | Lehký M. | 0.40-m f/5 + CCD G2-1600 + BVRI |
| PS Per | | | | | | | |
| 56609.45900 | 0.0008 | II | +0.0017 | CCD+R | 145/93 | Lehký M. | EQ6 + 0.25-m f/4 + CCD ST7 + R |
| 56643.51548 | 0.0003 | I | +0.0024 | CCD+R | 101/64 | Lehký M. | EQ6 + 0.25-m f/4 + CCD ST7 + R |
| | | | show all | | | | |
| V0432 Per | | | | | | | |
| 56584.38593 | 0.0004 | II | -0.0033 | CCD+V | 107/59 | Magris M. | Schmidt-Newton 0.25m F/4, CCD MX 716 |
| V0723 Per | | | | | | | |
| 56566.49395 | 0.0002 | I | -0.1067 | CCD+R | 407/310 | Lehký M. | EQ6 + 0.25-m f/4 + CCD ST7 + R |
| V0737 Per | | | | | | | |
| 56630.43171 | 0.0002 | II | -0.0018 | CCD+R | 166/98 | Lehký M. | EQ6 + 0.25-m f/4 + CCD ST7 + R |
| V0873 Per | | | | | | | |
| 56609.38301 | 0.0002 | II | -0.0011 | CCD+R | 59/32 | Lehký M. | 0.40-m f/5 + CCD G2-1600 + BVRI |
| 56609.38307 | 0.0002 | II | -0.0011 | CCD+I | 57/30 | Lehký M. | 0.40-m f/5 + CCD G2-1600 + BVRI |
| 56609.38309 | 0.0003 | II | -0.0010 | CCD+B | 54/28 | Lehký M. | 0.40-m f/5 + CCD G2-1600 + BVRI |
| 56609.38321 | 0.0001 | II | -0.0009 | CCD+V | 58/32 | Lehký M. | 0.40-m f/5 + CCD G2-1600 + BVRI |
| 56629.28852 | 0.0001 | I | -0.0020 | CCD+R | 39/22 | Lehký M. | 0.40-m f/5 + CCD G2-1600 + BVRI |
| 56629.28858 | 0.0001 | I | -0.0020 | CCD+V | 40/22 | Lehký M. | 0.40-m f/5 + CCD G2-1600 + BVRI |
| 56629.28863 | 0.0001 | I | -0.0019 | CCD+I | 39/22 | Lehký M. | 0.40-m f/5 + CCD G2-1600 + BVRI |
| 56629.28919 | 0.0001 | I | -0.0014 | CCD+B | 38/23 | Lehký M. | 0.40-m f/5 + CCD G2-1600 + BVRI |
| | | | show all | | | | |
| V0876 Per | | | | | | | |
| 56630.33446 | 0.0002 | II | -0.0013 | CCD+R | 346/55 | Hanžl D. | 0.2-m RL + CCD G2 8300 (DATEL telescope) |
| 56630.49456 | 0.0004 | I | -0.0012 | CCD+R | 346/225 | Hanžl D. | 0.2-m RL + CCD G2 8300 (DATEL telescope) |
| 56643.29328 | 0.0002 | I | -0.0009 | CCD+R | 444/105 | Hanžl D. | 0.2-m RL + CCD G2 8300 (DATEL telescope) |
| 56643.45239 | 0.0003 | II | -0.0018 | CCD+R | 444/296 | Hanžl D. | 0.2-m RL + CCD G2 8300 (DATEL telescope) |
| 56654.49216 | 0.0003 | I | -0.0007 | CCD+R | 188/152 | Hanžl D. | 0.2-m RL + CCD G2 8300 (DATEL telescope) |
| 56666.33024 | 0.0004 | I | -0.0012 | CCD+R | 194/30 | Hanžl D. | 0.2-m RL + CCD G2 8300 (DATEL telescope) |
| | | | show all | | | | |
| V0885 Per | | | | | | | |
| 56542.51073 | 0.0001 | I | -0.1382 | CCD+R | 114/74 | Lehký M. | EQ6 + 0.25-m f/4 + CCD ST7 + R |
| 56643.24399 | 0.0003 | II | ------- | CCD+R | 224/63 | Lehký M. | EQ6 + 0.25-m f/4 + CCD ST7 + R |
| 56643.39556 | 0.0003 | I | -0.1349 | CCD+R | 224/185 | Lehký M. | EQ6 + 0.25-m f/4 + CCD ST7 + R |
| | | | show all | | | | |
| V0887 Per | | | | | | | |
| 56729.33239 | 0.0007 | II | +0.0073 | CCD+Clear | 342/165 | Urbaník M. | ED 80/600,0.5x Reductor, G1 0300 |
| V0963 Per | | | | | | | |
| 56650.48052 | 0.0004 | I | -0.0116 | CCD+V | 76/53 | Šmelcer L. | Celestron 355/2460 + CCD G2 1600 |
| 56650.48197 | 0.0002 | I | -0.0101 | CCD+R | 75/51 | Šmelcer L. | Celestron 355/2460 + CCD G2 1600 |
| 56718.40625 | 0.0002 | I | -0.0113 | CCD+V | 58/24 | Šmelcer L. | Celestron 355/2460 + CCD G2 1600 |
| 56718.40625 | 0.0001 | I | -0.0113 | CCD+R | 66/28 | Šmelcer L. | Celestron 355/2460 + CCD G2 1600 |
| | | | show all | | | | |
| XZ Per | | | | | | | |
| 56654.46195 | 0.0001 | I | -0.0128 | CCD+V | 42/21 | Šmelcer L. | Celestron 355/2460 + CCD G2 1600 |
| 56654.46225 | 0.0001 | I | -0.0125 | CCD+R | 47/28 | Šmelcer L. | Celestron 355/2460 + CCD G2 1600 |
| 56692.46565 | 0.0001 | I | -0.0130 | CCD+V | 63/28 | Šmelcer L. | Celestron 355/2460 + CCD G2 1600 |
| 56692.46595 | 0.0001 | I | -0.0127 | CCD+R | 56/24 | Šmelcer L. | Celestron 355/2460 + CCD G2 1600 |
| | | | show all | | | | |
| CG Phe | | | | | | | |
| 56885.75022 | 0.0006 | I | ------- | CCD+B | 39/10 | M. Mašek, K. Hoňková, J. Juryšek | FRAM, Nikkor 300mm + G4-16000 |
| 56885.75069 | 0.0010 | I | ------- | CCD+V | 40/12 | M. Mašek, K. Hoňková, J. Juryšek | FRAM, Nikkor 300mm + G4-16000 |
| 56885.75118 | 0.0007 | I | ------- | CCD+R | 40/13 | M. Mašek, K. Hoňková, J. Juryšek | FRAM, Nikkor 300mm + G4-16000 |
| 56885.75313 | 0.0006 | I | ------- | CCD+I | 33/10 | M. Mašek, K. Hoňková, J. Juryšek | FRAM, Nikkor 300mm + G4-16000 |
| zet Phe | | | | | | | |
| 56585.65576 | 0.0011 | II | +0.0044 | DSLR | 179/109 | Benáček J. | Canon 600D |
| SW Phe | | | | | | | |
| 56886.65575 | 0.0015 | II | ------- | CCD+R | 42/19 | M. Mašek, K. Hoňková, J. Juryšek | FRAM, Nikkor 300mm + G4-16000 |
| ASAS J054134-6044.1 Pic | | | | | | | |
| 56883.76058 | 0.0013 | I | ------- | CCD+I | 36/7 | M. Mašek, K. Hoňková, J. Juryšek | FRAM, Nikkor 300mm + G4-16000 |
| 56883.76372 | 0.0007 | I | ------- | CCD+R | 43/8 | M. Mašek, K. Hoňková, J. Juryšek | FRAM, Nikkor 300mm + G4-16000 |
| 56883.76388 | 0.0009 | I | ------- | CCD+V | 40/8 | M. Mašek, K. Hoňková, J. Juryšek | FRAM, Nikkor 300mm + G4-16000 |
| 56883.76797 | 0.0015 | I | ------- | CCD+B | 36/9 | M. Mašek, K. Hoňková, J. Juryšek | FRAM, Nikkor 300mm + G4-16000 |
| WZ Pic | | | | | | | |
| 56895.84214 | 0.0003 | I | +0.0019 | CCD+I | 39/19 | M. Mašek, K. Hoňková, J. Juryšek | FRAM, 0.3m SCT + CCD G2-1600 |
| 56895.84265 | 0.0002 | I | +0.0024 | CCD+V | 51/27 | M. Mašek, K. Hoňková, J. Juryšek | FRAM, 0.3m SCT + CCD G2-1600 |
| 56895.84279 | 0.0002 | I | +0.0025 | CCD+R | 41/23 | M. Mašek, K. Hoňková, J. Juryšek | FRAM, 0.3m SCT + CCD G2-1600 |
| 56895.84294 | 0.0002 | I | +0.0027 | CCD+B | 58/30 | M. Mašek, K. Hoňková, J. Juryšek | FRAM, 0.3m SCT + CCD G2-1600 |
| | | | show all | | | | |
| AQ Psc | | | | | | | |
| 56563.46657 | 0.0001 | I | -0.0054 | DSLR | 79/36 | Mašek M. | N200/1000mm + Canon EOS 1000D |
| GT Psc | | | | | | | |
| 56564.44867 | 0.0003 | I | -0.0252 | CCD+Clear | 417/153 | Trnka J. | 200/1 000, ST-9E |
| HL Psc | | | | | | | |
| 56561.81729 | 0.0016 | II | ------- | CCD+R | 51/35 | M. Mašek, K. Hoňková, J. Juryšek | FRAM, 0.3m SCT + CCD G2-1600 |
| 56561.81796 | 0.0010 | II | ------- | CCD+B | 58/45 | M. Mašek, K. Hoňková, J. Juryšek | FRAM, 0.3m SCT + CCD G2-1600 |
| 56561.81990 | 0.0009 | II | ------- | CCD+V | 63/45 | M. Mašek, K. Hoňková, J. Juryšek | FRAM, 0.3m SCT + CCD G2-1600 |
| 56561.82427 | 0.0011 | II | ------- | CCD+I | 48/35 | M. Mašek, K. Hoňková, J. Juryšek | FRAM, 0.3m SCT + CCD G2-1600 |
| HN Psc | | | | | | | |
| 56584.33150 | 0.0032 | I | -0.0087 | CCD+Clear | 223/113 | Jacobsen J. | EQ6, 80 mm refractor, MX916 |
| SU Psc | | | | | | | |
| 56569.40775 | 0.0005 | I | +0.0017 | CCD+Clear | 75/37 | Mazanec J. | N400, G2 402 |
| CzeV507 Pup | § | | | | | | |
| 56650.66642 | 0.0009 | I | ------- | CCD+V | 76/10 | M. Mašek, K. Hoňková, J. Juryšek | FRAM, 0.3m SCT + CCD G2-1600 |
| 56650.66736 | 0.0006 | I | ------- | CCD+I | 61/11 | M. Mašek, K. Hoňková, J. Juryšek | FRAM, 0.3m SCT + CCD G2-1600 |
| 56691.74055 | 0.0005 | I | ------- | CCD+V | 88/77 | M. Mašek, K. Hoňková, J. Juryšek | FRAM, 0.3m SCT + CCD G2-1600 |
| 56691.74062 | 0.0020 | I | ------- | CCD+I | 76/68 | M. Mašek, K. Hoňková, J. Juryšek | FRAM, 0.3m SCT + CCD G2-1600 |
| 56691.74704 | 0.0004 | I | ------- | CCD+R | 97/82 | M. Mašek, K. Hoňková, J. Juryšek | FRAM, 0.3m SCT + CCD G2-1600 |
| 56691.74953 | 0.0006 | I | ------- | CCD+B | 90/77 | M. Mašek, K. Hoňková, J. Juryšek | FRAM, 0.3m SCT + CCD G2-1600 |
| VY Pup | | | | | | | |
| 56738.30657 | 0.0004 | I | +0.0029 | CCD+Clear | 80/47 | Urbaník M. | ED 80/600,0.5x Reductor, G1 0300 |
| GSC 07875-00125 Sco | | | | | | | |
| 56857.57018 | 0.0013 | I | ------- | CCD+V | 22/14 | M. Mašek, K. Hoňková, J. Juryšek | FRAM, Nikkor 300mm + G4-16000 |
| 56857.57742 | 0.0010 | I | ------- | CCD+B | 33/17 | M. Mašek, K. Hoňková, J. Juryšek | FRAM, Nikkor 300mm + G4-16000 |
| 56857.57768 | 0.0015 | I | ------- | CCD+I | 25/16 | M. Mašek, K. Hoňková, J. Juryšek | FRAM, Nikkor 300mm + G4-16000 |
| 56857.57899 | 0.0013 | I | ------- | CCD+R | 20/10 | M. Mašek, K. Hoňková, J. Juryšek | FRAM, Nikkor 300mm + G4-16000 |
| V0499 Sco | | | | | | | |
| 56813.60919 | 0.0011 | I | +0.0541 | CCD+B | 51/31 | M. Mašek, K. Hoňková, J. Juryšek | FRAM, Nikkor 300mm + G4-16000 |
| 56813.61127 | 0.0007 | I | +0.0562 | CCD+R | 44/27 | M. Mašek, K. Hoňková, J. Juryšek | FRAM, Nikkor 300mm + G4-16000 |
| 56813.61152 | 0.0013 | I | +0.0564 | CCD+V | 39/28 | M. Mašek, K. Hoňková, J. Juryšek | FRAM, Nikkor 300mm + G4-16000 |
| 56813.61164 | 0.0008 | I | +0.0566 | CCD+I | 44/28 | M. Mašek, K. Hoňková, J. Juryšek | FRAM, Nikkor 300mm + G4-16000 |
| | | | show all | | | | |
| V0569 Sco | | | | | | | |
| 56874.65366 | 0.0015 | I | -0.0012 | CCD+B | 38/15 | M. Mašek, K. Hoňková, J. Juryšek | FRAM, Nikkor 300mm + G4-16000 |
| 56874.65393 | 0.0007 | I | -0.0009 | CCD+R | 30/14 | M. Mašek, K. Hoňková, J. Juryšek | FRAM, Nikkor 300mm + G4-16000 |
| 56874.65500 | 0.0007 | I | +0.0001 | CCD+V | 39/16 | M. Mašek, K. Hoňková, J. Juryšek | FRAM, Nikkor 300mm + G4-16000 |
| | | | show all | | | | |
| V0581 Sco | | | | | | | |
| 56812.75316 | 0.0001 | I | +0.0231 | CCD+Clear | 266/196 | C. Quiñones | Reflector Schmidt-Cassegrain 14\" + CCD |
| V0590 Sco | | | | | | | |
| 56857.60393 | 0.0010 | I | +0.0077 | CCD+I | 25/16 | M. Mašek, K. Hoňková, J. Juryšek | FRAM, Nikkor 300mm + G4-16000 |
| 56857.60408 | 0.0008 | I | +0.0079 | CCD+R | 28/17 | M. Mašek, K. Hoňková, J. Juryšek | FRAM, Nikkor 300mm + G4-16000 |
| 56857.60451 | 0.0012 | I | +0.0083 | CCD+B | 31/19 | M. Mašek, K. Hoňková, J. Juryšek | FRAM, Nikkor 300mm + G4-16000 |
| 56857.60523 | 0.0015 | I | +0.0090 | CCD+V | 15/8 | M. Mašek, K. Hoňková, J. Juryšek | FRAM, Nikkor 300mm + G4-16000 |
| | | | show all | | | | |
| V1305 Sco | | | | | | | |
| 56770.76801 | 0.0009 | II | ------- | CCD+I | 43/25 | M. Mašek, K. Hoňková, J. Juryšek | FRAM, Nikkor 300mm + G4-16000 |
| 56770.76874 | 0.0005 | II | ------- | CCD+V | 40/24 | M. Mašek, K. Hoňková, J. Juryšek | FRAM, Nikkor 300mm + G4-16000 |
| 56770.76882 | 0.0004 | II | ------- | CCD+R | 44/27 | M. Mašek, K. Hoňková, J. Juryšek | FRAM, Nikkor 300mm + G4-16000 |
| 56770.77004 | 0.0005 | II | ------- | CCD+B | 39/23 | M. Mašek, K. Hoňková, J. Juryšek | FRAM, Nikkor 300mm + G4-16000 |
| 56813.57903 | 0.0007 | II | ------- | CCD+B | 53/24 | M. Mašek, K. Hoňková, J. Juryšek | FRAM, Nikkor 300mm + G4-16000 |
| 56813.57986 | 0.0009 | II | ------- | CCD+I | 45/21 | M. Mašek, K. Hoňková, J. Juryšek | FRAM, Nikkor 300mm + G4-16000 |
| 56813.58084 | 0.0007 | II | ------- | CCD+R | 45/22 | M. Mašek, K. Hoňková, J. Juryšek | FRAM, Nikkor 300mm + G4-16000 |
| 56813.58233 | 0.0010 | II | ------- | CCD+V | 35/21 | M. Mašek, K. Hoňková, J. Juryšek | FRAM, Nikkor 300mm + G4-16000 |
| AU Ser | | | | | | | |
| 56787.36781 | 0.0001 | I | -0.0007 | CCD+Clear | 177/74 | Urbaník M. | ED 80/600,0.5x Reductor, G1 0300 |
| CC Ser | | | | | | | |
| 56800.32931 | 0.0001 | I | +0.0085 | CCD+V | 283/135 | Öğmen Y. | 14"LX200R + SBIG ST-8XME |
| NSVS 10653195 Ser | | | | | | | |
| 56813.44475 | 0.0001 | I | ------- | CCD+R | 188/90 | Šmelcer L. | Newton 254/1200 + CCD G2 402 |
| 56818.48973 | 0.0002 | I | ------- | CCD+V | 95/68 | Šmelcer L. | Newton 254/1200 + CCD G2 402 |
| 56818.49134 | 0.0002 | I | ------- | CCD+R | 120/92 | Šmelcer L. | Newton 254/1200 + CCD G2 402 |
| 56824.37893 | 0.0001 | I | ------- | CCD+R | 173/34 | Šmelcer L. | Newton 254/1200 + CCD G2 402 |
| 56825.50053 | 0.0001 | II | ------- | CCD+R | 209/192 | Šmelcer L. | Newton 254/1200 + CCD G2 402 |
| V0384 Ser | | | | | | | |
| 56864.38763 | 0.0003 | II | +0.0044 | CCD+R | 103/43 | CCD group in Upice | Cassegrain 300/3500 + FLI |
| V0385 Ser | | | | | | | |
| 56723.62809 | 0.0005 | I | +0.0200 | CCD+Clear | 76/41 | Mašek M. | N150/600mm + CCD Meade DSI |
| WW Sex | | | | | | | |
| 56683.53867 | 0.0003 | I | +0.0031 | CCD+B | 39/16 | Lehký M. | 0.40-m f/5 + CCD G2-1600 + BVRI |
| 56683.53871 | 0.0003 | I | +0.0031 | CCD+I | 41/16 | Lehký M. | 0.40-m f/5 + CCD G2-1600 + BVRI |
| 56683.53898 | 0.0003 | I | +0.0034 | CCD+V | 37/15 | Lehký M. | 0.40-m f/5 + CCD G2-1600 + BVRI |
| 56683.54012 | 0.0004 | I | +0.0045 | CCD+R | 36/13 | Lehký M. | 0.40-m f/5 + CCD G2-1600 + BVRI |
| | | | show all | | | | |
| WY Sex | | | | | | | |
| 56706.52176 | 0.0001 | I | +0.0046 | CCD+Clear | 37/17 | Mašek M. | N150/600mm + CCD Meade DSI |
| 56729.33878 | 0.0009 | I | +0.0023 | CCD+Clear | 187/39 | Jacobsen J. | EQ6, 80 mm refractor, SX-M9 |
| | | | show all | | | | |
| UU Sge | | | | | | | |
| 56854.38704 | 0.0002 | I | +0.0015 | CCD+R | 57/32 | Lehký M. | 0.40-m f/5 + CCD G2-1600 + BVRI |
| AH Tau | | | | | | | |
| 56609.51103 | 0.0001 | I | -0.0027 | CCD+I | 29/15 | Lehký M. | 0.40-m f/5 + CCD G2-1600 + BVRI |
| 56609.51118 | 0.0001 | I | -0.0026 | CCD+R | 27/14 | Lehký M. | 0.40-m f/5 + CCD G2-1600 + BVRI |
| 56609.51136 | 0.0001 | I | -0.0024 | CCD+V | 29/15 | Lehký M. | 0.40-m f/5 + CCD G2-1600 + BVRI |
| 56609.51173 | 0.0002 | I | -0.0020 | CCD+B | 27/14 | Lehký M. | 0.40-m f/5 + CCD G2-1600 + BVRI |
| 56610.34362 | 0.0001 | II | -0.0015 | CCD+I | 33/17 | Lehký M. | 0.40-m f/5 + CCD G2-1600 + BVRI |
| 56610.34368 | 0.0002 | II | -0.0014 | CCD+R | 33/17 | Lehký M. | 0.40-m f/5 + CCD G2-1600 + BVRI |
| 56610.34371 | 0.0002 | II | -0.0014 | CCD+V | 33/19 | Lehký M. | 0.40-m f/5 + CCD G2-1600 + BVRI |
| 56610.34498 | 0.0003 | II | -0.0001 | CCD+B | 34/19 | Lehký M. | 0.40-m f/5 + CCD G2-1600 + BVRI |
| 56692.34625 | 0.0001 | I | -0.0032 | DSLR | 374/225 | Walter F. | RF Comet finder 20/137, Canon 350 D |
| | | | show all | | | | |
| CF Tau | | | | | | | |
| 56654.30986 | 0.0007 | II | -0.0254 | CCD+V | 94/56 | Šmelcer L. | Celestron 355/2460 + CCD G2 1600 |
| 56654.31182 | 0.0003 | II | -0.0234 | CCD+I | 121/59 | Šmelcer L. | Celestron 355/2460 + CCD G2 1600 |
| 56654.31288 | 0.0005 | II | -0.0224 | CCD+R | 123/62 | Šmelcer L. | Celestron 355/2460 + CCD G2 1600 |
| | | | show all | | | | |
| CU Tau | | | | | | | |
| 56597.43474 | 0.0003 | I | -0.0188 | CCD+V | 86/16 | Šmelcer L. | Newton 254/1200 + CCD G2 402 |
| 56597.43481 | 0.0003 | I | -0.0187 | CCD+R | 91/17 | Šmelcer L. | Newton 254/1200 + CCD G2 402 |
| 56643.22050 | 0.0004 | I | -0.0238 | CCD+V | 55/10 | Šmelcer L. | Celestron 355/2460 + CCD G2 1600 |
| 56643.22344 | 0.0004 | I | -0.0209 | CCD+R | 67/11 | Šmelcer L. | Celestron 355/2460 + CCD G2 1600 |
| 56692.31232 | 0.0002 | I | -0.0231 | DSLR | 375/150 | Walter F. | RF Comet finder 20/137, Canon 350 D |
| | | | show all | | | | |
| EQ Tau | | | | | | | |
| 56592.39127 | 0.0001 | II | +0.0041 | CCD+Clear | 137/75 | Urbaník M. | ED 80/600,0.5x Reductor, G1 0300 |
| GQ Tau | | | | | | | |
| 56630.32623 | 0.0001 | I | -0.0146 | CCD+V | 80/44 | Šmelcer L. | Celestron 355/2460 + CCD G2 1600 |
| 56630.32668 | 0.0001 | I | -0.0142 | CCD+R | 82/43 | Šmelcer L. | Celestron 355/2460 + CCD G2 1600 |
| | | | show all | | | | |
| GR Tau | | | | | | | |
| 56692.32429 | 0.0002 | I | -0.0019 | CCD+Clear | 279/159 | Urbaník M. | ED 80/600,0.5x Reductor, G1 0300 |
| RZ Tau | | | | | | | |
| 56574.45690 | 0.0002 | I | +0.0111 | CCD+Clear | 47/16 | Urbaník M. | ED 80/600,0.5x Reductor, G1 0300 |
| 56630.36600 | 0.0003 | II | +0.0116 | CCD+Clear | 156/86 | Urbaník M. | ED 80/600,0.5x Reductor, G1 0300 |
| 56639.30205 | 0.0006 | I | +0.0107 | CCD+Clear | 58/26 | J. Mravik, J. Grnja | SW 150/750 + EQ6 + SBIG ST7 |
| | | | show all | | | | |
| SV Tau | | | | | | | |
| 56671.33529 | 0.0002 | I | -0.0092 | CCD+Clear | 311/131 | Urbaník M. | ED 80/600,0.5x Reductor, G1 0300 |
| V1128 Tau | | | | | | | |
| 56644.34326 | 0.0002 | I | -0.0033 | CCD+Clear | 131/93 | Urbaník M. | ED 80/600,0.5x Reductor, G1 0300 |
| V1239 Tau | | | | | | | |
| 56597.57408 | 0.0004 | I | +0.0056 | CCD+Clear | 84/57 | Mazanec J. | N400, G2 402 |
| V1332 Tau | | | | | | | |
| 56559.60052 | 0.0006 | I | +0.0067 | CCD+V | 61/26 | Magris M. | Schmidt-Newton 0.25m F/4, CCD MX 716 |
| V1370 Tau | | | | | | | |
| 56664.38308 | 0.0002 | I | +0.0012 | CCD+R | 137/24 | F. Scaggiante, D. Zardin | Newton 410/1710 ccd Audine KAF402ME |
| WY Tau | | | | | | | |
| 56730.27798 | 0.0002 | I | -0.0008 | CCD+R | 82/31 | Lehký M. | EQ6 + 0.25-m f/4 + CCD ST7 + R |
| 56731.31662 | 0.0001 | II | -0.0013 | CCD+R | 134/73 | Lehký M. | EQ6 + 0.25-m f/4 + CCD ST7 + R |
| | | | show all | | | | |
| BU Tri | | | | | | | |
| 56566.40906 | 0.0007 | I | +0.0033 | CCD+R | 138/41 | Lehký M. | 0.40-m f/5 + CCD G2-1600 + BVRI |
| 56566.40998 | 0.0011 | I | +0.0043 | CCD+I | 131/32 | Lehký M. | 0.40-m f/5 + CCD G2-1600 + BVRI |
| 56566.55520 | 0.0010 | II | +0.0025 | CCD+I | 131/89 | Lehký M. | 0.40-m f/5 + CCD G2-1600 + BVRI |
| 56566.55611 | 0.0010 | II | +0.0034 | CCD+R | 138/100 | Lehký M. | 0.40-m f/5 + CCD G2-1600 + BVRI |
| | | | show all | | | | |
| BV Tri | | | | | | | |
| 56566.48424 | 0.0022 | II | +0.0005 | CCD+R | 137/75 | Lehký M. | 0.40-m f/5 + CCD G2-1600 + BVRI |
| 56566.48826 | 0.0018 | II | +0.0045 | CCD+I | 143/85 | Lehký M. | 0.40-m f/5 + CCD G2-1600 + BVRI |
| | | | show all | | | | |
| BX Tri | | | | | | | |
| 56557.75664 | 0.0008 | II | -0.0013 | CCD+R | 22/9 | M. Mašek, K. Hoňková, J. Juryšek | FRAM, 0.3m SCT + CCD G2-1600 |
| 56557.75689 | 0.0006 | II | -0.0010 | CCD+I | 23/10 | M. Mašek, K. Hoňková, J. Juryšek | FRAM, 0.3m SCT + CCD G2-1600 |
| 56557.75773 | 0.0008 | II | -0.0002 | CCD+V | 22/10 | M. Mašek, K. Hoňková, J. Juryšek | FRAM, 0.3m SCT + CCD G2-1600 |
| 56643.38475 | 0.0003 | I | -0.0002 | CCD+V | 96/29 | Šmelcer L. | Celestron 355/2460 + CCD G2 1600 |
| 56643.38500 | 0.0003 | I | +0.0001 | CCD+R | 117/41 | Šmelcer L. | Celestron 355/2460 + CCD G2 1600 |
| 56643.47753 | 0.0004 | II | -0.0034 | CCD+R | 117/93 | Šmelcer L. | Celestron 355/2460 + CCD G2 1600 |
| 56643.47983 | 0.0009 | II | -0.0011 | CCD+V | 96/81 | Šmelcer L. | Celestron 355/2460 + CCD G2 1600 |
| 56684.22318 | 0.0002 | I | -0.0005 | CCD+R | 132/8 | Šmelcer L. | Newton 254/1200 + CCD G2 402 |
| 56684.22350 | 0.0005 | I | -0.0002 | CCD+V | 142/7 | Šmelcer L. | Newton 254/1200 + CCD G2 402 |
| 56684.31754 | 0.0004 | II | -0.0022 | CCD+V | 142/63 | Šmelcer L. | Newton 254/1200 + CCD G2 402 |
| 56684.31868 | 0.0004 | II | -0.0010 | CCD+R | 132/64 | Šmelcer L. | Newton 254/1200 + CCD G2 402 |
| 56684.41445 | 0.0006 | I | -0.0019 | CCD+V | 142/126 | Šmelcer L. | Newton 254/1200 + CCD G2 402 |
| 56684.41506 | 0.0002 | I | -0.0013 | CCD+R | 132/117 | Šmelcer L. | Newton 254/1200 + CCD G2 402 |
| 56692.31250 | 0.0003 | I | -0.0019 | CCD+V | 110/46 | Šmelcer L. | Celestron 355/2460 + CCD G2 1600 |
| 56692.31319 | 0.0002 | I | -0.0012 | CCD+R | 108/54 | Šmelcer L. | Celestron 355/2460 + CCD G2 1600 |
| 56693.27692 | 0.0002 | I | -0.0007 | CCD+R | 121/44 | Šmelcer L. | Celestron 355/2460 + CCD G2 1600 |
| 56693.27794 | 0.0003 | I | +0.0003 | CCD+V | 128/47 | Šmelcer L. | Celestron 355/2460 + CCD G2 1600 |
| 56693.37013 | 0.0007 | II | -0.0035 | CCD+V | 128/102 | Šmelcer L. | Celestron 355/2460 + CCD G2 1600 |
| 56693.37020 | 0.0004 | II | -0.0034 | CCD+R | 121/98 | Šmelcer L. | Celestron 355/2460 + CCD G2 1600 |
| 56702.23248 | 0.0003 | II | -0.0024 | CCD+V | 51/9 | Šmelcer L. | Celestron 355/2460 + CCD G2 1600 |
| 56702.23304 | 0.0003 | II | -0.0018 | CCD+R | 74/13 | Šmelcer L. | Celestron 355/2460 + CCD G2 1600 |
| 56702.32950 | 0.0001 | I | -0.0020 | CCD+V | 51/48 | Šmelcer L. | Celestron 355/2460 + CCD G2 1600 |
| 56702.32982 | 0.0003 | I | -0.0017 | CCD+R | 74/68 | Šmelcer L. | Celestron 355/2460 + CCD G2 1600 |
| 56703.29315 | 0.0002 | I | -0.0015 | CCD+V | 57/10 | Šmelcer L. | Celestron 355/2460 + CCD G2 1600 |
| 56703.29361 | 0.0001 | I | -0.0011 | CCD+R | 54/11 | Šmelcer L. | Celestron 355/2460 + CCD G2 1600 |
| 56709.26421 | 0.0003 | I | -0.0022 | CCD+R | 79/22 | Šmelcer L. | Celestron 355/2460 + CCD G2 1600 |
| 56709.26449 | 0.0005 | I | -0.0019 | CCD+V | 65/19 | Šmelcer L. | Celestron 355/2460 + CCD G2 1600 |
| 56709.35722 | 0.0007 | II | -0.0052 | CCD+V | 65/58 | Šmelcer L. | Celestron 355/2460 + CCD G2 1600 |
| 56709.35882 | 0.0005 | II | -0.0036 | CCD+R | 79/73 | Šmelcer L. | Celestron 355/2460 + CCD G2 1600 |
| 56712.24857 | 0.0004 | II | -0.0033 | CCD+V | 158/25 | Šmelcer L. | Celestron 355/2460 + CCD G2 1600 |
| 56712.34635 | 0.0003 | I | -0.0022 | CCD+V | 158/127 | Šmelcer L. | Celestron 355/2460 + CCD G2 1600 |
| 56714.27370 | 0.0004 | I | -0.0012 | CCD+V | 32/22 | Šmelcer L. | Celestron 355/2460 + CCD G2 1600 |
| 56718.31937 | 0.0003 | I | -0.0009 | CCD+V | 58/39 | Šmelcer L. | Celestron 355/2460 + CCD G2 1600 |
| 56718.31943 | 0.0001 | I | -0.0008 | CCD+R | 59/42 | Šmelcer L. | Celestron 355/2460 + CCD G2 1600 |
| 56719.28220 | 0.0005 | I | -0.0012 | CCD+V | 35/14 | Šmelcer L. | Celestron 355/2460 + CCD G2 1600 |
| 56719.28297 | 0.0003 | I | -0.0005 | CCD+R | 29/13 | Šmelcer L. | Celestron 355/2460 + CCD G2 1600 |
| 56721.30225 | 0.0007 | II | -0.0036 | CCD+V | 50/40 | Šmelcer L. | Celestron 355/2460 + CCD G2 1600 |
| 56721.30440 | 0.0006 | II | -0.0014 | CCD+R | 59/46 | Šmelcer L. | Celestron 355/2460 + CCD G2 1600 |
| 56726.31174 | 0.0004 | II | -0.0026 | CCD+V | 44/36 | Šmelcer L. | Celestron 355/2460 + CCD G2 1600 |
| 56726.31282 | 0.0004 | II | -0.0015 | CCD+R | 50/43 | Šmelcer L. | Celestron 355/2460 + CCD G2 1600 |
| | | | show all | | | | |
| RV Tri | | | | | | | |
| 56566.51227 | 0.0002 | I | -0.0052 | CCD+R | 165/104 | Lehký M. | 0.40-m f/5 + CCD G2-1600 + BVRI |
| 56566.51228 | 0.0002 | I | -0.0052 | CCD+I | 164/99 | Lehký M. | 0.40-m f/5 + CCD G2-1600 + BVRI |
| | | | show all | | | | |
| DK Tuc | | | | | | | |
| 56586.66415 | 0.0061 | I | +0.0019 | DSLR | 104/32 | Benáček J. | Canon 600D |
| AA UMa | | | | | | | |
| 56725.38916 | 0.0001 | I | +0.0010 | CCD+R | 47/25 | Poddaný S. | Meade LX 200 40,6 cm, SBIG ST10XME |
| 56725.38923 | 0.0001 | I | +0.0011 | CCD+V | 49/26 | Poddaný S. | Meade LX 200 40,6 cm, SBIG ST10XME |
| 56725.38944 | 0.0001 | I | +0.0013 | CCD+I | 45/25 | Poddaný S. | Meade LX 200 40,6 cm, SBIG ST10XME |
| | | | show all | | | | |
| BH UMa | | | | | | | |
| 56799.45848 | 0.0006 | I | +0.0729 | CCD+Clear | 399/265 | Urbaník M. | ED 80/600,0.5x Reductor, G1 0300 |
| DW UMa | | | | | | | |
| 56692.52060 | 0.0002 | I | -0.0001 | CCD+R | 193/33 | Lehký M. | EQ6 + 0.25-m f/4 + CCD ST7 + R |
| 56692.65670 | 0.0001 | I | -0.0006 | CCD+R | 193/140 | Lehký M. | EQ6 + 0.25-m f/4 + CCD ST7 + R |
| 56718.47584 | 0.0002 | I | -0.0001 | CCD+R | 112/79 | Lehký M. | EQ6 + 0.25-m f/4 + CCD ST7 + R |
| | | | show all | | | | |
| EQ UMa | | | | | | | |
| 56683.70529 | 0.0010 | II | -0.1293 | CCD+R | 162/116 | Hanžl D. | 0.2-m RL + CCD G2 8300 (DATEL telescope) |
| 56754.43967 | 0.0006 | I | -0.1282 | CCD+I | 225/162 | Mazanec J. | N400, G2 402 |
| 56754.44086 | 0.0006 | I | -0.1270 | CCD+Clear | 227/164 | Mazanec J. | N400, G2 402 |
| 56754.44106 | 0.0003 | I | -0.1268 | CCD+R | 246/163 | Mazanec J. | N400, G2 402 |
| | | | show all | | | | |
| HX UMa | | | | | | | |
| 56703.40990 | 0.0006 | I | -0.0022 | CCD+V | 91/55 | Benáček J. | Newton 400 mm, G2-0402 |
| 56703.41000 | 0.0006 | I | -0.0021 | CCD+R | 90/54 | Benáček J. | Newton 400 mm, G2-0402 |
| 56703.41065 | 0.0005 | I | -0.0014 | CCD+I | 97/60 | Benáček J. | Newton 400 mm, G2-0402 |
| | | | show all | | | | |
| IW UMa | | | | | | | |
| 56718.60281 | 0.0009 | II | -0.0014 | CCD+R | 263/204 | Hanžl D. | 0.2-m RL + CCD G2 8300 (DATEL telescope) |
| KM UMa | | | | | | | |
| 56666.57881 | 0.0001 | I | -0.0123 | CCD+R | 22/8 | Lehký M. | 0.40-m f/5 + CCD G2-1600 + BVRI |
| 56666.57903 | 0.0005 | I | -0.0121 | CCD+I | 29/7 | Lehký M. | 0.40-m f/5 + CCD G2-1600 + BVRI |
| 56666.57951 | 0.0003 | I | -0.0116 | CCD+V | 29/10 | Lehký M. | 0.40-m f/5 + CCD G2-1600 + BVRI |
| 56666.57989 | 0.0006 | I | -0.0112 | CCD+B | 25/7 | Lehký M. | 0.40-m f/5 + CCD G2-1600 + BVRI |
| 56717.41472 | 0.0006 | II | -0.0204 | CCD+R | 42/15 | Lehký M. | 0.40-m f/5 + CCD G2-1600 + BVRI |
| 56717.41780 | 0.0010 | II | -0.0173 | CCD+B | 49/19 | Lehký M. | 0.40-m f/5 + CCD G2-1600 + BVRI |
| 56717.41860 | 0.0008 | II | -0.0165 | CCD+I | 49/20 | Lehký M. | 0.40-m f/5 + CCD G2-1600 + BVRI |
| 56717.41873 | 0.0010 | II | -0.0163 | CCD+V | 46/16 | Lehký M. | 0.40-m f/5 + CCD G2-1600 + BVRI |
| 56745.39754 | 0.0002 | I | -0.0104 | DSLR | 352/250 | Walter F. | RF Comet finder 20/137, Canon 350 D |
| | | | show all | | | | |
| LP UMa | | | | | | | |
| 56692.58667 | 0.0005 | I | +0.0304 | CCD+R | 193/90 | Lehký M. | EQ6 + 0.25-m f/4 + CCD ST7 + R |
| 56718.46133 | 0.0008 | II | +0.0285 | CCD+R | 116/72 | Lehký M. | EQ6 + 0.25-m f/4 + CCD ST7 + R |
| | | | show all | | | | |
| NSVS 02502726 UMa | | | | | | | |
| 56747.30090 | 0.0001 | I | ------- | CCD+I | 207/13 | Šmelcer L. | Celestron 355/2460 + CCD G2 1600 |
| 56747.30131 | 0.0001 | I | ------- | CCD+R | 202/16 | Šmelcer L. | Celestron 355/2460 + CCD G2 1600 |
| 56747.58000 | 0.0002 | II | ------- | CCD+R | 202/186 | Šmelcer L. | Celestron 355/2460 + CCD G2 1600 |
| 56747.58037 | 0.0002 | II | ------- | CCD+I | 207/186 | Šmelcer L. | Celestron 355/2460 + CCD G2 1600 |
| NSVS 697648 Uma | | | | | | | |
| 56729.33738 | 0.0005 | I | ------- | CCD+V | 43/20 | Vrašťák M. | 0,24m f/5 RL+CCD G2-1600, pointer 80/400+G1-300 |
| 56729.33579 | 0.0004 | I | ------- | CCD+R | 42/17 | Vrašťák M. | 0,24m f/5 RL+CCD G2-1600, pointer 80/400+G1-300 |
| 56729.33792 | 0.0004 | I | ------- | CCD+I | 41/17 | Vrašťák M. | 0,24m f/5 RL+CCD G2-1600, pointer 80/400+G1-300 |
| OT UMa | | | | | | | |
| 56729.35120 | 0.0003 | I | -0.0684 | CCD+R | 43/23 | Vrašťák M. | 0,24m f/5 RL+CCD G2-1600, pointer 80/400+G1-300 |
| 56729.35145 | 0.0003 | I | -0.0681 | CCD+I | 41/22 | Vrašťák M. | 0,24m f/5 RL+CCD G2-1600, pointer 80/400+G1-300 |
| 56729.35190 | 0.0003 | I | -0.0677 | CCD+V | 43/24 | Vrašťák M. | 0,24m f/5 RL+CCD G2-1600, pointer 80/400+G1-300 |
| | | | show all | | | | |
| QQ UMa | | | | | | | |
| 56730.41180 | 0.0003 | I | +0.0489 | CCD+R | 317/195 | Hanžl D. | 0.2-m RL + CCD G2 8300 (DATEL telescope) |
| QT UMa | | | | | | | |
| 56728.35996 | 0.0002 | II | +0.0011 | CCD+V | 22/7 | Corfini G. | GSO 200/800 SBIG STT1603ME |
| 56728.36017 | 0.0001 | II | +0.0013 | CCD+V | 22/7 | Corfini G. | GSO 200/800 CCD-UAI |
| | | | show all | | | | |
| TY UMa | | | | | | | |
| 56737.35745 | 0.0001 | I | +0.0172 | CCD+V | 44/23 | Vrašťák M. | 0,24m f/5 RL+CCD G2-1600, pointer 80/400+G1-300 |
| 56737.35763 | 0.0001 | I | +0.0174 | CCD+I | 43/24 | Vrašťák M. | 0,24m f/5 RL+CCD G2-1600, pointer 80/400+G1-300 |
| 56737.35775 | 0.0001 | I | +0.0175 | CCD+R | 44/24 | Vrašťák M. | 0,24m f/5 RL+CCD G2-1600, pointer 80/400+G1-300 |
| | | | show all | | | | |
| UX UMa | | | | | | | |
| 56794.31985 | 0.0001 | I | -0.0033 | CCD+Clear | 184/88 | Öğmen Y. | 14"LX200R + SBIG ST-8XME |
| V0337 UMa | | | | | | | |
| 56712.37414 | 0.0001 | II | +0.0043 | CCD+R | 62/35 | Benáček J. | Newton 400 mm, G2-0402 |
| 56726.36792 | 0.0003 | I | +0.0050 | CCD+R | 270/142 | Hanžl D. | 0.2-m RL + CCD G2 8300 (DATEL telescope) |
| | | | show all | | | | |
| V0356 UMa | | | | | | | |
| 56719.53106 | 0.0007 | I | -0.0066 | CCD+R | 200/155 | Hanžl D. | 0.2-m RL + CCD G2 8300 (DATEL telescope) |
| V0358 UMa | | | | | | | |
| 56727.52201 | 0.0005 | I | -0.0186 | CCD+R | 292/84 | Hanžl D. | 0.2-m RL + CCD G2 8300 (DATEL telescope) |
| AS Vel | | | | | | | |
| 56594.83421 | 0.0013 | I | -0.0069 | CCD+R | 43/33 | M. Mašek, K. Hoňková, J. Juryšek | FRAM, Nikkor 300mm + G4-16000 |
| FO Vir | | | | | | | |
| 56745.57461 | 0.0016 | I | -0.0254 | CCD+Clear | 125/39 | Hladík B. | RF 200/8, ATIK 320E, CG-4 |
| GSC 00330-01394 Vir | | | | | | | |
| 56726.66121 | 0.0003 | I | +0.0148 | CCD+Clear | 70/58 | Mašek M. | N150/600mm + CCD Meade DSI |
| GR Vir | | | | | | | |
| 56805.54039 | 0.0004 | I | +0.0110 | CCD+I | 44/23 | M. Mašek, K. Hoňková, J. Juryšek | FRAM, Nikkor 300mm + G4-16000 |
| 56805.54088 | 0.0005 | I | +0.0115 | CCD+B | 45/22 | M. Mašek, K. Hoňková, J. Juryšek | FRAM, Nikkor 300mm + G4-16000 |
| 56805.54102 | 0.0003 | I | +0.0116 | CCD+R | 40/17 | M. Mašek, K. Hoňková, J. Juryšek | FRAM, Nikkor 300mm + G4-16000 |
| 56805.54295 | 0.0004 | I | +0.0136 | CCD+V | 22/13 | M. Mašek, K. Hoňková, J. Juryšek | FRAM, Nikkor 300mm + G4-16000 |
| | | | show all | | | | |
| HW Vir | | | | | | | |
| 56718.53218 | 0.0000 | I | -0.0004 | CCD+R | 118/58 | Šmelcer L. | Celestron 355/2460 + CCD G2 1600 |
| 56718.53225 | 0.0000 | I | -0.0003 | CCD+V | 94/42 | Šmelcer L. | Celestron 355/2460 + CCD G2 1600 |
| 56718.59042 | 0.0001 | II | -0.0001 | CCD+R | 118/113 | Šmelcer L. | Celestron 355/2460 + CCD G2 1600 |
| 56718.59054 | 0.0005 | II | -0.0000 | CCD+V | 94/88 | Šmelcer L. | Celestron 355/2460 + CCD G2 1600 |
| 56723.55118 | 0.0000 | I | -0.0003 | CCD+Clear | 137/92 | Mašek M. | N150/600mm + CCD Meade DSI |
| | | | show all | | | | |
| IK Vir | | | | | | | |
| 56725.53700 | 0.0002 | I | -0.0058 | CCD+Clear | 80/41 | Mašek M. | N150/600mm + CCD Meade DSI |
| | | | | | | | |
| NSVS 10441882 Vir | | | | | | | |
| 56832.41989 | 0.0002 | II | ------- | CCD+R | 71/36 | Šmelcer L. | Newton 254/1200 + CCD G2 402 |
| 56846.36977 | 0.0001 | II | ------- | CCD+R | 109/49 | Šmelcer L. | Newton 254/1200 + CCD G2 402 |
| | | | | | | | |
| BK Vul | | | | | | | |
| 56579.27929 | 0.0004 | I | -0.0121 | CCD+V | 31/23 | Šmelcer L. | Newton 254/1200 + CCD G2 402 |
| | | | | | | | |
| BO Vul | | | | | | | |
| 56588.36515 | 0.0001 | I | -0.0045 | CCD+V | 57/22 | Šmelcer L. | Celestron 280/1765 + CCD ST7 |
| 56588.36521 | 0.0001 | I | -0.0045 | CCD+R | 60/25 | Šmelcer L. | Celestron 280/1765 + CCD ST7 |
| 56629.22860 | 0.0001 | I | -0.0043 | CCD+V | 134/39 | Šmelcer L. | Newton 254/1200 + CCD G2 402 |
| 56629.22865 | 0.0001 | I | -0.0043 | CCD+R | 122/40 | Šmelcer L. | Newton 254/1200 + CCD G2 402 |
| | | | show all | | | | |
| BU Vul | | | | | | | |
| 56563.41313 | 0.0004 | II | +0.0032 | CCD+R | 114/58 | Šmelcer L. | Newton 254/1200 + CCD G2 402 |
| 56563.41647 | 0.0004 | II | +0.0066 | CCD+V | 108/60 | Šmelcer L. | Newton 254/1200 + CCD G2 402 |
| 56565.39681 | 0.0001 | I | -0.0051 | DSLR | 270/125 | Walter F. | RF Comet finder 20/137, Canon 350 D |
| 56569.37966 | 0.0001 | I | -0.0052 | CCD+R | 243/171 | Šmelcer L. | Newton 254/1200 + CCD G2 402 |
| 56878.33922 | 0.0002 | I | -0.0091 | DSLR | 76/31 | Walter F. | RF Comet finder 20/137, Canon 350 D |
| | | | show all | | | | |
| GP Vul | | | | | | | |
| 56887.45672 | 0.0004 | II | -0.0053 | CCD+R | 93/16 | Šmelcer L. | Celestron 280/1765 + CCD G2 4000 |
| 56887.45710 | 0.0003 | II | -0.0049 | CCD+V | 85/18 | Šmelcer L. | Celestron 280/1765 + CCD G2 4000 |
| 56887.45840 | 0.0004 | II | -0.0036 | CCD+Clear | 82/17 | Šmelcer L. | Celestron 280/1765 + CCD G2 4000 |
| | | | show all | | | | |
| IM Vul | | | | | | | |
| 56855.45542 | 0.0003 | I | +0.0162 | CCD+I | 76/40 | Šmelcer L. | Celestron 280/1765 + CCD G2 4000 |
| 56855.45625 | 0.0003 | I | +0.0171 | CCD+R | 80/46 | Šmelcer L. | Celestron 280/1765 + CCD G2 4000 |
| 56855.45631 | 0.0003 | I | +0.0171 | CCD+V | 77/42 | Šmelcer L. | Celestron 280/1765 + CCD G2 4000 |
| 56855.45687 | 0.0002 | I | +0.0177 | CCD+Clear | 77/44 | Šmelcer L. | Celestron 280/1765 + CCD G2 4000 |
| | | | show all | | | | |

## Section 2

**Table 2 – List of observers and corresponding amount of obtained times of minima**

| # | Observer(s) name(s) | Minima times total |
|---|---|---|
| 1 | Lehký M. | 482 |
| 2 | Šmelcer L. | 438 |
| 3 | Mašek M., Hoňková K., Juryšek J. | 159 |
| 4 | Mazanec J. | 90 |
| 5 | Hanžl D. | 76 |
| 6 | Urbaník M. | 37 |
| 7 | Magris M. | 31 |
| 8 | Vrašťák M. | 29 |
| 9 | Walter F. | 23 |
| 10 | Mašek M. | 19 |
| 11 | Hladík B. | 13 |
| 12 | Medulka T. | 11 |
| 13 | Bílek F. | 10 |
| 14 | Trnka J. | 9 |
| 15 | Jacobsen J. | 8 |
| 16 | Benáček J. | 7 |
| 17 | Kuchťák B. | 6 |
| 18 | Audejean M. | 4 |
| 19 | Öğmen Y. | 4 |
| 20 | Zíbar M. | 4 |
| 22 | Fatka P. | 3 |
| 23 | Marchi F. | 3 |
| 24 | Poddaný S. | 3 |
| 25 | Quiñones C., Tapia L. | 3 |
| 26 | Scaggiante F., Zardin D. | 3 |
| 27 | Corfini G. | 2 |
| 28 | Lehký M., Hajek P. | 2 |
| 29 | Lomoz F. | 2 |
| 30 | Mravik J., Grnja J. | 2 |
| 31 | Campos F. | 1 |
| 32 | Čaloud J. | 1 |
| 33 | Esseiva N. | 1 |
| 34 | Hájek P., Lehký M. | 1 |
| 35 | Jakš S., Horník M. | 1 |
| 36 | Mašek M., Filip, J. | 1 |
| 37 | Mašek M., Uhlář R. | 1 |
| 38 | Mina, F., Artola, R., Zalazar, J. | 1 |
| 39 | Müller D., Walter F. | 1 |
| 40 | Pintr P. | 1 |
| 41 | Walter F., Divišová L. | 1 |

**Table 3 - List of stars without GCVS designation**

| Star | Const. | Coordinates (RA DEC) |
|---|---|---|
| ASAS J030827-1609.1 | Eri | 03 08 27 -16 09 06 |
| ASAS J054134-6044.1 | Pic | 05 41 34 -60 44 06 |
| ASAS J102420-1833.4 | Hya | 10 24 20 -18 33 24 |
| ASAS J102528-1911.0 | Hya | 10 25 28 -19 11 00 |
| ASAS J102612-2000.7 | Hya | 10 26 12 -20 00 42 |
| ASAS J105115-6032.1 | Car | 10 51 15 -60 32 06 |
| ASAS J114951-6101.1 | Cen | 11 49 51 -61 01 06 |
| ASAS J115108-6355.1 | Cen | 11 51 08 -63 55 06 |
| ASAS J175135-0622.4 | Oph | 17 51 35 -06 22 24 |
| ASAS J204836-4609.7 | Ind | 20 48 36 -46 09 42 |
| CzeV161 | Lac | 22 27 04 +44 45 59 |
| CzeV242 | Lyn | 08 24 46 +40 31 32 |
| CzeV267 | And | 02 04 28 +46 46 58 |
| CzeV475 | Lac | 22 25 02 +53 57 15 |
| CzeV507 | Pup | 08 26 56 -39 02 36 |
| GJ 3236 | Cas | 03 37 14 +69 10 50 |
| GSC 00330-01394 | Vir | 14 56 16 +04 02 24 |
| GSC 01965-00735 | Leo | 09 36 31 +28 20 23 |
| GSC 02016-00444 | Boo | 14 47 25 +22 50 11 |
| GSC 02594-00971 | Her | 17 04 20 +33 12 50 |
| GSC 03557-01334 | Cyg | 19 49 08 +46 07 05 |
| GSC 04833-1209 | Mon | 07 56 48 -00 39 60 |
| GSC 07875-00125 | Sco | 16 47 51 -41 58 42 |
| GSC 08976-03898 | Cen | 11 42 01 -62 33 16 |
| GSC 08977-00474 | Cen | 11 47 36 -63 10 26 |
| GSC 08977-01661 | Cen | 11 51 54 -62 39 54 |
| NSVS 01031772 | Cam | 13 45 34 +79 23 48 |
| NSVS 01286630 | Dra | 18 47 09 +78 42 34 |
| NSVS 02502726 | UMa | 08 44 11 +54 23 47 |
| NSVS 10123419 | Cnc | 08 42 06 +21 26 12 |
| NSVS 10441882 | Vir | 13 30 25 +13 49 32 |
| NSVS 10653195 | Ser | 16 07 28 +12 13 58 |
| NSVS 1622436 | Cas | 00 24 16 +60 34 54 |
| NSVS 1750812 | Per | 01 34 57 +54 16 35 |
| NSVS 2690221 | Dra | 13 07 21 +65 06 45 |
| NSVS 697648 | UMa | 08 17 38 +66 14 12 |
| ROTSE1 J170438.01+330348.3 | Her | 17 04 38 +33 03 48 |
| ROTSE1 J173413.59+440118.5 | Her | 17 34 14 +44 01 19 |
| SvkV020 | And | 00 36 35 +42 18 19 |
| USNO-B1.0 1135-0102876 | Gem | 06 11 56 +23 30 30 |
| USNO-B1.0 1197-0128756 | Aur | 06 31 09 +29 45 19 |
| VSX J062606.6+275559 | Aur | 06 26 06 +27 55 59 |
| VSX J201516.1+645805 | Dra | 20 15 16 +64 58 06 |

**Table 4 – List of DSLR observers**

| Observer name | Minima times totals |
|---|---|
| F. Walter | 19 |
| T. Medulka | 11 |
| M. Mašek | 5 |
| J. Benáček | 3 |
| F. Walter, D. Müller | 1 |
| F. Walter, L. Divišová | 1 |
| S. Jakš, M. Horník | 1 |
| P. Pintr | 1 |

## Notes on individual stars

*A list of variable stars from the CzeV and SvkV catalogue.*

There is a list of stars with known and previously published light elements, which have been discovered by the Czech and Slovak observers. New minima of these stars are presented in Table 1.

**CzeV267 And**
An EW type eclipsing binary at coordinates RA (2000) = 02 04 28.76, DEC (2000) = +46 46 58.8, 15.8 - 16.5 mag (Clear), M = 2455806.5350+0.2954*E (*J.Trnka*, VSX).

**SvkV020 And**
An EA/RS type eclipsing binary at coordinates RA (2000) = 00 36 35.70, DEC (2000) = +42 18 19.0, 11.5 – 11.8 mag, M = 2455473.839 + 1.18552*E (M. Vrašťák, 2011).

**CzeV161 Lac**
An EW type eclipsing binary at coordinates RA (2000) = 22 27 04.28, DEC (2000) = +44 45 59.7, 14.84 – 15.38 mag, M = 2455068.50679 + 0.2576355*E (M. Lehký, 2009).

**CzeV242 Lyn**
An EW type eclipsing binary at coordinates RA (2000) = 08 24 45.87, DEC (2000) +40 31 31.8 discovered by L. Šmelcer in 2011. The light elements M = 2454500.51600 + 0.298226*E were described by A. V. Khruslov (Khruslov, 2013).

**CzeV507 Pup**
An EA type eclipsing binary at coordinates RA (2000) = 08 26 55.60, DEC (2000) = -39 02 36.2, 13.00 – 14.05 (V) mag, M = 2456650.66450 + 0.746804*E (*M. Mašek, K. Hoňková, J. Juryšek*, VSX).

*A list of selected interesting stars*

**GJ 3236 Cas**
In the Fig. 1, there is a light curve of interesting eclipsing eruptive star GJ 3236 Cas. L. Šmelcer has captured brightening just before the minimum. There are 27 times of minima of this star in the Table 1.

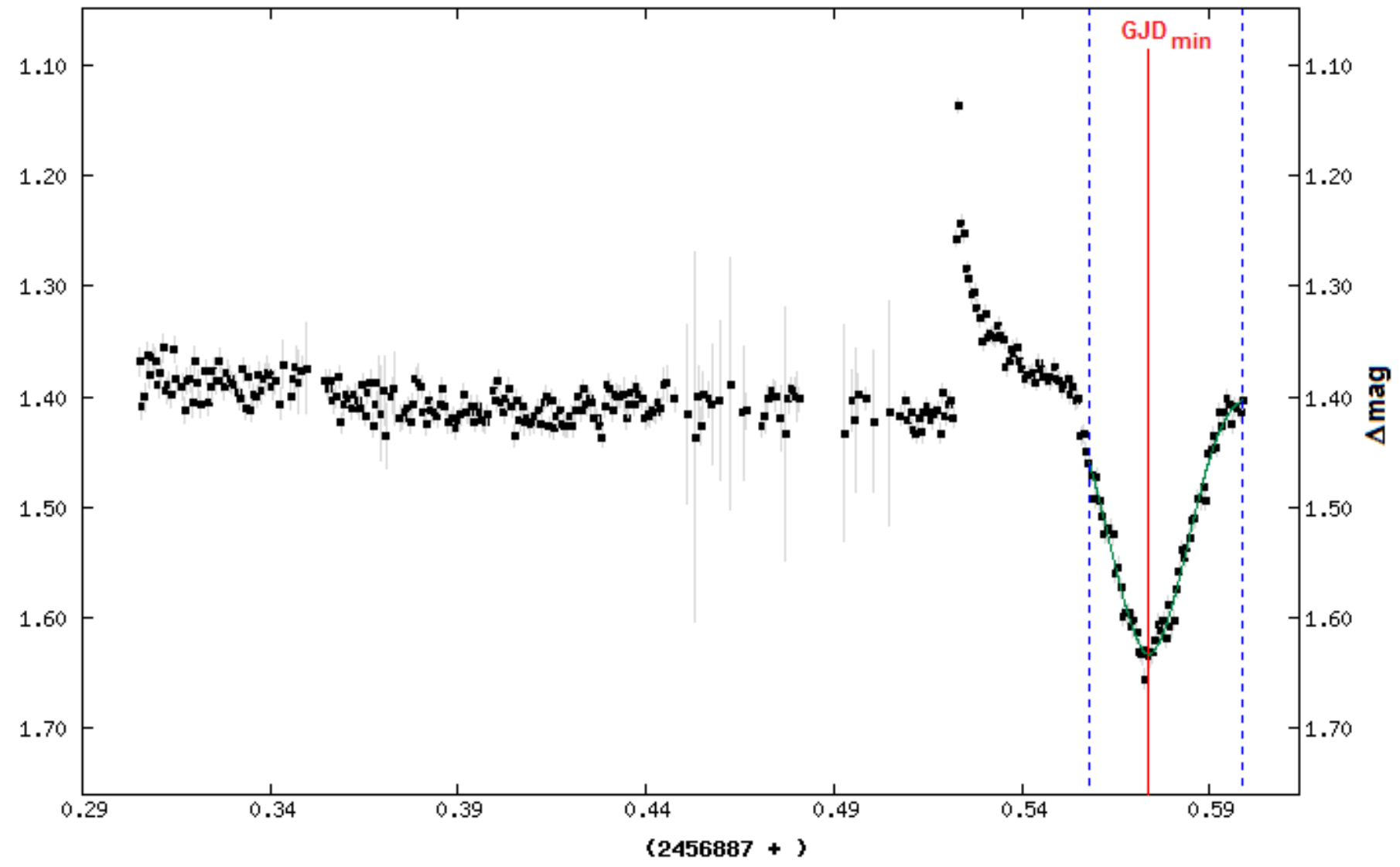

**Figure 1.** Light curve with eruption and eclipse of GJ 3236 Cas (L. Šmelcer).

## Aknowledgements

We are grateful to all the observers for their contributions, and we would like to give special thanks to Luboš Brát for technical support of the server var.astro.cz and Anton Paschke for maintaining the O-C gateway which was used to get light elements of observed stars. This research was made using the SIMBAD database, maintained by CDS, Strasbourg, France. We often used the International Variable Star Index (VSX) database, operated at AAVSO, Cambridge, Massachusetts, USA. The operation of the robotic telescope FRAM is supported by the EU grant GLORIA (No. 283783 in FP7-Capacities program) and by the grant of the Ministry of Education of the Czech Republic (MSMT-CR LG13007).